\documentclass[conference]{IEEEtran}
\usepackage[linesnumbered,ruled,vlined]{algorithm2e}
\usepackage{xspace}
\usepackage{amsthm,amssymb,amsmath,epsfig,graphics,enumerate,float,subfigure}
\usepackage{epstopdf}
\usepackage{balance}
\usepackage{times}
\usepackage{color}
\usepackage{setspace, url}
\def\BibTeX{{\rm B\kern-.05em{\sc i\kern-.025em b}\kern-.08em
		T\kern-.1667em\lower.7ex\hbox{E}\kern-.125emX}}

\newcommand{\ignore}[1]{}
\newcommand{\nop}[1]{}
\newcommand{\eat}[1]{}
\newcommand{\kw}[1]{{\ensuremath{\mathsf{#1}}}\xspace}
\newcommand{\kwnospace}[1]{{\ensuremath {\mathsf{#1}}}}
\newcommand{\stitle}[1]{\vspace{1ex} \noindent{\bf #1}}
\long\def\comment#1{}

\newtheorem{definition}{Definition}

\newtheorem{example}{Example}
\newtheorem{observation}{Observation}

\newtheorem{theorem}{Theorem}

\newcommand{\hbz}{\kwnospace{h}\textrm{-}\kwnospace{LB}\textrm{+}\kw{UB}}
\newcommand{\khc}{\kw{KHC}}
\newcommand{\khcs}{\kw{KHCS}}
\newcommand{\khcore}{\kw{KHCore}}
\newcommand{\khcoresamp}{\kw{KHCoreSamp}}
\newcommand{\updatenbr}{\kw{UpdateHNbr}}
\newcommand{\bitmapupdate}{\kw{BmUpdateHNBr}}
\newcommand{\updatehnbrsamp}{\kw{UpdateHNbrSamp}}
\newcommand{\core}{\kw{core}}
\newcommand{\dgcore}{$(k,h)$-core\xspace}
\newcommand{\hdegree}{$h$-hop degree\xspace}
\newcommand{\hdegrees}{$h$-hop degrees\xspace}
\newcommand{\hneighbor}{$h$-hop neighborhood\xspace}

\newcommand{\reach}{\kw{Reach}}
\newcommand{\bitmap}{\kw{bitmap}}
\newcommand{\bitmaps}{\kw{bitmaps}}
\newcommand{\mydiv}{\kw{div}}
\newcommand{\mymod}{\kw{mod}}
\newcommand{\myand}{\kw{and}}
\newcommand{\select}{\kw{sec}}
\newcommand{\rate}{\kw{rate}}
\newcommand{\cnt}{\kw{cnt}}
\newcommand{\infy}{\kw{INF}}

\newcommand{\urlsnap}{\url{http://snap.stanford.edu/data}}
\newcommand{\urlkonect}{\url{http://konect.uni-koblenz.de}}
\newcommand{\urlnet}{\url{http://networkrepository.com}}

\newcommand{\caas}{\kw{Ca}-\kw{As}}
\newcommand{\comamazon}{\kw{Amazon}}
\newcommand{\douban}{\kw{Douban}}
\newcommand{\hyves}{\kw{Hyves}}
\newcommand{\socLJ}{\kw{SocLJ}}
\newcommand{\socytb}{\kw{Socytb}}
\newcommand{\itcnr}{\kw{Cnr2000}}
\newcommand{\bioce}{\kw{BioCE}}
\newcommand{\bioworm}{\kw{BioWorm}}
\newcommand{\socpokec}{\kw{Pokec}}
\newcommand{\soceps}{\kw{SocEps}}
\newcommand{\flic}{\kw{Flickr}}

\begin{document}
	\title{Scaling Up Distance-generalized Core Decomposition\\
	}

	\author{\IEEEauthorblockN{Qiangqiang Dai\textsuperscript{$*$}, Rong-Hua Li\textsuperscript{$*$}, Lu Qin\textsuperscript{$\dagger$}, Guoren Wang\textsuperscript{$*$}, Weihua Yang\textsuperscript{$\ddagger$}, Zhiwei Zhang\textsuperscript{$*$}, Ye Yuan\textsuperscript{$*$}}
		\IEEEauthorblockA{\textit{\textsuperscript{$*$}Beijing Institute of Technology, Beijing, China; \textsuperscript{$\dagger$}University of Technology Sydney, Australia;}\\
			\textit{\textsuperscript{$\ddagger$}Taiyuan University of Technology, Taiyuan, China}\\
			qiangd66@gmail.com; lironghuabit@126.com; Lu.Qin@uts.edu.au; wanggrbit@126.com; \\yangweihua@tyut.edu.cn; cszwzhang@comp.hkbu.edu.hk; yuan-ye@bit.edu.cn}
	}	
	

	\maketitle

	\begin{abstract}
		Core decomposition is a fundamental operator in network analysis. In this paper, we study a problem of computing distance-generalized core decomposition on a network. A distance-generalized core, also termed $(k, h)$-core, is a maximal subgraph in which every vertex has at least $k$ other vertices at distance no larger than $h$. The state-of-the-art algorithm for solving this problem is based on a peeling technique which iteratively removes the vertex (denoted by $v$) from the graph that has the smallest \hdegree. The \hdegree of a vertex $v$ denotes the number of other vertices that are reachable from $v$ within $h$ hops. Such a peeling algorithm, however, needs to frequently recompute the \hdegrees of $v$'s neighbors after deleting $v$, which is typically very costly for a large $h$. To overcome this limitation, we propose an efficient peeling algorithm based on a novel \hdegree updating technique. Instead of recomputing the \hdegrees, our algorithm can dynamically maintain the \hdegrees for all vertices via exploring a very small subgraph, after peeling a vertex. We show that such an \hdegree updating procedure can be efficiently implemented by an elegant \bitmap technique. In addition, we also propose a sampling-based algorithm and a parallelization technique to further improve the efficiency. Finally, we conduct extensive experiments on 12 real-world graphs to evaluate our algorithms. The results show that, when $h\ge 3$, our exact and sampling-based algorithms can achieve up to $10\times$ and $100\times$ speedup over the state-of-the-art algorithm, respectively.
	\end{abstract}

	\section{Introduction} \label{sec:introduction}
	Many real-world networks such as social networks, biological networks, and collaboration networks often contain cohesive subgraph structures. Finding cohesive subgraphs from a network is a fundamental problem in networks analysis which has attracted much attention in recent years \cite{11iorcliquerelaxations,11generalizedcoredecomposition,12vldbtruss, 15sigmodPlexes,19sigmodBonchiKS}. A variety of cohesive subgraph models have been proposed, such as maximal clique \cite{73BKalgmaximalclique,11todsmaximalclique}, $k$-plex \cite{78JMSplex,15sigmodPlexes}, $k$-truss \cite{05trusses,12vldbtruss,14sigmodtrusscommunity}, and $k$-core \cite{83kcoredef}. Among of them, $k$-core is the most appealing model, because it can be computed in linear time \cite{03omalgkcore}. However, computing cohesive subgraphs based on the other models is often very costly. As a consequence, the $k$-core model has been widely used in many application domains, including community discovery \cite{11asunamcorecommunities,14sigmodcommunitylocalserach}, network topology analysis \cite{07pnasmodelingcore}, protein complex modeling \cite{03gipredictionprotein,03bmcbimolecularcomplexes}, and network visualization \cite{05nipsfingerprintingusingcore} \cite{12icdevisualizingmotifs}.
	
	The $k$-core of a graph $G$ is defined as a maximal subgraph in which every vertex has a degree at least $k$ within that subgraph. Although it is commonly used in practice, the $k$-core model sometimes cannot detect cohesive subgraphs. For example, let us consider a graph shown in Fig.~\ref{fig:example}. Intuitively, the subgraph induced by the vertices $\{v_8, v_9, \cdots, v_{14}\}$ is a cohesive subgraph. Such a cohesive subgraph, however, cannot be identified by the $k$-core model. This is because the entire graph is 2-core, and we cannot distinguish the cohesive subgraph and the entire graph based on different $k$ values using the $k$-core model.
	
	To overcome this limitation, Bonchi et al.\ \cite{19sigmodBonchiKS} recently proposed a distance-generalized $k$-core concept, called $(k,h)$-core, where $k$ and $h$ ($h\ge 1$) are two integer parameters. Specifically, the $(k,h)$-core is a maximal subgraph in which every vertex has at least $k$ other vertices with distance at most $h$ within that subgraph. As indicated in \cite{19sigmodBonchiKS}, such a distance-generalized $k$-core model can detect cohesive subgraphs that cannot be found by the traditional $k$-core model. Reconsider the graph in Fig.~\ref{fig:example}.  Suppose that $h=2$. We can easily verify that the subgraph induced by $\{v_8,v_9, \cdots, v_{14}\}$ is a $(6,2)$-core, while the entire graph is a $(4, 2)$-core. Therefore, we are able to apply the $(k,h)$-core model to identify the cohesive subgraph induced by $\{v_8,v_9, \cdots, v_{14}\}$.
	
	In this paper, we focus on the problem of computing all $(k, h)$-cores on a graph $G$ for a given parameter $h$. Such a problem is also called $(k, h)$-core decomposition. The $(k, h)$-core decomposition has many applications in practice. As shown in \cite{19sigmodBonchiKS}, the $(k, h)$-core decomposition can be used to speed up the computation of finding the maximum $h$-club on a graph; It can also be used to find a good approximation for the distance-generalized densest subgraph problem.
	
	To compute the $(k, h)$-core decomposition, Bonchi et al.\ \cite{19sigmodBonchiKS} proposed a peeling algorithm which iteratively removes the vertex that has the smallest \hdegree until all vertices are deleted. Here the \hdegree of a vertex $v$ is defined as the number of other vertices that are reachable from $v$ within $h$ hops. The defect of such a peeling algorithm is that it needs to recompute the \hdegrees for all vertices in $v$'s \hneighbor when peeling a vertex $v$, which is often costly for a large $h$. Here the \hneighbor of $v$, denoted by $N_v^h(G)$, is a set of other vertices that are reachable from $v$ within $h$ hops. Bonchi et al.\ \cite{19sigmodBonchiKS} also developed an improved algorithm with several lower and upper bounding techniques to alleviate such \hdegree re-computation costs. However, as shown in our experiments, such an improved peeling algorithm is still very costly for $h \ge 3$ on large graphs, because the algorithm may still need to frequently recompute the \hdegrees.
	
	To circumvent this issue, we propose an efficient peeling algorithm, called \khcore, based on a novel \hdegree updating technique. Specifically, when peeling a vertex $v$, we prove that the \hdegree for each vertex in $N_v^h(G)$ can be updated by exploring a small subgraph induced by  $N_v^h(G)$. Based on this key result, we devise the \khcore algorithm which does not recompute the \hdegrees for all vertices in $N_v^h(G)$, but it updates the \hdegrees for every vertex in $N_v^h(G)$ by only accessing a small subgraph induced by $N_v^h(G)$, thus it is very efficient in practice. We also develop an elegant \bitmap technique to implement the \hdegree updating procedure which not only improves the efficiency, but it also reduces the space usage of our algorithm. In addition, a sampling-based algorithm is also presented to further improve the efficiency. To scale to larger graphs, we also propose a parallelization strategy to parallelize our algorithms for $(k, h)$-core decomposition. Finally, we conduct extensive experiments using 12 real-world datasets to evaluate the proposed algorithms. The results show that, if $h\ge 3$, our exact and sampling-based algorithms (with a sampling rate $r=0.1$) using the \bitmap technique can achieve up to $10\times$ and $100\times$ acceleration over the state-of-the-art algorithm. The results also show that the proposed sampling-based algorithm is very accurate. The average accuracy of our sampling-based algorithm is no less than 98\% on most graphs with a sampling rate $r = 0.1$, when $h \ge 3$. To summarize, the main contributions of this paper are as follows.
	\begin{itemize}
		\item {\bf A new algorithm.} We propose a new peeling algorithm, called \khcore, for $(k,h)$-core decomposition. The appealing feature of \khcore is that it can update the \hdegrees for all vertices in $N_v^h(G)$ when peeling a vertex $v$ by exploring a small subgraph induced by $N_v^h(G)$, without recomputing the \hdegrees for all vertices in $N_v^h(G)$.
		\item {\bf Optimization techniques.} We develop a \bitmap technique, a sampling-based algorithm, and a parallelization strategy to improve the efficiency and scalability of \khcore.
		\item {\bf Extensive experiments.} We make use of 12 large real-world datasets to evaluate our algorithms, and the results demonstrate the efficiency and scalability of our algorithms.
        \item {\bf Reproducibility.} For reproducibility purpose, we release the source code of this paper at \url{https://github.com/BITDataScience/khcore}.
	\end{itemize}
	
	\stitle{Organization.} The rest of this paper is organized as follows. Section~\ref{sec:problemsatement} describes the $(k,h)$-core model and the problem statement. Section~\ref{sec:exist-solution} introduces existing algorithms for $(k,h)$-core decomposition. All our algorithms are presented in Section~\ref{sec:ouralgorithm}. The experimental results are reported in Section~\ref{sec:experiment}. Finally, we survey the related work and conclude this paper in Section~\ref{sec:relatedwork} and Section~\ref{sec:conclusion} respectively.
	
	\section{Problem Statement} \label{sec:problemsatement}
	
	\begin{figure}[t!]
		\begin{center}
			\begin{tabular}[t]{c}
				\includegraphics[width=0.6\columnwidth]{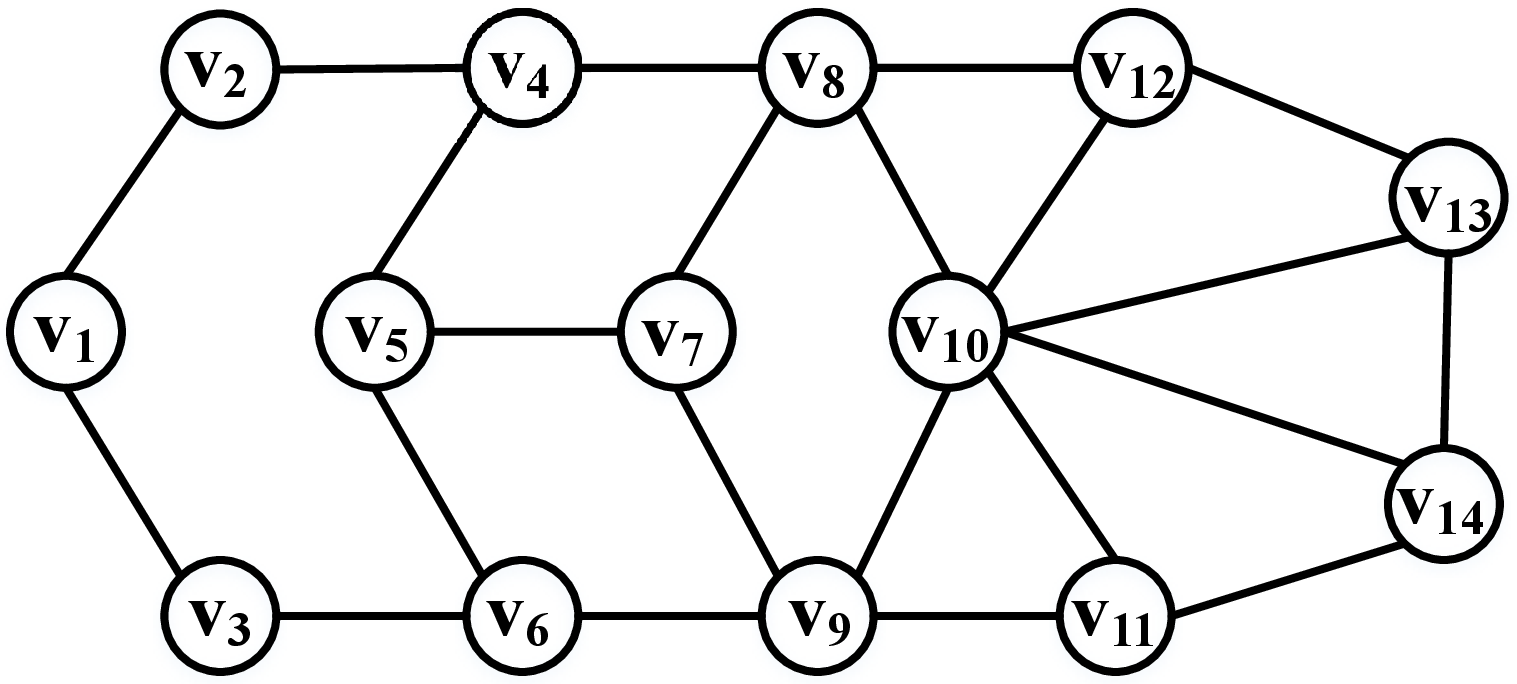}
			\end{tabular}
		\end{center}
		\vspace*{-0.4cm}
		\caption{Running example}
		\vspace*{-0.5cm}
		\label{fig:example}
	\end{figure}
	
	In this paper, we focus on an undirected and unweighted graph $G=(V,E)$, where $V$ is the set of vertices and $E $ is the set of edges. Let  $n = |V|$ and $m = |E|$ be the number of vertices and edges respectively. For each vertex $v$, the neighborhood of $v$, denoted by $N_v(G)$, is defined as $N_{v}(G) \triangleq \{u \in V|(v,u) \in E \}$. The degree of a vertex $v$ in $G$, denoted by  $d_{v}(G)$, is the cardinality of $N_{v}(G)$, i.e., $d_{v}(G) = |N_{v}(G)|$. Let $G(S) = (S, E(S))$ be an induced subgraph of $G$ if $S \subseteq V$ and $E(S) = \{(u,v)|(u,v) \in E, u \in S, v \in S\} $. According to \cite{83kcoredef}, a $k$-core of a graph $G$ is defined as follows.
	
	\begin{definition}[$k$-core] \label{def:core}
		Given a graph $G$, the $k$-core of $G$, denoted by $C_{k}$, is a maximal subgraph of $G$ in which every vertex has a degree at least $k$, i.e., $\forall v \in C_{k},\; d_{v}(C_{k}) \geq k $.
	\end{definition}
	Based on Definition~\ref{def:core}, the core number of a vertex $v$, denoted by $\core(v)$, is the largest integer $k$ such that there is a $k$-core containing $v$. Denote by $k_{\max}$ the maximum $k$ value such that a $k$-core of $G$ exists, i.e., the maximum core number. It is easy to verify that the $k$-cores satisfy a containment property, i.e.,  $C_{k+1} \subset C_{k}$ for all $1\le k<k_{\max}$. The core decomposition of $G$ is a problem of computing the core numbers for all vertices in $G$. Note that the core decomposition of a graph $G$ can be computed in linear time by a classic peeling algorithm \cite{03omalgkcore}, which iteratively removes the minimum-degree node in $G$ using an elegant bin-sort data structure.
	
	Similar to the definition of $k$-core, Bonchi et al.\ \cite{19sigmodBonchiKS} recently introduced a distance-generalized $k$-core notion, called $(k, h)$-core,  based on the $h$-hop degrees of the vertices. Specifically, we denote by $dis_{G}(u,v)$ the shortest-path distance between $u$ and $v$ in $G$. Given a positive integer $h$, the $h$-hop neighborhood of a vertex $v$  in $G$ is defined as $N_{v}^{h}(G) \triangleq \{u|u\ne v, u \in V, dis_{G}(u,v) \leq h \}$. The $h$-hop degree of a vertex $v$ in $G$, denoted by $d_{v}^{h}(G)$, is the cardinality of $N_{v}^{h}(G)$, i.e., $d_{v}^{h}(G) = |N_{v}^{h}(G)|$.
	\begin{definition}[($k$,$h$)-core]
		\label{def:k-h-core} Given a graph $G$ and two integers $k$ and $h$ ($h>0$), the $(k, h)$-core of $G$ is a maximal subgraph $C_{k}^{h}$ such that every vertex $v$ in $C_{k}^{h}$ has an $h$-hop degree at least $k$, i.e., $\forall v \in C_{k}^{h},\;d_{v}^{h}(C_{k}^{h}) \geq k $.
	\end{definition}
	
	It is worth noting that in Definition~\ref{def:k-h-core}, the $h$-hop degree for each vertex in $C_{k}^{h}$ is defined on the subgraph $C_{k}^{h}$ (not on the original graph $G$). When $h=1$, we can easily show that the $(k,h)$-core is the same as the traditional $k$-core.
	
	As shown in \cite{19sigmodBonchiKS}, the ($k$,$h$)-core of a graph $G$ is unique for any positive integer $h$. For a positive integer $h$, the $(k, h)$-core number of a vertex $v$, denoted by $\core_h(v)$, is the largest integer $k$ such that there is a $(k, h)$-core containing $v$. Let $k_{\max}^h$ be the maximum $k$ value such that a $(k, h)$-core of $G$ exists, i.e., the maximum $(k, h)$-core number of $G$. Then, similar to the traditional $k$-cores, the $(k, h)$-cores of $G$ also satisfy a containment property, i.e., $C_{k+1}^{h} \subset C_{k}^{h}$ for all $1\le k<k_{\max}^h$.
	
		\begin{example}
			Consider the graph $G$ in  Fig.~\ref{fig:example}. Clearly, the entire graph is a 2-core, as all vertices in this graph have degrees no less than 2. Suppose that $h=2$. Then, we can see that the subgraph $G(S)$ induced by $S=\{v_8, v_9, \cdots, v_{14}\}$ is a $(6,2)$-core. This is because each vertex in $G(S)$ has an \hdegree no less than 6, and there is no other subgraph that contains $G(S)$ and satisfies the \hdegree constraint (i.e., every vertex has an \hdegree no less than 6). Similarly, we can easily check that the subgraph induced by $\{v_4, v_5, \cdots, v_{14}\}$ is a $(5, 2)$-core, and the entire graph is a $(4, 2)$-core. Given $h=2$, the \dgcore numbers of $\{v_1, v_2, v_3\}$, $\{v_4, v_5, v_6, v_7\}$, and $\{v_8, v_9, \cdots, v_{14}\}$ are 4,  5, 6, respectively.
		\end{example}
	
	For a positive integer $h$, the distance-generalized core decomposition of $G$ is a problem of determining the $(k, h)$-core numbers for all vertices in $G$.  Below, we formally define our problem.
	
	\stitle{Problem statement.} Given a graph $G$ and a positive integer $h$, our goal is to compute the $(k, h)$-core number for each vertex in $G$.
	
	\section{Existing Solutions} \label{sec:exist-solution}
	In this section, we introduce several existing solutions proposed in \cite{19sigmodBonchiKS} to compute the $(k,h)$-core decomposition. Similar to the traditional core decomposition algorithm, the $(k,h)$-core decomposition algorithm proposed in \cite{19sigmodBonchiKS} is also based on a \emph{peeling} idea. In particular, the peeling algorithm iteratively removes the vertex with the smallest $h$-hop degree and sets the \dgcore number as its \hdegree at the time of removal. The detailed procedure of the peeling algorithm is shown in Algorithm~\ref{alg:basic-algorithm}.
	
	The algorithm first computes the \hdegree for each vertex $v \in V$ (line 3), and uses a bucketing array $B$ to maintain all the vertices in $V$ that have the same \hdegree (line 4).  Then, the algorithm iteratively deletes the vertices in $V$ based on the non-decreasing order of the \hdegrees of the vertices (lines~5-12). Specifically,  in the $k$-th iteration, the algorithm sequentially removes each vertex $v$ in $B[k]$ (the \hdegrees of $v$ is equal to $k$) and sets its \dgcore numbers as $k$ (lines~6-8). After that, the algorithm updates the \hdegrees of the vertices in $v$'s \hneighbor ($N_{v}^{h}(G)$), because the \hdegrees of the vertices in $N_{v}^{h}(G)$ may need to update after removing $v$. For each $u \in N_{v}^{h}(G)$, the algorithm first recomputes the \hdegree of $u$ in the reduced subgraph $G(V\setminus \{v\})$ (line~10), and then moves $u$ into $B[\max\{k,d_{u}^{h}(G(V \setminus \{v\}))\}]$ if necessary. It is easy to see that the number of iterations of the algorithm is at most $n$, as the \hdegrees of the vertices in $G$ are bounded by $n$.  The time complexity of Algorithm~\ref{alg:basic-algorithm} is  $O(n\tilde n(\tilde n+\tilde m))$ \cite{19sigmodBonchiKS}, where $\tilde n$ and $\tilde m$ are the number of vertices and edges of the largest subgraph induced by the $h$-hop neighborhood of a vertex in $V$, respectively.
	
	As analyzed in \cite{19sigmodBonchiKS}, the most time-consuming step in Algorithm~\ref{alg:basic-algorithm} is to recompute the \hdegrees of all the vertices in $N_{v}^{h}(G)$ when deleting a vertex $v$. To speed up the algorithm, Bonchi et al.\ \cite{19sigmodBonchiKS} proposed two improved algorithms based on lower and upper bounding techniques, called $h$-LB and $h$-LB+UB respectively. In particular, the $h$-LB algorithm first estimates the lower bound of the \dgcore number for each vertex. Then, based on the lower bounds, the $h$-LB algorithm can avoid a number of useless \hdegree re-computations for the vertices whose lower bounds are no less than the \hdegree of the current removed vertex \cite{19sigmodBonchiKS}. The $h$-LB+UB algorithm also leverages an upper bound of the \dgcore number for each vertex to further improve the efficiency. Specifically, the algorithm first applies the upper bounds of vertices to partition the graph into several nested subgraphs. Then, the algorithm invokes $h$-LB to compute $(k,h)$-cores in the induced subgraph $G(V[i])$ following a top-down manner, where $V[i]$ denotes a set of vertices with upper bounds no less than $i$. As shown in \cite{19sigmodBonchiKS},  the $h$-LB+UB algorithm is the state-of-the-art algorithm for computing the \dgcore decomposition.
	
	\begin{algorithm}[t]
		\caption{The basic peeling algorithm \cite{19sigmodBonchiKS}}
		\label{alg:basic-algorithm}
		\KwIn{a graph $G=(V,E)$ and a positive integer $h$}
		\KwOut{$\core_h(v)$ for all $v\in V$}
		Initialize $B[v]\gets \emptyset $ for each $v\in V$\;
		\For {$v \in V$} {
			Compute $d_{v}^{h}(G)$\;
			$B[d_{v}^{h}(G)] \gets B[d_{v}^{h}(G)] \cup \{v\} $\;
		}
		\For {$k=1$ to $n$} {
			\While {$B[k] \neq \emptyset$} {
				Pick and remove a vertex $v$ from $B[k]$\;
				$\core_h(v) \gets k$\;
				\For {$u \in N_{v}^{h}(G) $} {
					Compute $d_{u}^{h}(G(V \setminus \{v\}))$\;
					Move $u$ to $B[\max\{k,d_{u}^{h}(G(V \setminus \{v\}))\}]$\;
				}
				$V \gets V \setminus \{v\} $\;
			}
		}
		{\bf return } $\core_h(v)$ for all $v\in V$\;
	\end{algorithm}
	
	\stitle{Limitations of the existing solutions.} Although the $h$-LB+UB algorithm is more efficient than the basic peeling algorithm, it is still very costly for handling medium-sized graphs given that $h \ge 3$. For example, as reported in \cite{19sigmodBonchiKS}, the $h$-LB+UB algorithm takes nearly one hour to compute the \dgcore decomposition on the social network \douban (154,908 vertices and 327,162 edges) when $h=4$. The main defect of the $h$-LB+UB algorithm is that the algorithm still needs to frequently recompute the \hdegrees of the vertices when peeling a vertex. For a relatively large $h$ value (e.g., $h\ge 3$), the time overheads for recomputing \hdegrees can be very high on large graphs. To circumvent this issue, in the following sections, we will propose several efficient algorithms which can dynamically update the \hdegrees of the vertices when peeling a vertex, instead of recomputing the \hdegrees. Due to the efficient \hdegree updating technique, the proposed algorithms are much faster than the state-of-the-art $h$-LB+UB algorithm as confirmed in our experiments.
	
	\section{The proposed algorithms} \label{sec:ouralgorithm}
	In this section, we propose several efficient $(k,h)$-core decomposition algorithms based on a novel \hdegree updating technique. Below, we first introduce the basic version of our $(k,h)$-core decomposition algorithm. Then, we will develop a \bitmap technique to improve the time and space overheads of our basic algorithm. Finally, we will propose a more efficient sampling-based algorithm, as well as a parallelization technique to further improve the efficiency and scalability of the $(k,h)$-core decomposition algorithms.
	
	\subsection{The basic \hdegree updating algorithm} \label{subsec:basic-alg}
	Recall that the most time-consuming step in Algorithm~\ref{alg:basic-algorithm} is to recompute the \hdegrees of the vertices in $N_v^h(G)$ after peeling $v$ (lines~9-10 of Algorithm~\ref{alg:basic-algorithm}). To alleviate the computational costs, we propose a novel \hdegree updating technique based on the following key observations.
	
	Note that when deleting $v$, only the vertices in $N_v^h(G)$ may need to update their \hdegrees. For any vertex $u\notin N_v^h(G)$, its \hdegree keeps unchanged after removing $v$. For a vertex $u\in N_v^h(G)$, the question is how can we efficiently update the \hdegree of $u$ after deleting $v$,  without recomputing its \hdegree on $G(V\setminus \{v\})$ (i.e., $d_u^h(G(V\setminus \{v\}))$). Clearly, after deleting $v$, the \hdegree of $u$ may reduce by more than 1 if $h>1$. In order to derive the exact gap between $d_u^h(G)$ and $d_u^h(G(V\setminus \{v\}))$, it is sufficient to consider the vertices in $N_v^{h-s}(G) \cup \{v\}$, where $s=dis_G(u,v)$ is the shortest-path distance between $u$ and $v$ in $G$ ($s\le h$). Below, we give two key observations. 
	
	\begin{observation}
		\label{Obs:1}Given a positive integer $h \in \mathbb{N^{+}}$ and a vertex $u \in N_{v}^{h}(G) $, we have $S_u = N_u^{h}(G) \setminus (N_v^{h-s}(G) \cup \{v\}) \subseteq N_u^{h}(G(V\setminus\{v\}))$ for $s\le h$.
	\end{observation}
	
	\begin{proof}
	Clearly, for any vertex $w \in S_u$, we have $dis_G(w,v) > h-s$ by definition. To prove the observation, we consider two disjoint subsets of $S_u$: $A=\{w|w\in S_u, dis_G(w,v)>h\}$ and $B=\{w|w\in S_u,h-s < dis_G(w,v)\leq h\}$. First, we claim that for any vertex $w \in A$, we have $w \in N_u^{h}(G(V\setminus\{v\}))$. Since $w\in S_u \subseteq N_u^h(G)$, we have $dis_G(w, u) \le h < dis_G(w, v)$. That is to say, there does not exist any shortest path between $u$ and $w$ that passes through $v$. Therefore, after deleting $v$ from $G$, the shortest-path distance between $w$ and $u$ does not affect, indicating that $dis_{G\setminus \{v\}}(w, u) \le h $. Second, for any vertex $w\in B$, we have $dis_G(u,w) < dis_G(u,v) + dis_G(v,w)$. This is because  $dis_G(u,w) \leq h$, $dis_G(u,v) = s $ and $dis_G(v,w) > h-s$. Therefore, any shortest-path between $u$ and $w$ does not pass through $v$, which suggests that $dis_{G(V\setminus \{v\})}(w, u) \le h $. 
	\end{proof}
	
	Based on the Observation \ref{Obs:1}, we can see that only the vertices in $N_v^{h-s}(G) \cup \{v\}$ may affect the \hdegree of $u$ after deleting $v$ for any $u\in N_v^h(G)$. Below, we show that any vertex $w$ in $N_v^{h-s}(G)\cup \{v\}$ that satisfies $dis_{G(V\setminus \{v\})}(u,w) > h$ must be excluded in $N_u^h(G(V\setminus\{v\})$.
	
	\begin{observation}\label{Obs:2}
		Given a positive integer $h \in \mathbb{N^{+}}$ and a vertex $u \in N_{v}^{h}(G) $, we define $F_u \triangleq \{w|w\in N_v^{h-s}(G), \;dis_{G(V\setminus \{v\})}(u,w) > h\}$. Then, we have $N_u^h(G) \setminus N_u^h(G(V\setminus\{v\}))= \{v\} \cup F_u$.
	\end{observation}
	\begin{proof}
	  Clearly, the vertex $v$ is contained in $N_u^h(G) \setminus N_u^h(G(V\setminus\{v\}))$. On the one hand, for any vertex $w\ne v$ and $w  \in N_u^h(G) \setminus N_u^h(G(V\setminus\{v\}))$, we have $dis_{G}(u, w)\le h$ and $ dis_{G(V\setminus\{v\})}(u, w) $$>h $. Therefore, the shortest path from $u$ to $w$ in $G$ must pass through $v$. Since $dis_{G}(u, v) = s$, we have $dis_G(v, w) \le h-s$. In other words, $w \in N_v^{h-s}(G)$ which indicates that $w \in F_u$ holds. On the other hand, for any vertex  $w\ne v$ and $w \in F_u$, $w \notin N_u^h(G(V\setminus\{v\}))$ clearly holds (by the definition of $F_u$). Since $w\in N_v^{h-s}(G)$ and $dis_{G}(u, v) = s$, we have $dis_{G}(u, w) \le h$ by triangle inequality. Hence, we obtain that $w \in N_u^h(G)$. This completes the proof.
	\end{proof}
	
	Based on the Observation \ref{Obs:2}, we can obtain that $d_u^h(G)-d_u^h(G(V\setminus\{v\})) = 1+|F_u|$. As a result, the key to update the \hdegree of a vertex $u$ after removing $v$ is to identify the set $F_u$. Since the set $N_v^{h-s}(G)$ can be easily derived by $N_v^h(G)$, the challenge is how can we efficiently compute $dis_{G(V\setminus \{v\})}(u, w)$ on the graph after removing $v$. Below, we prove an interesting result which indicates that the shortest-path distance $dis_{G(V\setminus \{v\})}(u, w)$ can be computed on the subgraph induced by $N_v^h(G)$ if $dis_{G(V\setminus \{v\})}(u, w) \le h$.
	\begin{theorem}
		\label{thm:h-neighbor} Given a positive integer $h \in \mathbb{N^{+}}$, all shortest-paths between $u\in N_v^{h}(G)$ and $w \in N_v^{h-s}(G)$ on $G(V\setminus\{v\})$ that satisfy $dis_{G(V\setminus \{v\})}(u,w)\leq h$ are contained in the induced subgraph $G(N_v^{h}(G))$, where $s=dis_G(u,v)$. In other words, for any shortest path $P=(u,...,w_i,...,w)$ between $u$ and $w$ on $G(V\setminus\{v\})$, we have $w_i \in N_v^{h}(G)$ for all $w_i \in P$.
	\end{theorem}
	\begin{proof}
	Suppose, to the contrary, that there exists a shortest-path $P=(u,...,w',...,w)$ between $u \in N_v^{h}(G)$ and $w \in N_v^{h-s}(G)$ on $G(V\setminus\{v\})$ that satisfies $w' \notin N_v^{h}(G)$. By this assumption, we have  $dis_{G(V\setminus\{v\})}(u,w) = dis_{G(V\setminus\{v\})}(u,w') + dis_{G(V\setminus\{v\})}(w',w)$. Then, $dis_G(v, w') - dis_G(v, u) \le dis_G(u, w') \le dis_{G(V\setminus\{v\})}(u,w')$ holds by triangle inequality. Since $w' \notin N_v^{h}(G)$ (by assumption), we have $dis_G(v, w')>h$. Thus, we have $h-s < dis_{G(V\setminus\{v\})}(u,w')$. Similarly, we have $dis_G(v, w')-dis_G(v, w) \le dis_G( w', w) \le dis_{G(V\setminus\{v\})}(w',w)$.  Therefore, we get that $s= h-(h-s) < dis_{G(V\setminus\{v\})}(w',w)$. Putting it all together, we can derive that $h < dis_{G(V\setminus\{v\})}(u,w) $ which is a contradiction.
	\end{proof}
	
	%

	Let $\bar F_u \triangleq  \{ w|w\in N_v^{h-s}(G), dis_{G(V\setminus \{v\})}(u,w) \le h\}=N_v^{h-s}(G)\setminus F_u$. By Theorem~\ref{thm:h-neighbor}, $\bar F_u$ can be determined on the subgraph induced by $N_v^h(G)$. As a result, we are also able to compute $|F_u|$ on the induced subgraph $G(N_v^h(G))$ (not on the entire graph $G(V\setminus\{v\})$). In other words, we only need to explore a small subgraph $G(N_v^h(G))$ to maintain the \hdegrees for all vertices in $N_v^h(G)$ after removing $v$, without recomputing the \hdegree for every vertex in $N_v^h(G)$.
	
	Based on such an efficient \hdegree updating technique, we propose a new $(k, h)$-core decomposition algorithm, called \khcore, which is shown in Algorithm~\ref{alg:khcore-algorithm}. Algorithm~\ref{alg:khcore-algorithm} is also a peeling algorithm which iteratively deletes the vertices with the minimum \hdegree (lines~3-13 in Algorithm~\ref{alg:khcore-algorithm}). The algorithm terminates when all vertices are deleted. However, unlike Algorithm~\ref{alg:basic-algorithm}, Algorithm~\ref{alg:khcore-algorithm} invokes a \updatenbr procedure (Algorithm~\ref{alg:update-hnbr}) to update the \hdegree for each vertex in $N_v^h(G)$ after removing $v$ based on the results shown in Theorem~\ref{thm:h-neighbor} (line~9). Below, we describe the detailed implementation of Algorithm~\ref{alg:update-hnbr}.
	
	In Algorithm~\ref{alg:update-hnbr}, we develop a new data structure, named \reach, to maintain the set of vertices that are reachable from $u \in N_{v}^{h}(G)$ within $h$ hops in the induced subgraph $G(N_{v}^{h}(G))$. Initially, for each $u\in N_v^h(G)$, if $dis_G(v, u) < h$, $\reach(u)=\{u\}$, and otherwise $\reach(u) = \emptyset$ (lines~2-6). This is because when $dis_G(v, u) = h$, the \hdegree of $u$ decreases by 1 after deleting $v$, and thus we do not need to maintain the \reach structure for $u$ in this case (i.e., $\reach(u) = \emptyset$). Then, we can make use of a dynamic programming (DP) procedure to identify all the vertices in $N_{v}^{h}(G)$ that are reachable from $u$ within $h$ hops (lines~7-12). In particular, the DP procedure is based on the following results. Let $R_u^s$ be the set of vertices that are reachable from $u$ within $s$ hops. Then, $R_u^{s+1}$ can be obtained by merging the sets $R_w^{s}$ for all $w \in N_u(G) \cup \{u\}$, i.e., $R_u^{s+1} = \bigcup_{w \in N_u(G)\cup\{u\}}R_w^s$. We can adopt the \reach structure to implement such a DP procedure which is shown in lines~7-12 of Algorithm~\ref{alg:update-hnbr}. Subsequently, Algorithm~\ref{alg:update-hnbr} applies the results in Theorem~\ref{thm:h-neighbor} to update the \hdegree for each $u\in N_v^h(G)$ (lines~13-17). The following example illustrates the detailed procedure of Algorithm~\ref{alg:khcore-algorithm} and Algorithm~\ref{alg:update-hnbr}.
	
	\begin{algorithm}[t]
		\caption{\khcore}
		\label{alg:khcore-algorithm}
		\KwIn{a graph $G=(V,E)$ and a positive integer $h$}
		\KwOut{$\core_h(v)$ for all $v\in V$}
		\For{$v \in V$}{ Compute $d_v^h(G)$;}
		\While{$V \neq \emptyset $}{
			$k \gets \arg\min_{v \in V}\{d_v^h(G)\}$\;
			$B \gets \{v|v \in V, d_v^h(G) = k\} $\;
			\While{$B \ne \emptyset $}{
				Pick and remove a vertex $v$ from $B$\;
				$\core_h(v) \gets k$\;
				$d^h(G ( V \setminus \{v\})) \gets $ \text{\updatenbr}($G, h, v$)\;
				\For{$u \in N_{v}^{h}(G)$}{
					\If{$d_u^h(G ( V \setminus \{v\}) ) \leq k$ \myand $u \notin B$}{
						$B \gets B \cup \{u\} $\;
					}
				}
				$V \gets V \setminus \{v\} $;
			}
		}	
	\end{algorithm}
	
	\begin{algorithm}[t]
		\caption{\updatenbr($G, h, v$)}
		\label{alg:update-hnbr}
		$G(R)=(R, E(R)) \gets $ the subgraph induced by $R=N_{v}^{h}(G)$\;
		\For{$u \in R$}{
			\If{$dis_G(v,u) < h$}{
				$\reach[0][u] \gets \{u\}$; $\reach[1][u] \gets \{u\} $\;
			}
			\Else{
				$\reach[0][u] \gets \emptyset$; $\reach[1][u] \gets \emptyset $\;
			}
		}
		$p \gets 1$; $q \gets 0$\;
		\For{$hop=1$ to $h$}{
			$q\gets p$; $p \gets 1 - p$\;
			\For{$(u,w) \in E(R)$}{
				$\reach[q][u] \gets$ $\reach[q][u] \cup \reach[p][w]$\;
				$\reach[q][w] \gets$ $\reach[q][w] \cup \reach[p][u]$\;
			}
		}
		\For{$u\in R$}{
			$s \gets dis_G(u,v)$; $d_u^h(G ( V \setminus \{v\})) \gets d_u^h(G) - 1$\;
			\For{$w\in R$ s.t. $dis_G(v,w) \leq h-s$}{
				\If{$w \notin \reach[q][u]$}{
					$d_u^h(G( V \setminus \{v\})) \gets d_u^h(G( V \setminus \{v\})) - 1$\;
				}
			}
		}
		\Return $d_u^h(G ( V \setminus \{v\}))$ for each vertex $u \in R$\;
	\end{algorithm}
	
		\begin{example}
			Consider the graph shown in Fig.~\ref{fig:example}. Assume that $h=2$. We can see that $v_{1}$ has the minimum 2-hop degree which is 4. When removing $v_{1}$, Algorithm~\ref{alg:khcore-algorithm} needs to invoke Algorithm~\ref{alg:update-hnbr} to update the 2-hop degrees for the vertices in $R = N_{v_{1}}^{2}(G)=\{v_{2}, v_{3}, v_{4},v_{6}\}$ (line~9 of Algorithm~\ref{alg:khcore-algorithm}). Specifically, Algorithm~\ref{alg:update-hnbr} initializes the \reach sets for all vertices in $R$ as follows: $\reach(v_{2})=\{v_{2}\}$, $\reach(v_{3})=\{v_{3}\}$, and $\reach(v_{4})=\reach(v_{6})=\emptyset$ (lines~2-6 of Algorithm~\ref{alg:update-hnbr}). Then, the algorithm performs the DP procedure to compute the \reach sets for all vertices in $R$ (lines~7-12 of Algorithm~\ref{alg:update-hnbr}). After that, we can get that $\reach(v_{2})=\{v_{2}\}$, $\reach(v_{3})=\{v_{3}\}$, $\reach(v_{4})=\{v_{2}\}$ and $\reach(v_{6})=\{v_{3}\}$, respectively. Then, based on the \reach sets, the algorithm updates the 2-hop degrees for the vertices in $R$ (lines~13-17 of Algorithm~\ref{alg:update-hnbr}). In particular, $d_{v_{2}}^2(G)$ decreases by 2 ($d_{v_2}^{2}(G(V\setminus \{v_1\})) = 3$), since $v_{3} \in N_{v_{1}}^{1}(G)$ is not included in $\reach(v_{2})$. Similarly, $d_{v_{3}}^{2}(G)$ decreases by 2 ($d_{v_3}^{2}(G(V\setminus \{v_1\})) = 3$), and both $d_{v_{4}}^{2}(G)$ and $d_{v_{6}}^{2}(G)$ decreases by 1 ($d_{v_4}^{2}(G(V\setminus \{v_1\})) = 7$ and $d_{v_6}^{2}(G(V\setminus \{v_1\})) = 7$). As a result, the vertices $\{v_{2}, v_{3}\}$ are also deleted after removing $v_1$, and the $(k,h)$-core numbers for $\{v_1, v_2, v_3\}$ are equal to 4.  In the next iteration of Algorithm~\ref{alg:khcore-algorithm},  $v_5$ has the minimum 2-hop degree. The algorithm uses Algorithm~\ref{alg:update-hnbr} to update the 2-hop degrees of the vertices in $N_{v_{5}}^{2}(G(V\setminus\{v_1\}))$. After that, we can derive that the vertices $\{v_{4}, v_{6}, v_{7}\}$ are also deleted after removing $v_5$ in this iteration. The $(k,h)$-core numbers for $\{v_{4}, v_{5},v_{6}, v_{7}\}$ are 5. In the last iteration, the algorithm will remove all vertices, and we can obtain that the $(k,h)$-core numbers for the vertices $\{v_8, \cdots, v_{14}\}$ are 6.
		\end{example}

	\stitle{Complexity analysis.} We start by analyzing the time complexity of Algorithm~\ref{alg:update-hnbr} as follows. First, Algorithm~\ref{alg:update-hnbr} takes $O(d_v^h(G))$ time to initialize the \reach structures. Then, the algorithm takes $O(h|E(R)|d_v^{h}(G))$ time to compute the \reach sets (lines~7-12). This is because the size of the \reach set is bounded by $d_v^h(G)$, and thus the set union operator can be computed in $O(d_v^{h}(G))$ time using some hash techniques. Finally, the time cost for updating the \hdegrees in line~13-17 is $O(d_v^h(G) \times d_v^{h-1}(G))$. Let $\tilde n$ and $\tilde m$ be the number of vertices and edges of the largest subgraph induced by the $h$-hop neighborhood of a vertex in $V$, respectively. Then, the worst-case time complexity of Algorithm~\ref{alg:update-hnbr} is bounded by $O(\tilde n^2 + h\tilde n \tilde m )$. Based on this, we can easily derive that the worst-case time complexity of Algorithm~\ref{alg:khcore-algorithm} is $O(n\tilde n^2 + n h\tilde n\tilde m )$, which is asymptotically the same as the time complexity of Algorithm~\ref{alg:basic-algorithm} (because $h$ is often a very small integer). For the space overhead, we need to maintain the \reach sets for all vertices in $N_v^h(G)$ when deleting a vertex $v$ which takes at most $O(d_v^h(G)^2) \le O(\tilde n^2)$ in total. Therefore, the space complexity of Algorithm~\ref{alg:khcore-algorithm} can be bounded by $O(m+n+\tilde n^2)$. Below, we propose a \bitmap technique to further improve the time and space overheads of our algorithm.

	\subsection{A bitmap optimization} \label{subsec:bitmap}
	Recall that in Algorithm~\ref{alg:update-hnbr}, we have a \reach structure for each vertex $u\in N_v^h(G)$ which maintains the set of vertices in $N_v^h(G)$ that are reachable from $u$ within $h$ hops. To improve the efficiency of the algorithm, we develop a \bitmap to implement such a \reach structure for each vertex $u\in N_v^h(G)$. Suppose without loss of generality that the vertices in $N_v^h(G)$ are labeled from $u_0$ to $u_{d_v^h(G)-1}$. For each vertex $u_i\in N_v^h(G)$, we create a \bitmap to represent the \reach structure of $u_i$. If $u_j$ ($j \ne i$, $j\in \{0, 1, \cdots, d_v^h(G)-1\}$) is reachable within $h$ hops from $u_i$ in the subgraph induced by $N_v^h(G)$, the $j$-th bit of $u_i$'s \bitmap is equal to 1, and otherwise it equals $0$. For example, if $u_i$'s \bitmap is $10101$, we can conclude that $u_i$ can reach $u_0$, $u_2$, and $u_4$ within $h$ hops in the induced graph $G(N_v^h(G))$. To merge two \reach sets, we can perform a \emph{bitwise-or} operator using two \bitmaps which is much more efficient than the traditional set-union operator. In this sense, the \bitmap technique is not only reduce the space usage, but it also improves the time overhead of our algorithm.
	
	\stitle{Implementation details.}  The detailed implementation of the \bitmap technique is outlined in Algorithm~\ref{alg:bitmap-update-hnbr}. Specifically, we make use of a set of 64-bit integers to represent a \bitmap $\reach(u_i)$ for each vertex $u_i \in N_v^h(G)$. In other words, the \bitmap of a vertex $u_i$ (i.e., $\reach(u_i)$) is an integer array. For any vertex $u_i$, if $u_j$ is reachable from $u_i$ within $h$ hops in $G(N_v^h(G))$, then we can compute the position of $u_j$ in $u_i$'s \bitmap array by $\mydiv(j, 64)=\left \lfloor \frac{j}{64}\right\rfloor$. In Algorithm~\ref{alg:bitmap-update-hnbr}, for each vertex $u_i\in N_v^h(G)$, we first initialize its \bitmap to 0 (line~1 of  Algorithm~\ref{alg:bitmap-update-hnbr}). Then, for each vertex $u_i$, we set the $i$-th bit of $u_i$'s \bitmap to 1 (lines~4-6), denoting that the \reach set of $u_i$ contains $u_i$ itself. Note that in Algorithm~\ref{alg:bitmap-update-hnbr}, the notation $\mymod(i, 64)$ means $i\%64$ (lines~5-6), which is used to determine the bit-position of $u_i$ in a \bitmap. After that, we perform the DP procedure to compute the \reach sets. Note that the process of merging two \reach sets is implemented by a \emph{bitwise-or} operator ( lines~11-13). Finally, Algorithm~\ref{alg:bitmap-update-hnbr} updates the \hdegrees for all vertices in $N_v^h(G)$ (lines~14-19). Notice that based on the \bitmap structure, we can use a \emph{bitwise-and} operator to determine whether a vertex $u_j\in N_v^{h-s}(G)$ is reachable from $u_i$ within $h$ hops {(lines~17-18)}. The following example illustrates the detailed procedure of our \bitmap technique.
	
	\begin{algorithm}[t]
		\caption{\bitmapupdate($G, h, v$)}
		\label{alg:bitmap-update-hnbr}
		$G(R)=(R, E(R)) \gets $ the subgraph induced by $R=N_{v}^{h}(G)$\;
		Initialize the \bitmaps (the $\reach$ arrays) for all $u_i\in R$ to 0\;
		$N_{v}^{h-1}(G) \gets \{u_i \in R|dis_G(v,u_i) < h\}$; $d\gets |N_{v}^{h-1}(G)|$\;
		\For{$u_i \in N_{v}^{h-1}(G)$}{
			$\reach[0][i][\mydiv(i,64)] \gets 1 \ll \mymod(i,64)$\;
			$\reach[1][i][\mydiv(i,64)] \gets 1 \ll \mymod(i,64)$\;
		}
		$p \gets 1$; $q \gets 0$\;
		\For {$hop=1$ to $h$} {
			$q\gets p$; $p \gets 1 - p$\;
			\For {$(u_i,u_j) \in E(R)$} {
				\For {$b = 0$ to $\mydiv(d,64)$}{
					{\small 
					$\reach[q][i][b] = \reach[q][i][b] \vee \reach[p][j][b]$\;
					$\reach[q][j][b] = \reach[q][j][b] \vee \reach[p][i][b]$\;
					}
				}
			}
		}
		\For {$u_i\in R$}{
			$s \gets dis_G(u_i,v)$; $d_{u_i}^h(G ( V \setminus \{v\}) ) \gets d_{u_i}^h(G) - 1$\;
			\For {$u_j\in R$ s.t. $dis_G(v,u_j) \leq h-s$}{
				$x \gets 1 \ll \mymod(j,64)$; $y \gets \reach[q][i][\mydiv(j,64)]$\;
				\If {$(x \land y)  = 0$}{
					$d_{u_i}^h(G ( V \setminus \{v\}) ) \gets d_{u_i}^h(G ( V \setminus \{v\}) ) - 1$\;
				}
				
			}
		}
		\Return $d_{u_i}^h(G ( V \setminus \{v\}) )$ for each vertex $u_i \in R $;
	\end{algorithm}
	
		\begin{table}[t!]
			\centering
			\caption{The \bitmaps of vertices in $N_{v_5}^2(G(V\setminus\{v_1, v_2, v_3\}))$ } \vspace*{-0.25cm} 
			\label{tab:bitmap}
			\begin{tabular}{c|c|c|c|c|c}
				\hline
				$N_{v_5}^2(G(V\setminus\{v_1, v_2, v_3\}))$ & $v_4$ & $v_6$ & $v_7$ & $v_8$ & $v_9$\cr\hline
				re-label  & $u_0$  & $u_1$  & $u_2$ & $u_3$ & $u_4$ \cr \hline \hline
				Initialization  & 1  & 2  & 4 & 0 & 0 \cr \hline 
				Iteration 1  & 1  & 2  & 4 & 5 & 6\cr \hline
				Iteration 2  & 5  & 6  & 7 & 5 & 6\cr \hline
			\end{tabular}
		\end{table}
		
		\begin{example}
			Reconsider the graph $G$ shown in Fig.~\ref{fig:example}. Suppose that $h=2$. After the first iteration, we can obtain a subgraph $G^\prime$  induced by the vertices $\{v_4, v_5, \cdots, v_{14}\}$. Then, let us consider the vertex $v_5$, which has the smallest \hdegree in $G^\prime$. We show the \bitmap structure of each vertex in $N_{v_5}^{2}(G^\prime)$ in Table~\ref{tab:bitmap}. First, we relabel the vertices in $N_{v_5}^{2}(G^\prime)=\{v_4, v_6, v_7, v_8, v_9\}$ by $\{u_0, \cdots, u_4\}$ (the second row of Table~\ref{tab:bitmap}). Since $dis_{G^\prime}(v_5, v_8)=h=2$ and $dis_{G^\prime}(v_5, v_9)=h=2$, the \bitmaps of $u_3$ and $u_4$ are initialized by 0. We can easily derive that the \bitmaps of $u_0$, $u_1$, and $u_2$, are initialized by $1, 2$, and $4$, respectively. Note that the set of edges in $N_{v_5}^{2}({G^\prime})$ is $\{(u_0,u_3), (u_2,u_3), (u_1,u_4), (u_2,u_4)\}$. In the first iteration (i.e., $hop=1$ in line~8 of Algorithm~\ref{alg:bitmap-update-hnbr}), the \bitmap of $u_3$ (i.e., $v_8$) is updated by $5$ (obtained by merging the \bitmaps of $u_0$ and $u_2$), and the \bitmap of $u_4$ is updated by 6 (obtained by merging the \bitmaps of $u_1$ and $u_2$). For the other vertices, their \bitmaps keep unchanged in the first iteration. Similarly, in the second iteration ($hop=2$), we can derive that the \bitmaps of $u_0$, $u_1$, and $u_2$ are updated by 5, 6, and 7 respectively.
		\end{example}
	
	\stitle{Complexity analysis.} Armed with the \bitmap technique, Algorithm~\ref{alg:bitmap-update-hnbr} can significantly reduce the set-union costs. In our basic \khcore algorithm (Algorithm~\ref{alg:update-hnbr}), the set-union operator can be done in $O(d_v^h(G))$ time (lines~10-12 of Algorithm~\ref{alg:update-hnbr}). However, by using the \bitmap technique, we can implement the set union operator by a \emph{bitwise-or} operator which takes $O(d_v^h(G)/64)$ time. In other words, the \bitmap technique can achieve around $64\times$ speedup for the set union computation. As a result, the total time costs of the \khcore algorithm with \bitmap technique can be bounded by $O(n\tilde n^2+nh\tilde n \tilde m/64)$. Since $h$ is typically smaller than 64, the time complexity of our algorithm is lower than that of Algorithm~\ref{alg:basic-algorithm} which is confirmed in our experiments.

	\stitle{Remark.} It is worth remarking that the lower and upper bounding techniques developed in \cite{19sigmodBonchiKS} can also be integrated into Algorithm~\ref{alg:khcore-algorithm}. However, we empirically find that such lower and upper bounding techniques cannot significantly improve the efficiency of our algorithm, thus in this work we mainly focus on our algorithms without using the lower and upper bounds developed in \cite{19sigmodBonchiKS}. Also, it is worth emphasizing that the \bitmap technique is an elegant implementation of our theoretical finding; it is not a general optimization technique and it cannot be used in the state-of-the-art algorithm \cite{19sigmodBonchiKS}. In the experiments, we will focus mainly on evaluating the proposed algorithms with the \bitmap implementation.
	
	
	%
	
	\subsection{A sampling-based algorithm}\label{subsec:sample-alg}
	To further improve the efficiency, we propose a sampling-based algorithm to compute the $(k,h)$-core decomposition. The key idea of the sampling-based algorithm is that when deleting a vertex $v$, it estimates the updated \hdegree for a vertex $u\in N_v^h(G)$ using the randomly sampled vertices (not all vertices in $N_v^h(G)$). Due to the less computation for updating the \hdegrees of vertices, the sampling-based approach can significantly reduce the time cost compared to the exact algorithm.
	
	The implementation details of the sampling-based algorithm are shown in Algorithm~\ref{alg:KHCore-Sampling}. First, the algorithm randomly selects $r|V|$ vertices from $V$ (line~2 of Algorithm~\ref{alg:KHCore-Sampling}), where $0<r<1$ denotes the sampling rate. Then, for each vertex $v$, the algorithm computes the number of selected vertices in the \hneighbor of $v$ (line~3), denoted by $\select[v]$. Based on $\select[v]$, the algorithm calculates the sampling rate for $v$ (line~4 of Algorithm~\ref{alg:KHCore-Sampling}), i.e., $\rate[v] = \select[v] / d_v^h(G)$. Similar to Algorithm~\ref{alg:khcore-algorithm}, the algorithm iteratively deletes the vertex that has the smallest \hdegree (lines~5-7). When removing a vertex $v$, it invokes Algorithm~\ref{alg:updatehnbrsampling} to update the \hdegrees of the vertices in $N_v^h(G)$ (line~6).
	
	In Algorithm~\ref{alg:updatehnbrsampling}, it first initializes the \bitmap structures for the vertices in $N_v^{h}(G)$ (lines~1-2 of Algorithm~\ref{alg:updatehnbrsampling}). Let $S$ be the set of sampled vertices. Then, the algorithm computes the \bitmaps for the vertices in $N_v^{h-1}(G)\cap S$ (lines~2-3). Note that for the vertices in $N_v^{h}(G)\setminus N_v^{h-1}(G)$, their \hdegrees decrease by 1 after deleting $v$, thus we do not need to maintain the \bitmaps for those vertices. Subsequently, for each $u_i \in N_v^h(G)$, the algorithm updates the \hdegree of $u_i$ based on the sampled vertices {(lines~4-11)}. Notice that it first updates $\select[u_i]$, and then uses $\select[u_i] / \rate[u_i]$ as an estimator for the updated $d_{u_i}^h(G)$ {(lines~10-11)}. The following illustrative example shows how our algorithm works.
		\begin{example}
			Suppose there is a vertex $v$ and its $h$-neighbors $N_v^{h}(G)=\{u_1, u_2,u_3,...\}$, the sampling $rate$ of vertices in $N_v^{h}(G)$ is $rate[u_1] = 0.3$, $rate[u_2] = 0.45$, $rate[u_3] = 0.25$, respectively. If the vertex $v$ is removed from the graph, the selected $h$-degrees of vertices in $N_v^{h}(G)$ are decreased to $6, 8, 7$, respectively. According to the proposed method, we estimate $h$-degrees of $\{u_1, u_2,u_3,...\}$ in graph $G$. The approximate $h$-degrees of $u_1$, $u_2$ and $u_3$ are decreased to $6\div rate[u_1] = 20$, $8 \div rate[u_2] = 17$ and $7\div rate[u_3] = 20$, respectively.
		\end{example}

	\begin{algorithm}[t]
		\caption{\khcoresamp}
		\label{alg:KHCore-Sampling}
		\KwIn{a graph $G=(V,E)$, a positive integer $h$, and a sampling rate $r$}
		\KwOut{$\core_h(v)$ for all $v \in V$}
		Lines~1-2 of Algorithm~\ref{alg:khcore-algorithm}\;
		$S \gets$ uniformly sampling $r|V|$ vertices from $V$\;
		$\select[v] \gets |\{u|u \in N_v^h(G),u \in S \}|$ for each $v\in V$\;
		$\rate[v] \gets \select[v] / d_v^h(G)$  for each $v\in V$\;
		Lines~3-8 of Algorithm~\ref{alg:khcore-algorithm}\;
		$d^h(G ( V \setminus \{v\})) \gets $ \text{\updatehnbrsamp}({$G, h, v, S, \select, \rate$})\;
		Lines~10-13 of Algorithm~\ref{alg:khcore-algorithm}\;
	\end{algorithm}

	\vspace*{-0.1cm}
	\stitle{Complexity analysis.} We first analyze the time complexity of Algorithm~\ref{alg:updatehnbrsampling}. Compared to Algorithm~\ref{alg:bitmap-update-hnbr}, Algorithm~\ref{alg:updatehnbrsampling} only need to maintain the \bitmaps for the sampled vertices $N_v^{h-1}(G)\cap S$. The cardinality of the set $N_v^{h-1}(G)\cap S$ can be bounded by $O(rd_v^h(G))\le O(r\tilde n)$. Similar to Algorithm~\ref{alg:bitmap-update-hnbr}, we can easily derive that the time complexity of  Algorithm~\ref{alg:bitmap-update-hnbr} is $O(r\tilde n^2+hr\tilde n \tilde m / 64)$, where $r <1$ is sampling rate. Based on this, the time complexity of Algorithm~\ref{alg:KHCore-Sampling} is $O(rn\tilde n^2+hrn\tilde n \tilde m / 64)$, which is lower than our exact algorithm by a factor $r$. For example, if $r=0.1$, the sampling-based algorithm can be one order of magnitude faster than the proposed exact algorithm, as confirmed in our experiment. For the space usage, we can easily derive that the complexity of the sampling-based algorithm is the same as that of the exact algorithm.
	
	\vspace*{-0.1cm}
	\subsection{Parallelization} \label{subsec:parallel}
	In this section, we explore how Algorithm~\ref{alg:khcore-algorithm} splits the computation in several sub-tasks which can be processed independently. Note that the parallelization strategy for Algorithm~\ref{alg:khcore-algorithm} and Algorithm~\ref{alg:KHCore-Sampling} is the same. Therefore, we focus mainly on developing parallelization strategy for Algorithm~\ref{alg:khcore-algorithm}.
	
	First, in lines~1-2 of Algorithm~\ref{alg:khcore-algorithm}, we can compute the \hdegree for each vertex in parallel, because the sub-tasks for computing \hdegrees are clearly independent.  Second, when deleting the vertices in the bucket $B$ (line~6 of Algorithm~\ref{alg:khcore-algorithm}), we can also process the vertices in parallel. However, the sub-task for deleting a vertex is not independent, but it depends on the former deleted vertices. To make all the sub-tasks independent, we can follow an increasing order by vertex ID to delete vertex. When processing a vertex $v_i$, we use a thread to update the \hdegrees of the vertices in $N_{v_i}^h(G)$ that either has a \hdegree no less than $d_{v_i}^h(G)$ or has a larger vertex ID. Based on this strategy, the sub-tasks for removing the vertices in the bucket $B$ are independent, and therefore we can safely process the vertices in $B$ in parallel. Note that in Algorithm~\ref{alg:bitmap-update-hnbr}, the procedure of updating the \hdegree of a vertex should be considered as an atomic operator (line~15 and line~19). In our experiments, we will show that the proposed parallel algorithms can achieve a very good speedup ratio over the corresponding sequential algorithms.
	
	\begin{algorithm}[t]
		\caption{\updatehnbrsamp($G, h, v, S, \select, \rate$)}
		\label{alg:updatehnbrsampling}
		Lines~1-2 of Algorithm~\ref{alg:bitmap-update-hnbr}; $R=N_v^h(G)$\;
		$\tilde N_v^{h-1}(G) \gets N_v^{h-1}(G) \cap S$; $d\gets |\tilde N_v^{h-1}(G)|$\;
		Lines~4-13 of Algorithm~\ref{alg:bitmap-update-hnbr}\;
		\For {$u_i\in R$}{
			$s \gets dis_G(u_i,v)$; $\cnt \gets 0 $\;
			{\bf if} {{$v \in S $}} {\bf then} $\cnt \gets 1$\;
			\For {$u_j \in R\cap S$ s.t. $dis_G(v,u_j) < h - s $} {
				$x \gets 1 \ll \mymod(j,64)$; $y \gets \reach[q][i][\mydiv(j,64)]$\;
				\lIf {$(x \land y) = 0$}{$\cnt \gets \cnt + 1$}
			}
			$\select[u_i]\gets \select[u_i] - \cnt$\;
			$d_{u_i}^h(G ( V \setminus \{v\}) ) \gets \select[u_i] / \rate[u_i]$\;
		}
		\Return $d_{u_i}^h(G ( V \setminus \{v\}) )$ for each vertex $u_i \in N_{v}^{h}(G) $\;
	\end{algorithm}

\section{Experiments} \label{sec:experiment}
In this section, we conduct extensive experiments to evaluate the efficiency and scalability of the proposed algorithms. Below, we first describe the experimental setup and then report our results.
\vspace*{-0.1cm}
\subsection{Experimental setup} \label{subsec:exp-setup}
We implement three sequential algorithms to compute the $(k,h)$-core decomposition: \khc, \khcs, and \hbz. The \khc and \khcs are our exact and sampling-based $(k,h)$-core decomposition algorithms respectively. Both \khc and \khcs are integrated with the \bitmap technique proposed in Section~\ref{subsec:bitmap}. The \hbz algorithm denotes the state-of-the-art $h$-LB+UB algorithm \cite{19sigmodBonchiKS}, which is served as a baseline in our experiments. For all these algorithms, we also implement the parallelized versions using OpenMP. All algorithms are implemented in C++. We conduct all experiments on a PC with two 2.3 GHz Xeon CPUs (16 cores in total) and 64GB memory running Ubuntu 16.4.

\stitle{Datasets.} We make use of 12 real-world datasets in our experiments. Table~\ref{tab:datasets} shows the detailed statistics of the datasets, where $d_{\max}$, $\Delta$ and $k_{\max}$ denote the maximum degree, the diameter and the maximum $k$-core number of the network. \kw{ca}-\kw{AstroPH}\footnote{\urlsnap} (\caas for short) is a collaboration network; \kw{com}-\kw{amazon}\footnotemark[1] (\comamazon) is a co-purchasing network; \douban\footnote{\urlkonect}, \hyves\footnotemark[2], \kw{soc}-\kw{LiveJournal}\footnotemark[1] (\socLJ), \kw{soc}-\kw{youtube}\footnote{\urlnet} (\socytb), \kw{soc}-\kw{pokec}\footnotemark[2] (\socpokec), and \kw{soc}-\kw{Epinions}\footnotemark[1] (\soceps) are social networks; \kw{flickrEdges}\footnotemark[2] (\flic) is a network of Flickr images sharing common metadata such as tags, groups, locations etc; \kw{bio}-\kw{CE}-\kw{CX} \footnotemark[3] (\bioce) and \kw{bio}-\kw{WormNet}-\kw{v3}\footnotemark[3] (\bioworm) are biological networks; \kw{italycnr}-\kw{2000}\footnotemark[3] (\itcnr) is a web graph.

\stitle{Parameters.} Both \khc and \hbz have only one parameter $h \in \mathbb{N^{+}}$, and the \khcs algorithm has an additional parameter $r$ which denotes the sampling rate. In our experiment, the parameter $h$ is selected from the interval $[2,5]$ (the same parameter setting also used in \cite{19sigmodBonchiKS}), because larger values are often not interesting in practice \cite{19sigmodBonchiKS}. For \khcs, the parameter $r$ is selected from the interval $[0.05, 0.8]$ with a default value of $r=0.1$, because \khcs performs very well on all datasets given that $r=0.1$.
\begin{table}[t!]
	\small
	\centering
	\caption{Datasets} 
	\label{tab:datasets}
	\begin{tabular}{c|c|c|c|c|c}
		\hline
		\bf Dataset & $|V|$ & $|E|$ & $d_{\max}$ & $\Delta$ & $k_{\max}$ \cr\hline \hline
		\bioce & 15,229 & 245,952 & 375 & 13 & 78 \cr \hline
		\bioworm & 16,347 & 762,822 & 1,272 & 12 & 164 \cr \hline
		\caas & 18,771 & 198,050 & 504 & 14 & 56 \cr \hline
		\soceps & 75,880 & 405,740 & 3,044 & 15 & 67 \cr \hline
		\flic & 105,939 & 2,316,948 & 5,425 & 9 & 573 \cr \hline
		\douban & 154,908 & 327,162 & 287 & 9 & 15 \cr \hline
		\itcnr & 325,557 & 2,738,969 & 18,236 & 34 & 83 \cr \hline
		\comamazon & 334,863 & 925,872 & 549 & 44 & 6 \cr \hline
		\socytb & 495,957 & 1,936,748 & 25,409 & 21 & 49 \cr \hline
		\hyves & 1,402,673 & 2,777,419 & 31,883 & 10 & 39 \cr \hline
		\socpokec & 1,632,803 & 22,301,964 & 14,854 & 14 & 47 \cr \hline
		\socLJ & 4,846,609 & 42,851,237 & 20,333 & 16 & 372 \cr \hline
	\end{tabular}
\vspace*{-0.2cm}
\end{table}

\begin{figure*}[t!] 
	\begin{center}
		\begin{tabular}[t]{c}
			\subfigure[{\small $h=2$}]{
				\includegraphics[width=0.9\columnwidth, height=4.3cm]{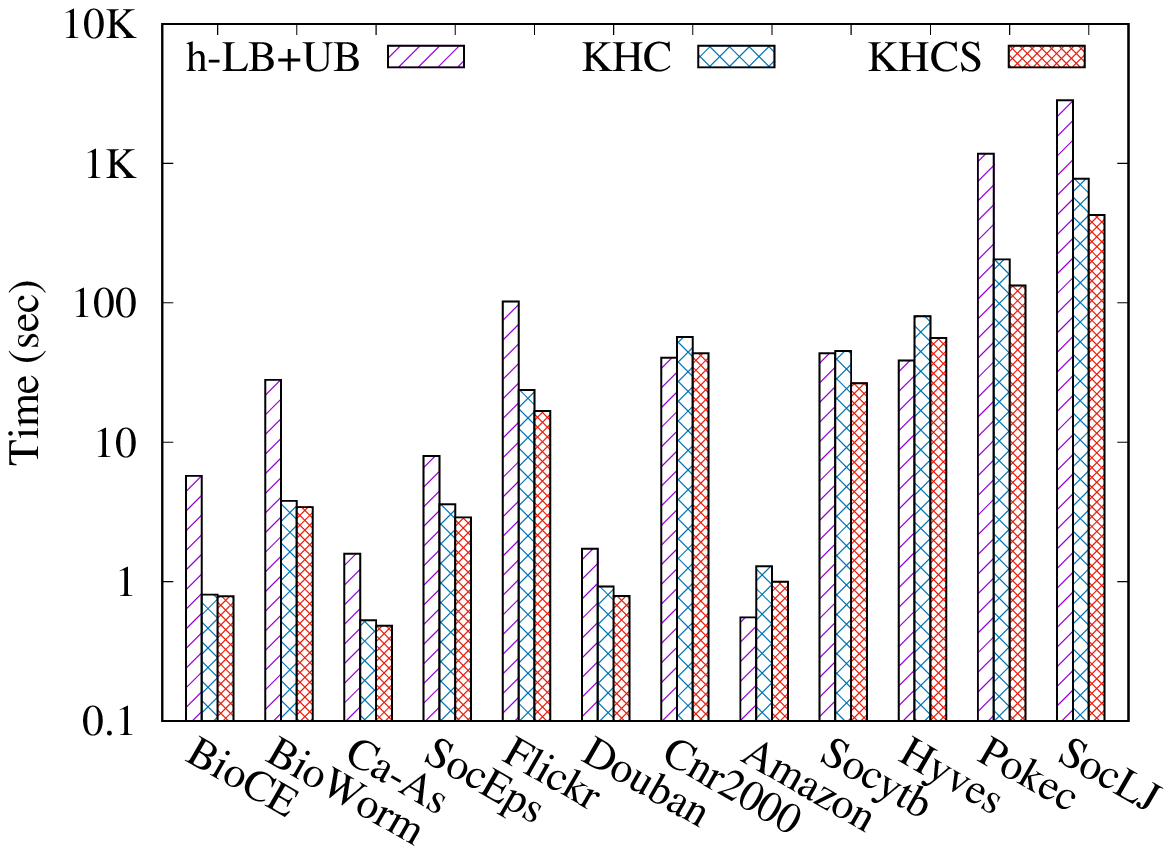}
			}
			\subfigure[{\small $h=3$}]{
				\includegraphics[width=0.9\columnwidth, height=4.3cm]{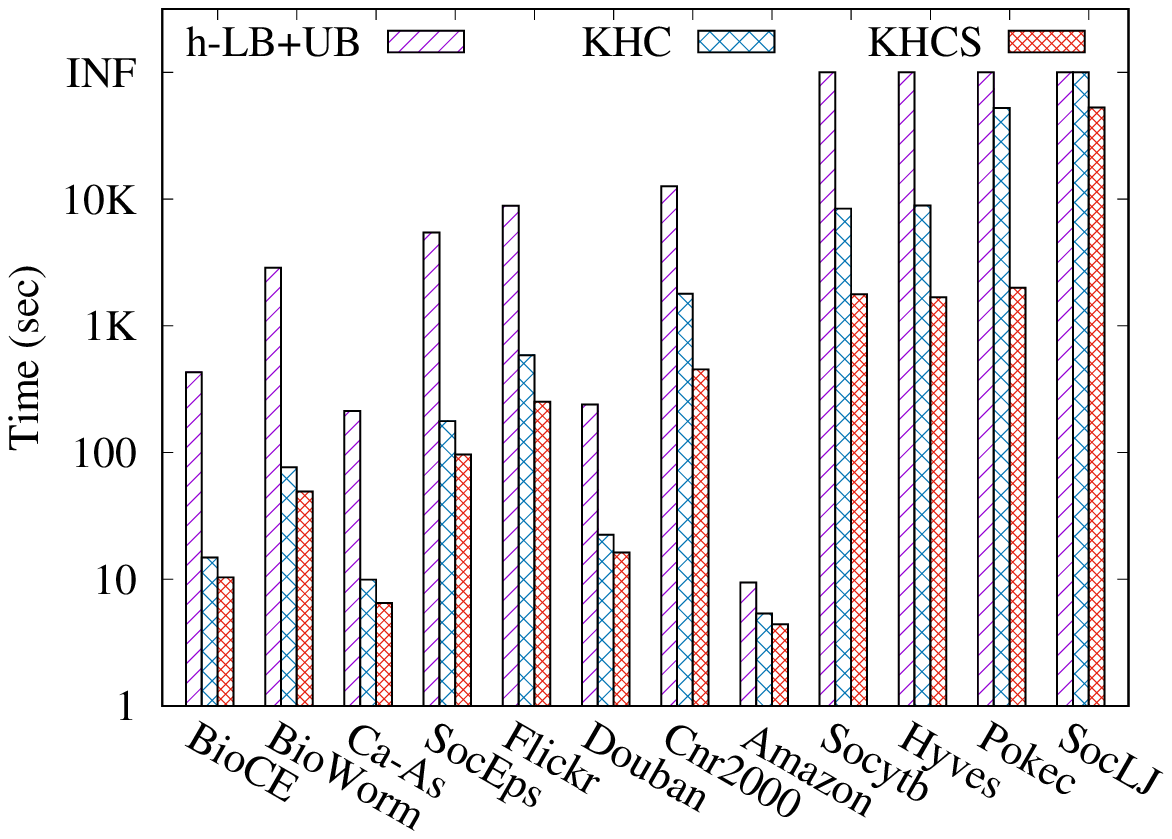}
			}\vspace*{-0.3cm}\\
			\subfigure[{\small $h=4$}]{
				\includegraphics[width=0.9\columnwidth, height=4.3cm]{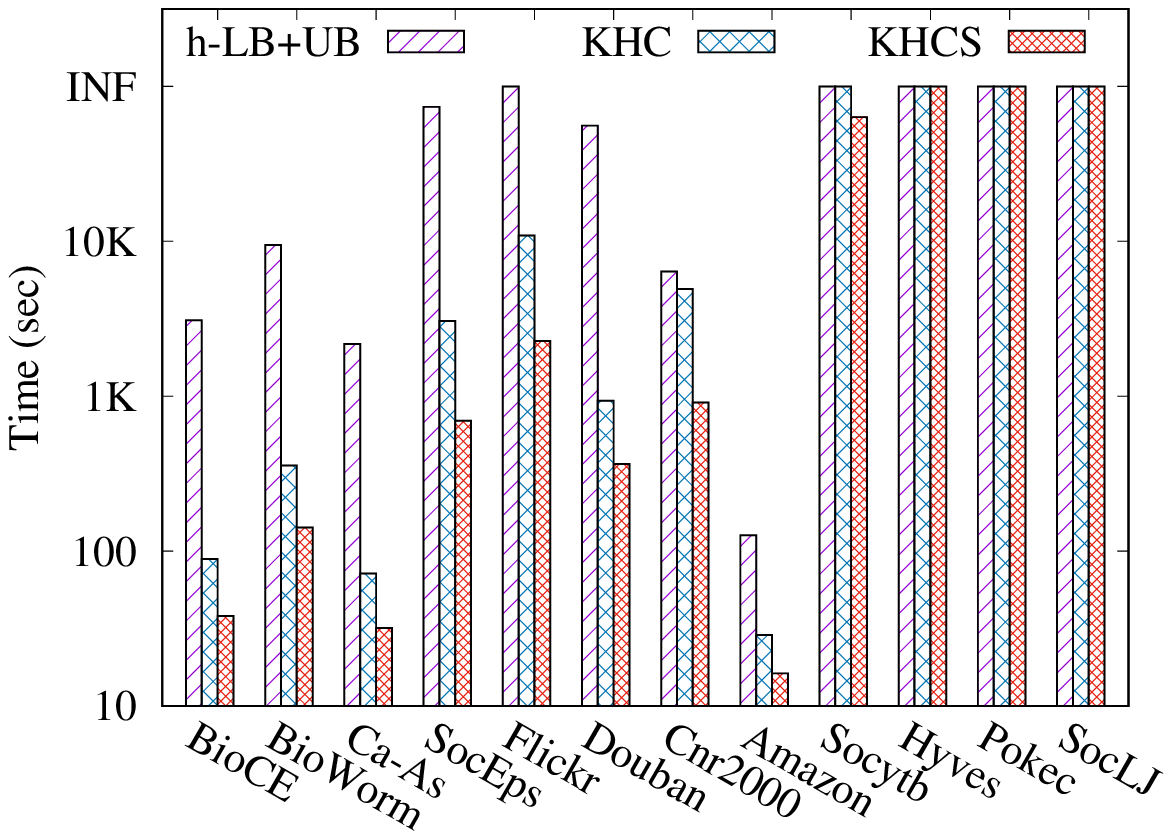}
			}
			\subfigure[{\small $h=5$}]{
				\includegraphics[width=0.9\columnwidth, height=4.3cm]{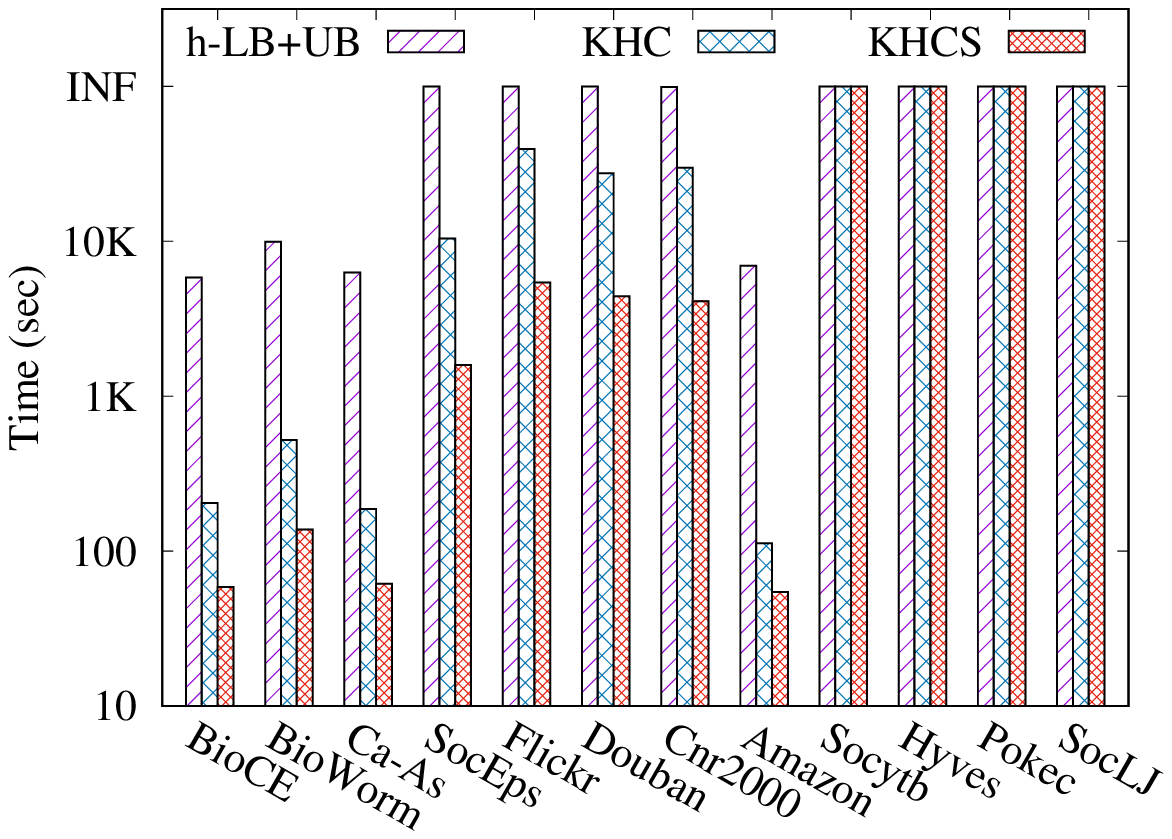}
			}
		\end{tabular}
	\end{center}
	\vspace*{-0.4cm}
	\caption{Runtime of different sequential algorithms on all datasets}
	\vspace*{-0.3cm}
	\label{fig:exp1:efficiency-vary-h}
\end{figure*}

\vspace*{-0.1cm}
\subsection{Experimental results} \label{subsec:exp-result}
\stitle{Exp-1: Efficiency of various sequential algorithms}. We start by comparing the efficiency of different sequential algorithms. Fig.~\ref{fig:exp1:efficiency-vary-h} shows the runtime of \hbz, \khc, and \khcs on all datasets. Note that in all experiments, \infy means that the algorithm does not terminate in 28 hours. From Fig.~\ref{fig:exp1:efficiency-vary-h}(a), we observe that \khc and \khcs significantly outperform the state-of-the-art \hbz algorithm on most datasets with $h=2$. We also notice that on some very sparse graphs, such as \comamazon and \hyves, \hbz is faster than \khc and \khcs. This is because, on very sparse graphs, the costs for recomputing the \hdegrees are very low with $h=2$. However, when $h\ge 3$ (Figs.~\ref{fig:exp1:efficiency-vary-h}(b-d)), we can clearly see that \khc and \khcs are substantially faster than \hbz on all datasets. For example, on \bioce, \khc is at least one order of magnitude faster than \hbz with $h\ge 3$. On larger datasets, such as \socpokec (more than 1.6 million vertices and 22 million edges), \hbz cannot terminate within 28 hours when $h=3$, while \khc takes around 52,000 seconds to compute all $(k,h)$-cores. When comparing \khc with \khcs, we find that \khcs (with the sampling rate $r=0.1$) is much more efficient than \khc given that $h \ge 3$. On some large graphs, \khcs is one order of magnitude faster than \khc when $h\ge 3$. For instance, on \socpokec, \khcs takes around 2,000 seconds to compute all $(k,h)$-cores when $h=3$, whereas the time overhead of \khc is around 52,000 seconds. In addition, when $h=5$ (Fig.~\ref{fig:exp1:efficiency-vary-h}(d)), \hbz cannot handle four medium-sized graphs, while our algorithms still work well on all eight medium-sized graphs. These results are consistent with our theoretical analysis in Section~\ref{sec:ouralgorithm}.

\begin{figure*}[t!]
	\begin{center}
		\begin{tabular}[t]{c}
		
			\subfigure[{\scriptsize $h=2$}]{
				\includegraphics[width=0.38\columnwidth, height=2.5cm]{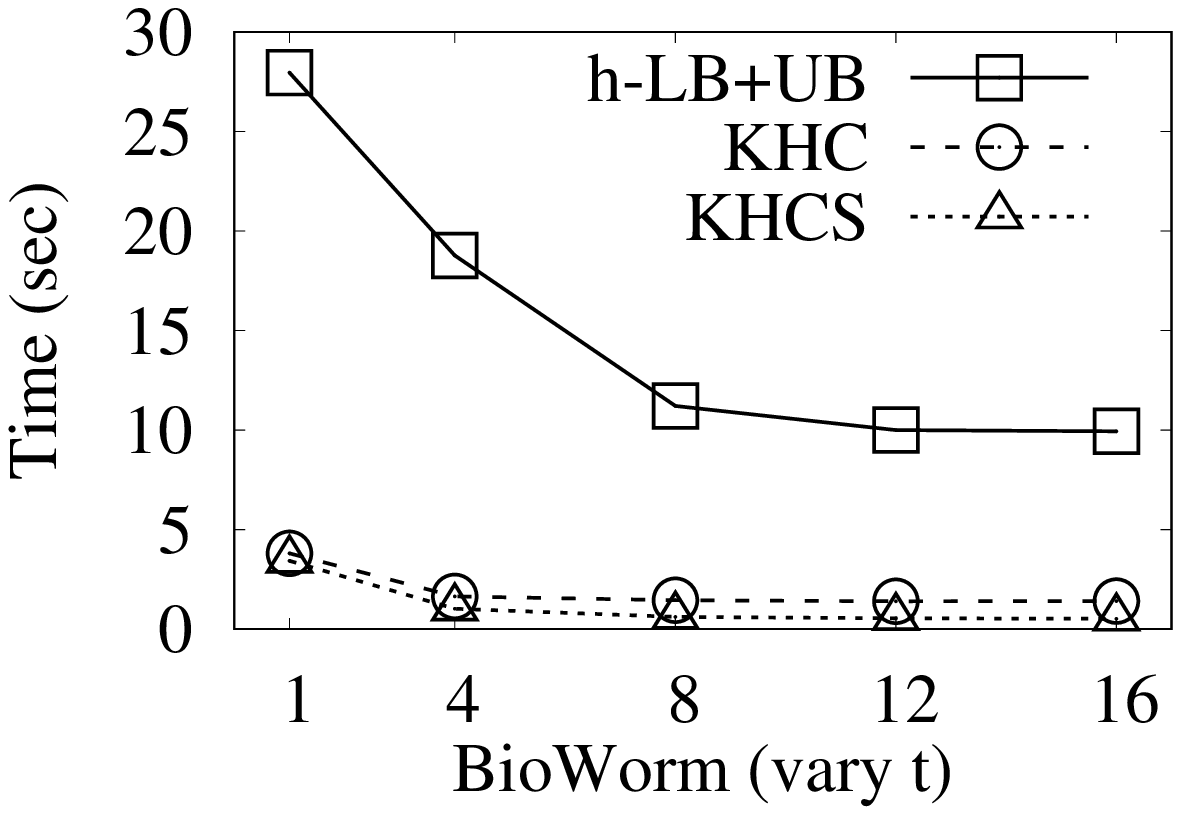}
			}
			\subfigure[{\scriptsize $h=2$}]{
				\includegraphics[width=0.38\columnwidth, height=2.5cm]{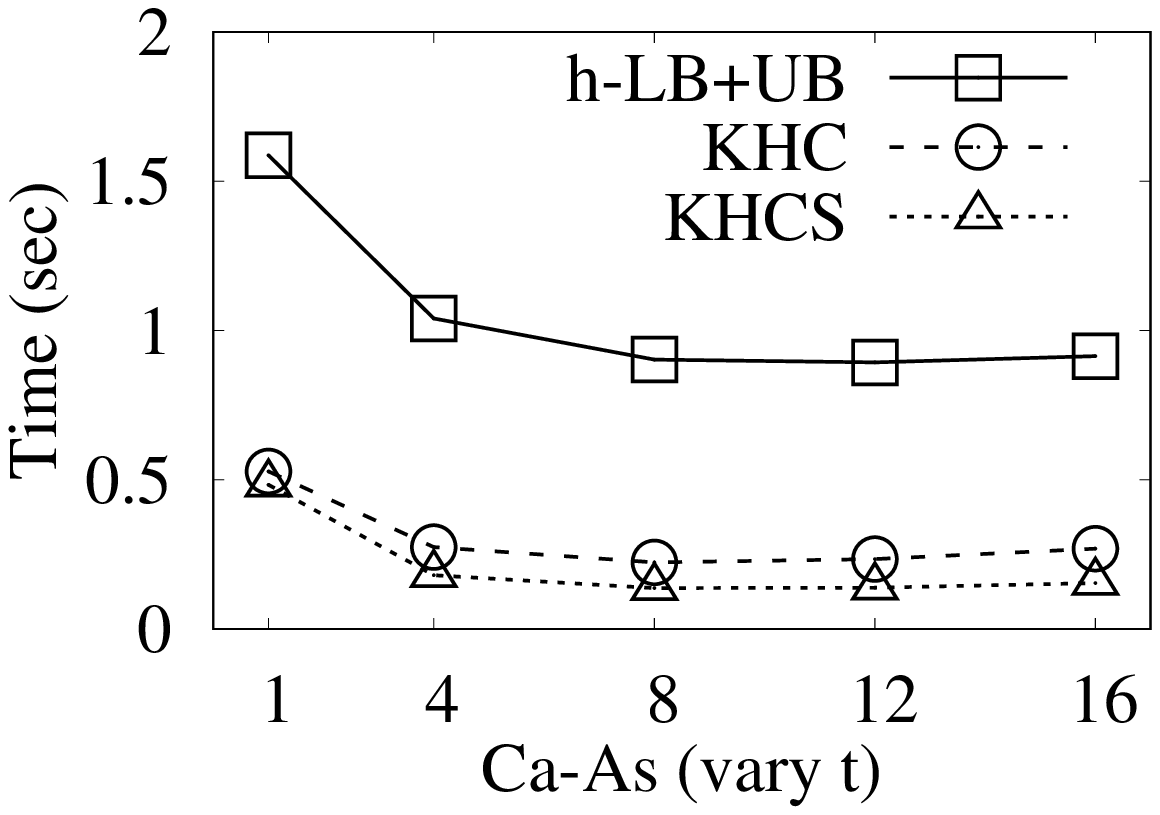}
			}
			\subfigure[{\scriptsize $h=2$}]{
				\includegraphics[width=0.38\columnwidth, height=2.5cm]{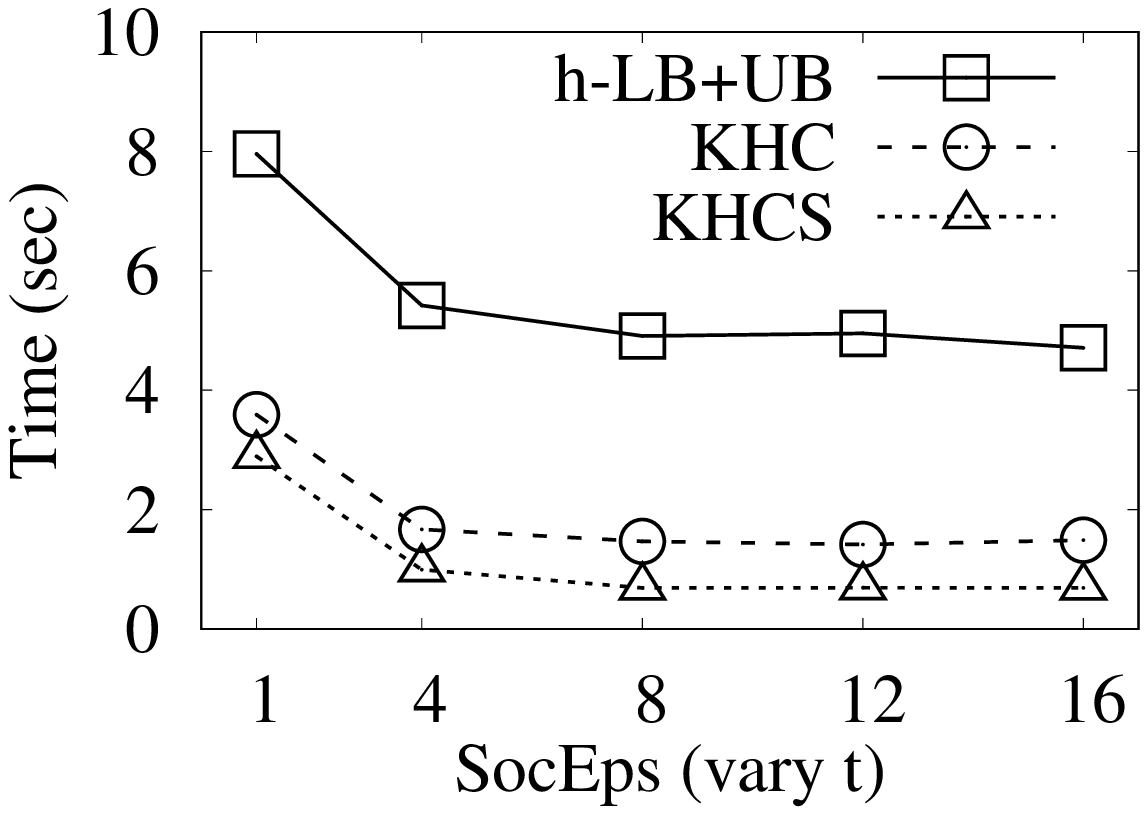}
			}
			\subfigure[{\scriptsize $h=2$}]{
				\includegraphics[width=0.38\columnwidth, height=2.5cm]{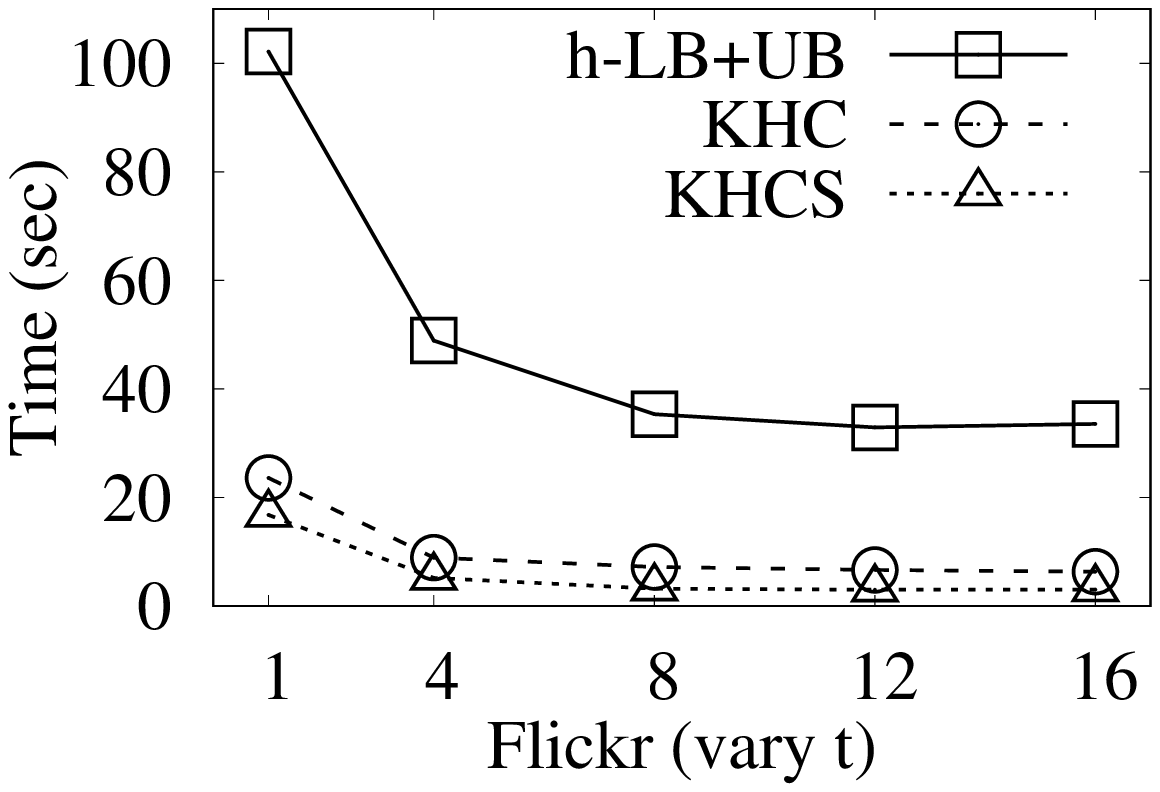}
			}
			\subfigure[{\scriptsize $h=2$}]{
				\includegraphics[width=0.38\columnwidth, height=2.5cm]{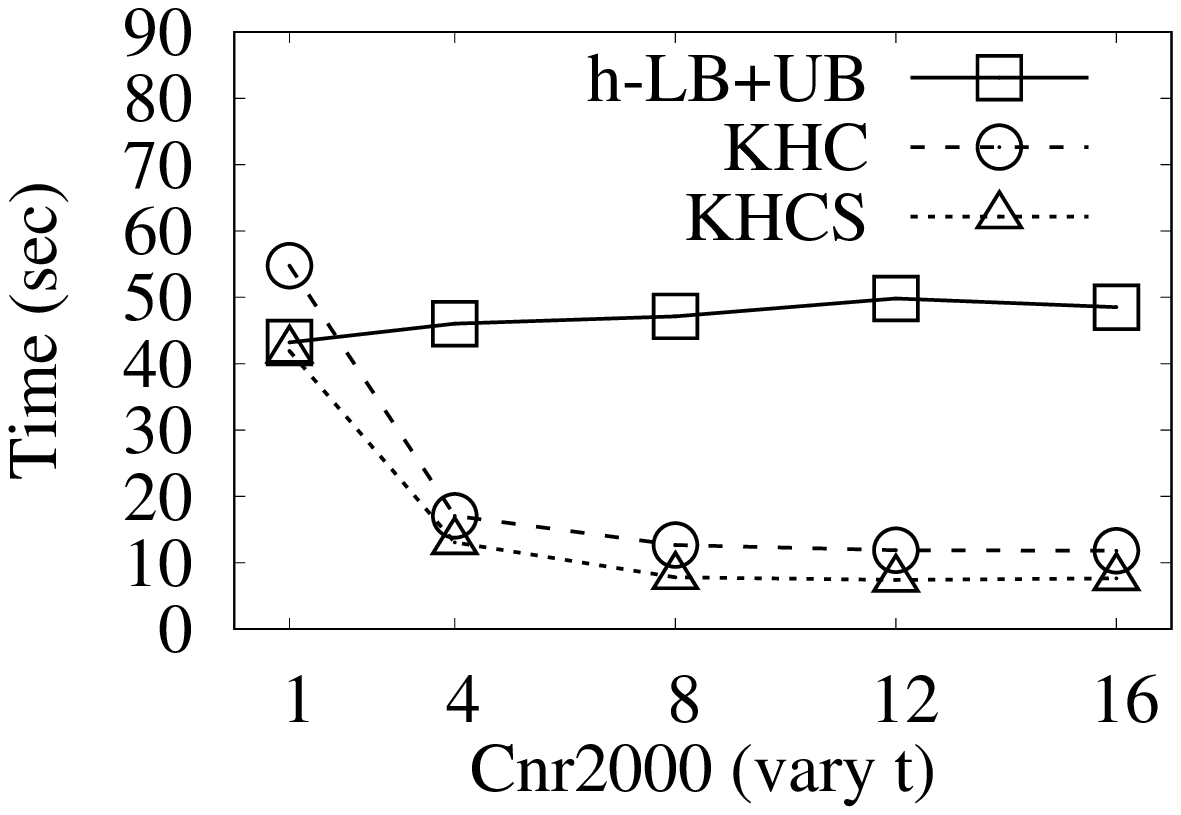}
			}\vspace*{-0.3cm}\\
			\subfigure[{\scriptsize $h=3$}]{
				\includegraphics[width=0.38\columnwidth, height=2.5cm]{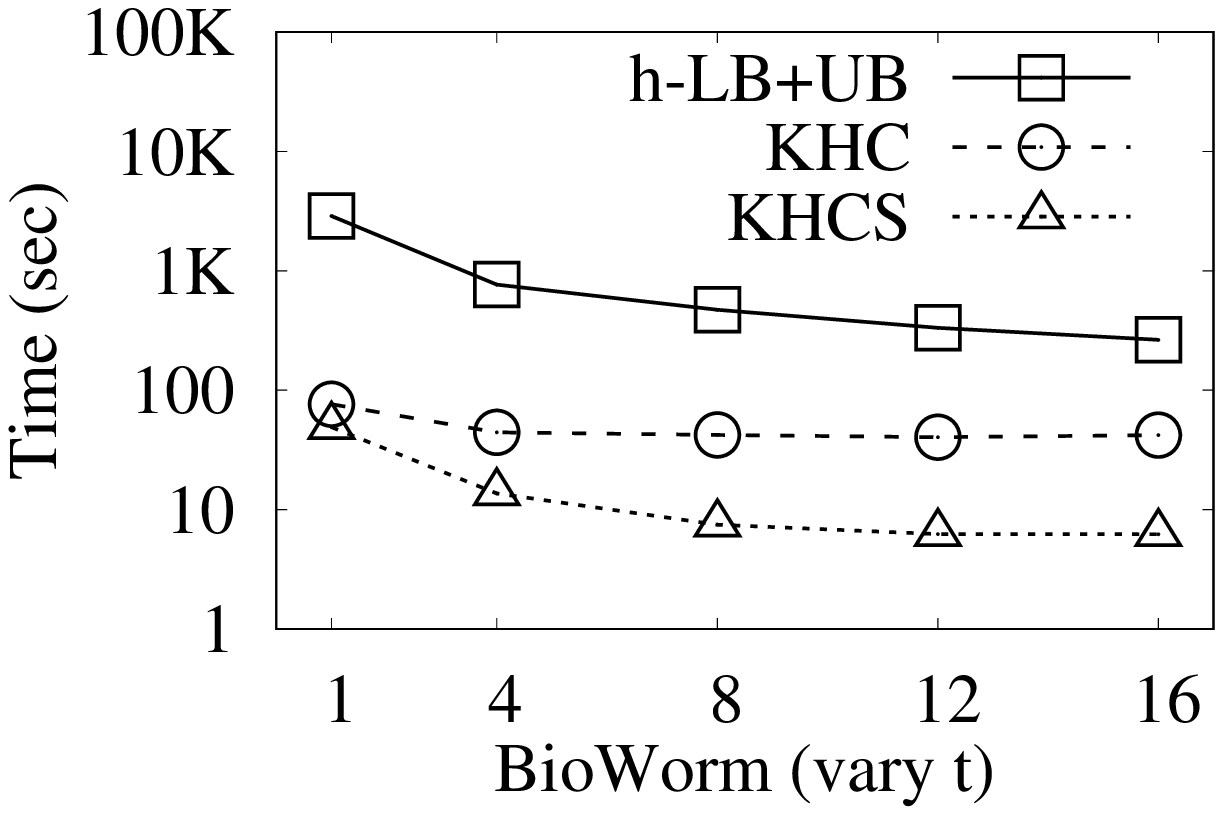}
			}
			\subfigure[{\scriptsize $h=3$}]{
				\includegraphics[width=0.38\columnwidth, height=2.5cm]{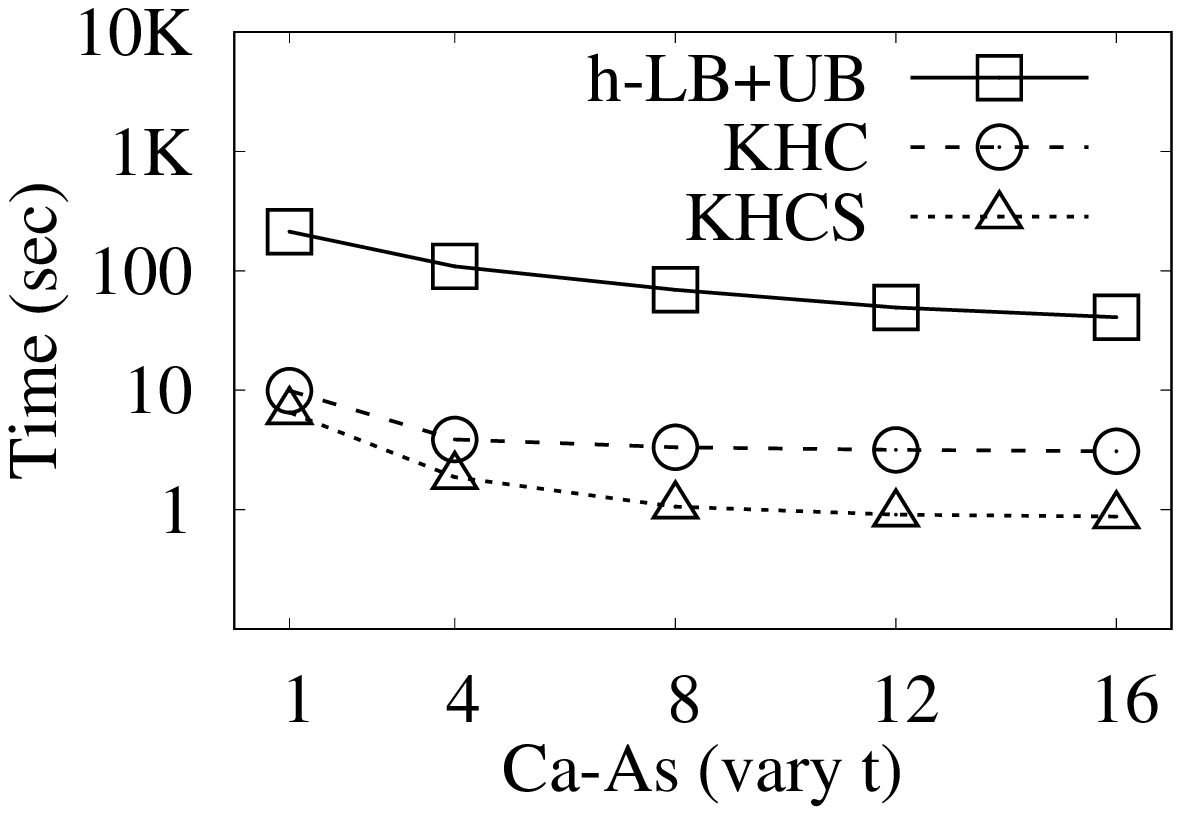}
			}
			\subfigure[{\scriptsize $h=3$}]{
				\includegraphics[width=0.38\columnwidth, height=2.5cm]{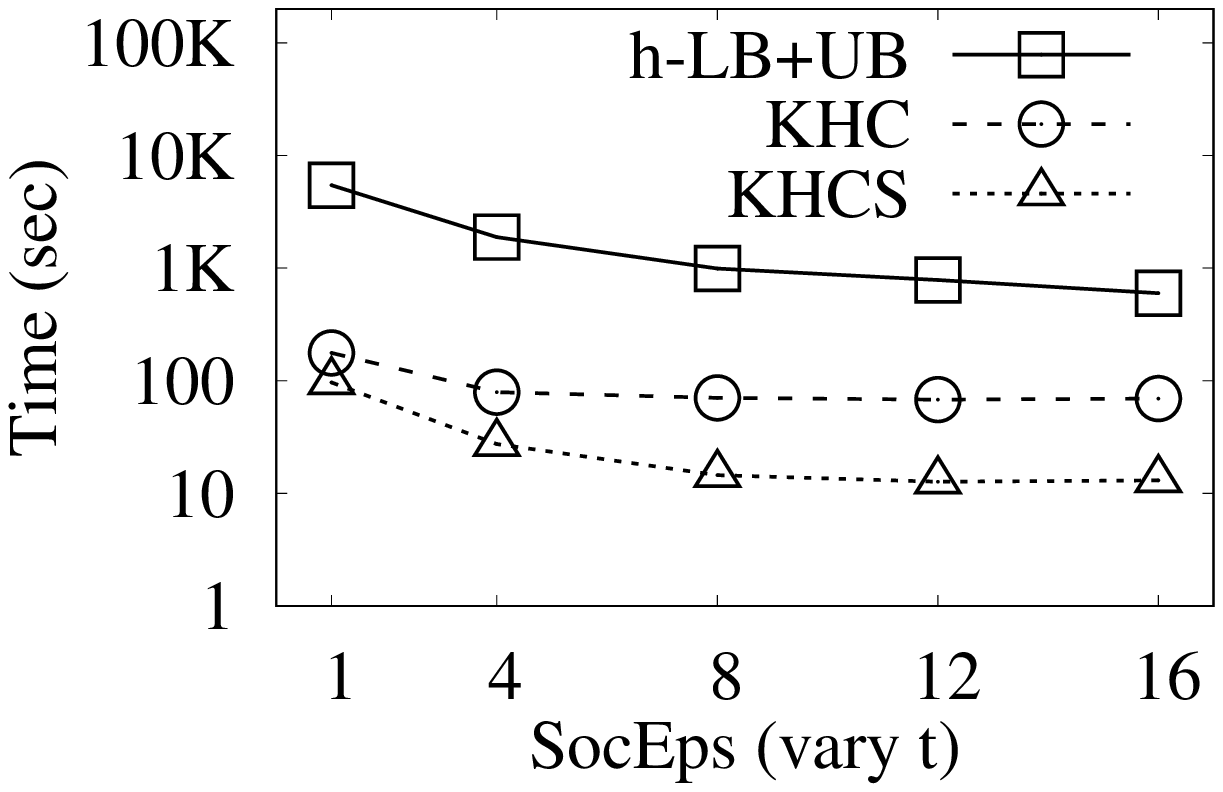}
			}
			\subfigure[{\scriptsize $h=3$}]{
				\includegraphics[width=0.38\columnwidth, height=2.5cm]{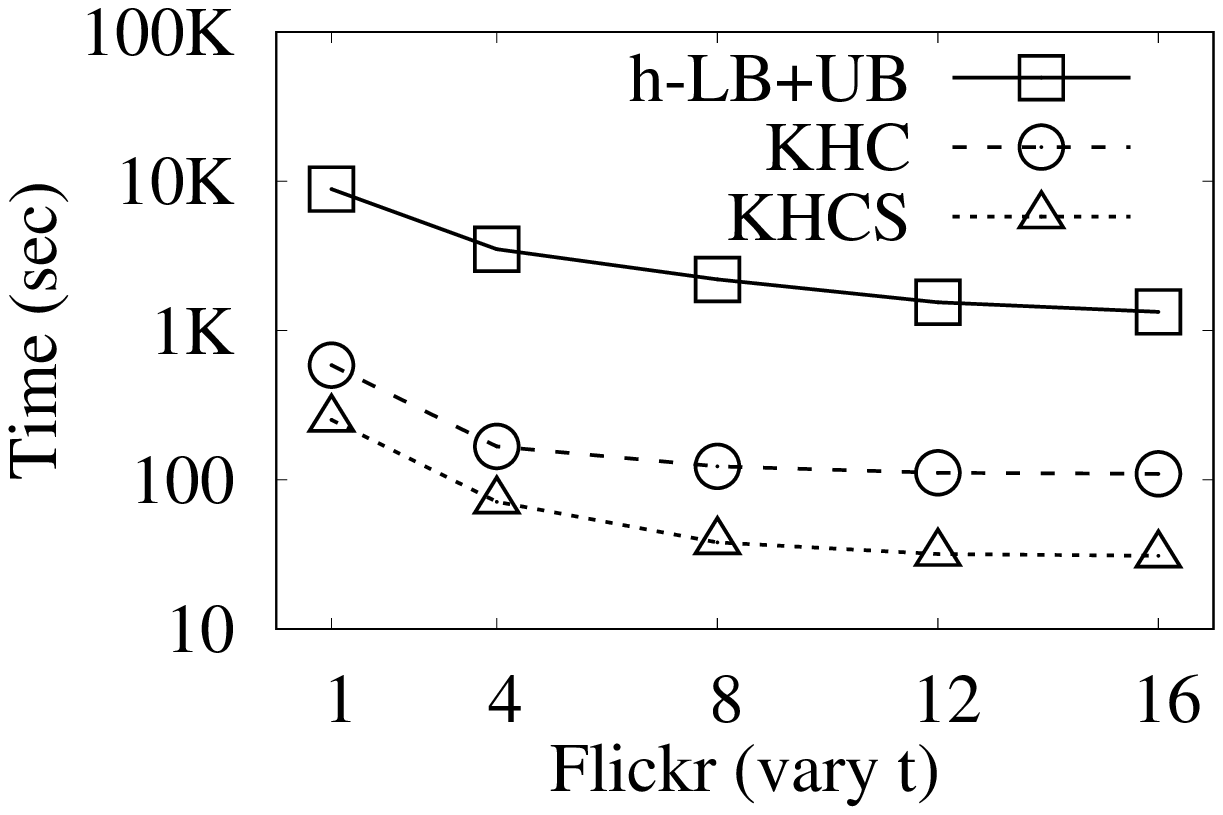}
			}
			\subfigure[{\scriptsize $h=3$}]{
				\includegraphics[width=0.38\columnwidth, height=2.5cm]{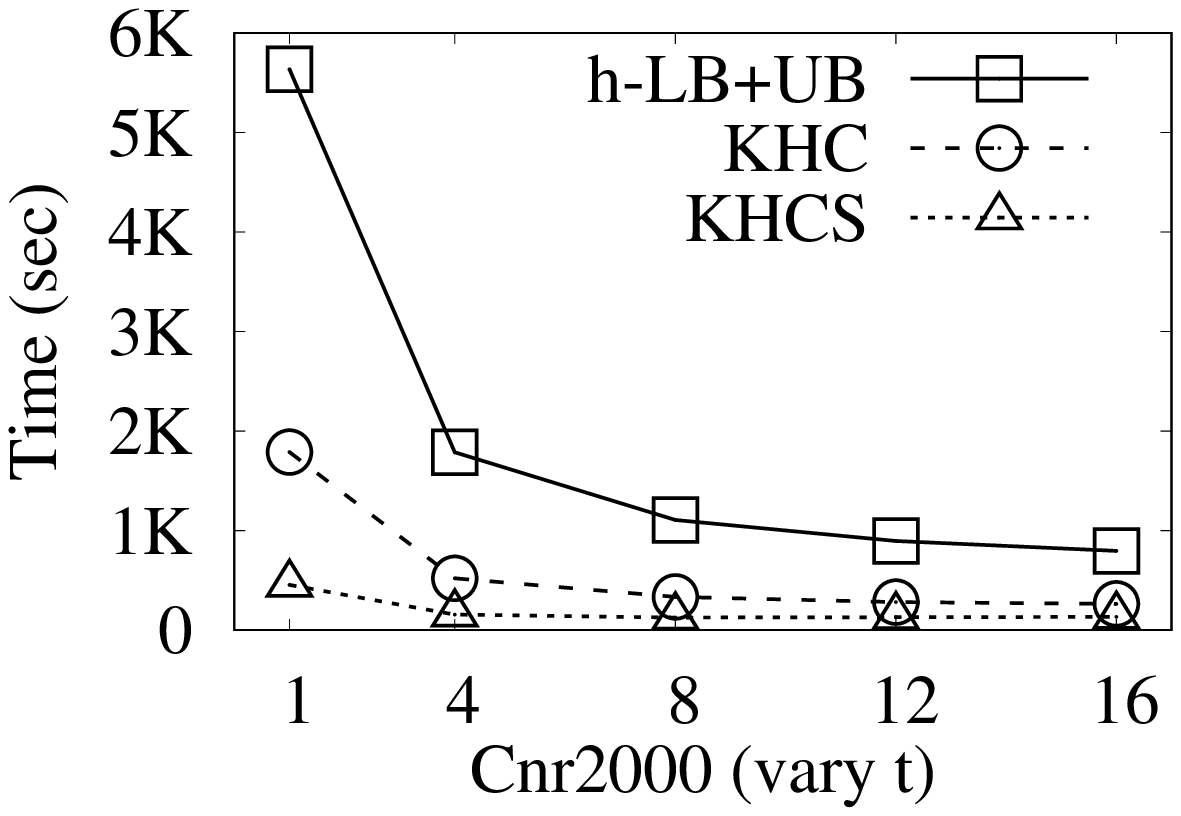}
			}\vspace*{-0.3cm}\\
		
			\subfigure[{\scriptsize $h=4$}]{
				\includegraphics[width=0.38\columnwidth, height=2.5cm]{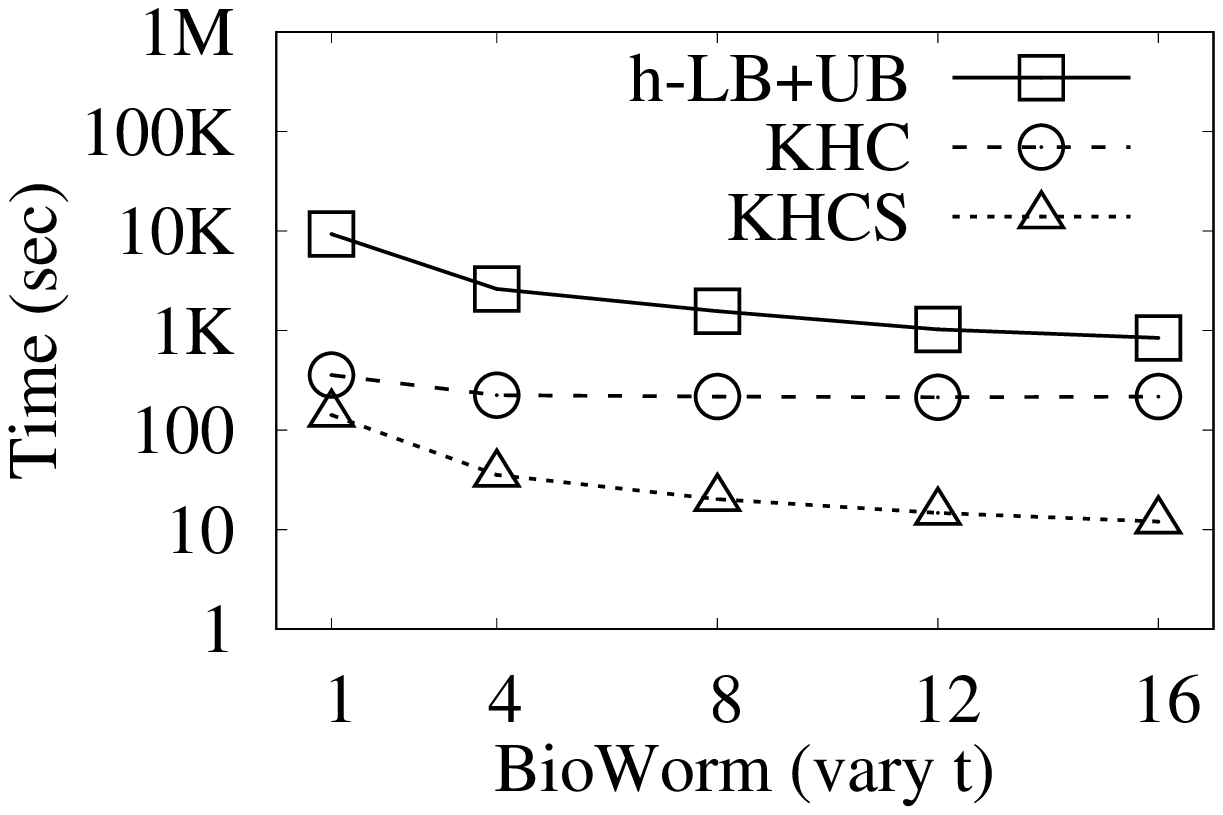}
			}
			\subfigure[{\scriptsize $h=4$}]{
				\includegraphics[width=0.38\columnwidth, height=2.5cm]{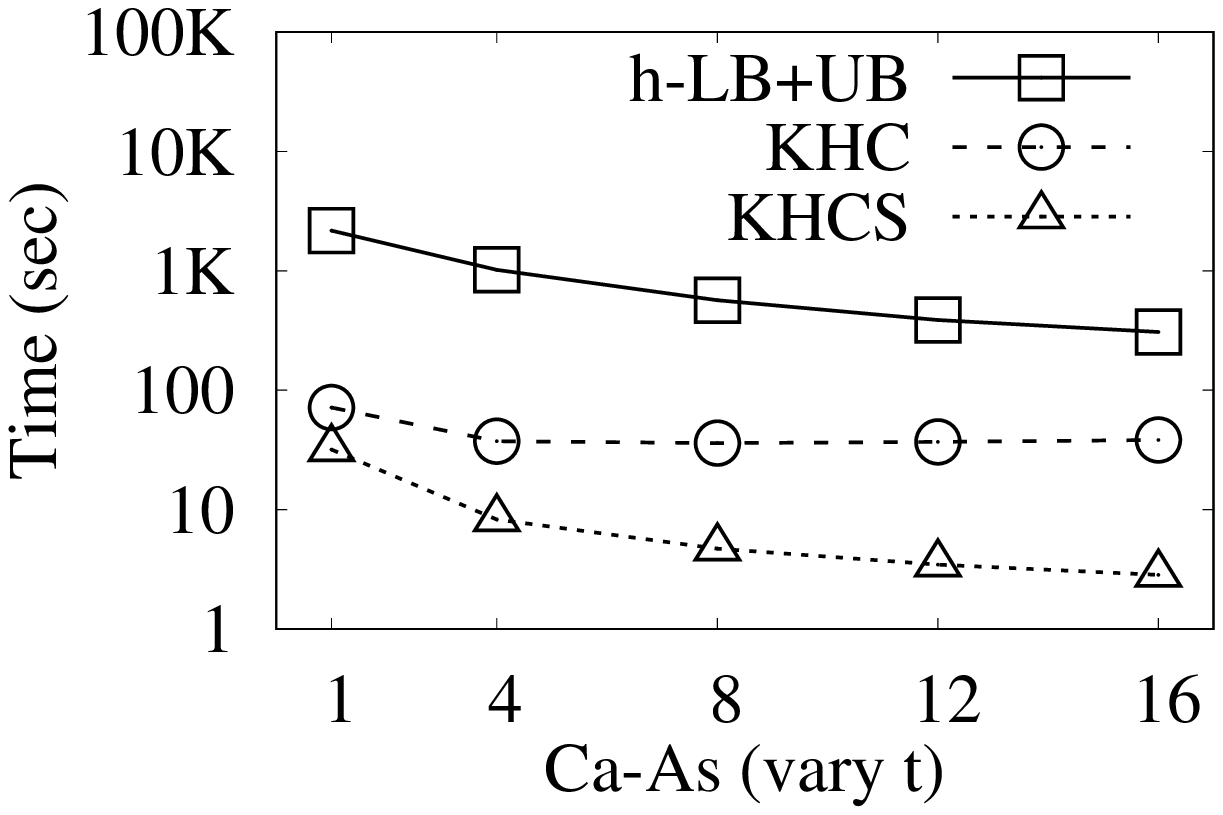}
			}\subfigure[{\scriptsize $h=4$}]{
				\includegraphics[width=0.38\columnwidth, height=2.5cm]{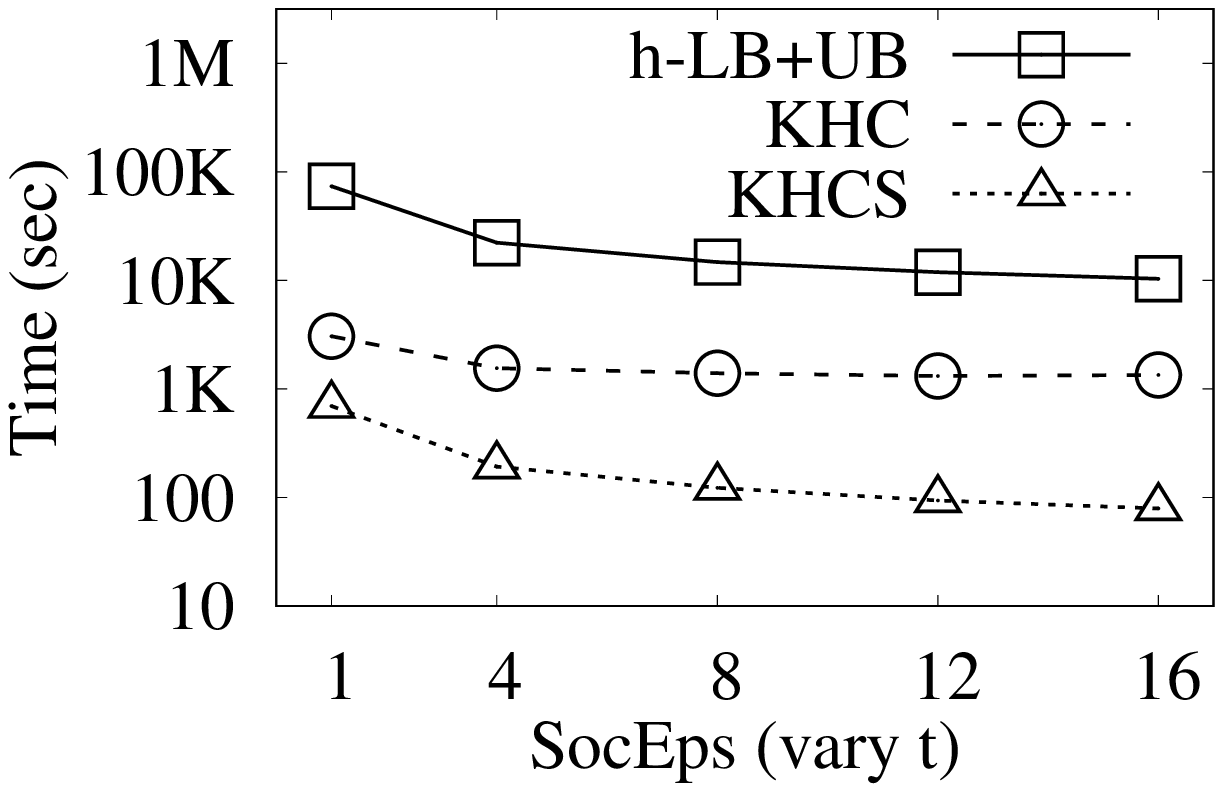}
			}
			\subfigure[{\scriptsize $h=4$}]{
				\includegraphics[width=0.38\columnwidth, height=2.5cm]{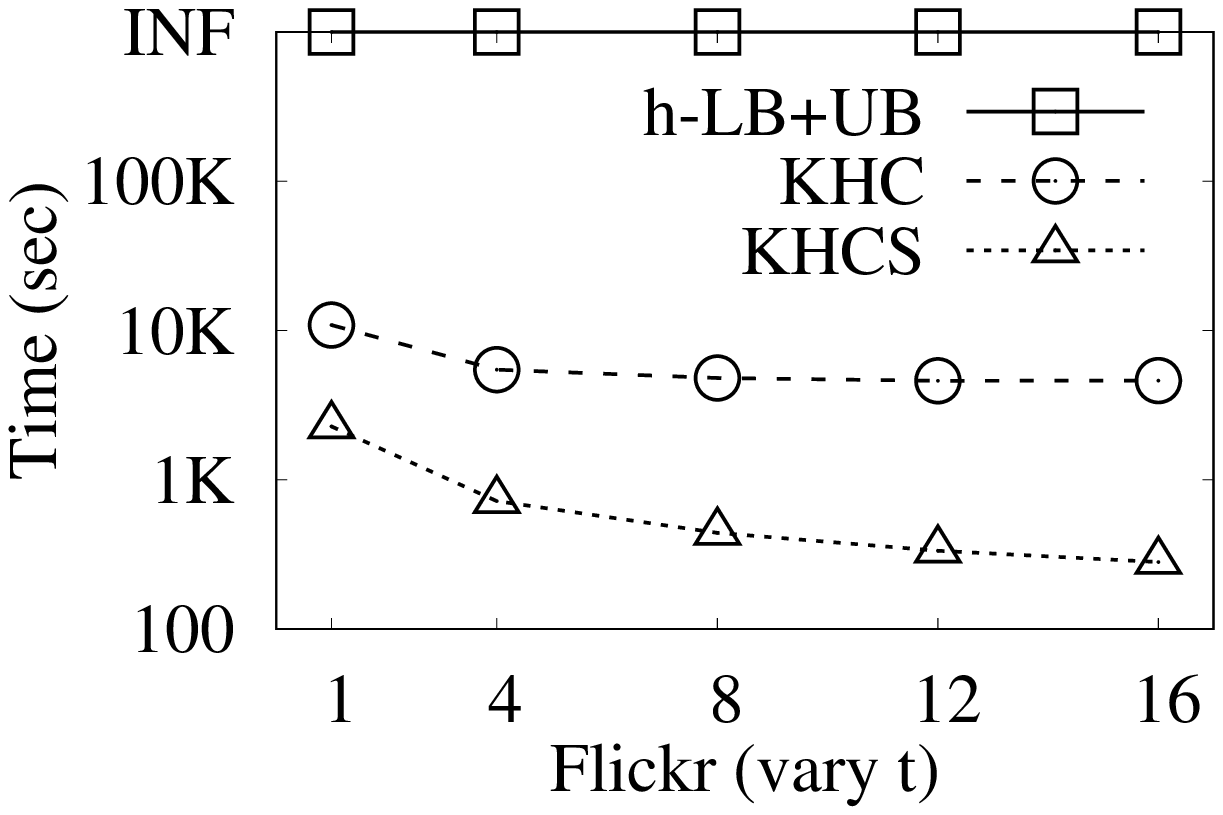}
			}
			\subfigure[{\scriptsize $h=4$}]{
				\includegraphics[width=0.38\columnwidth, height=2.5cm]{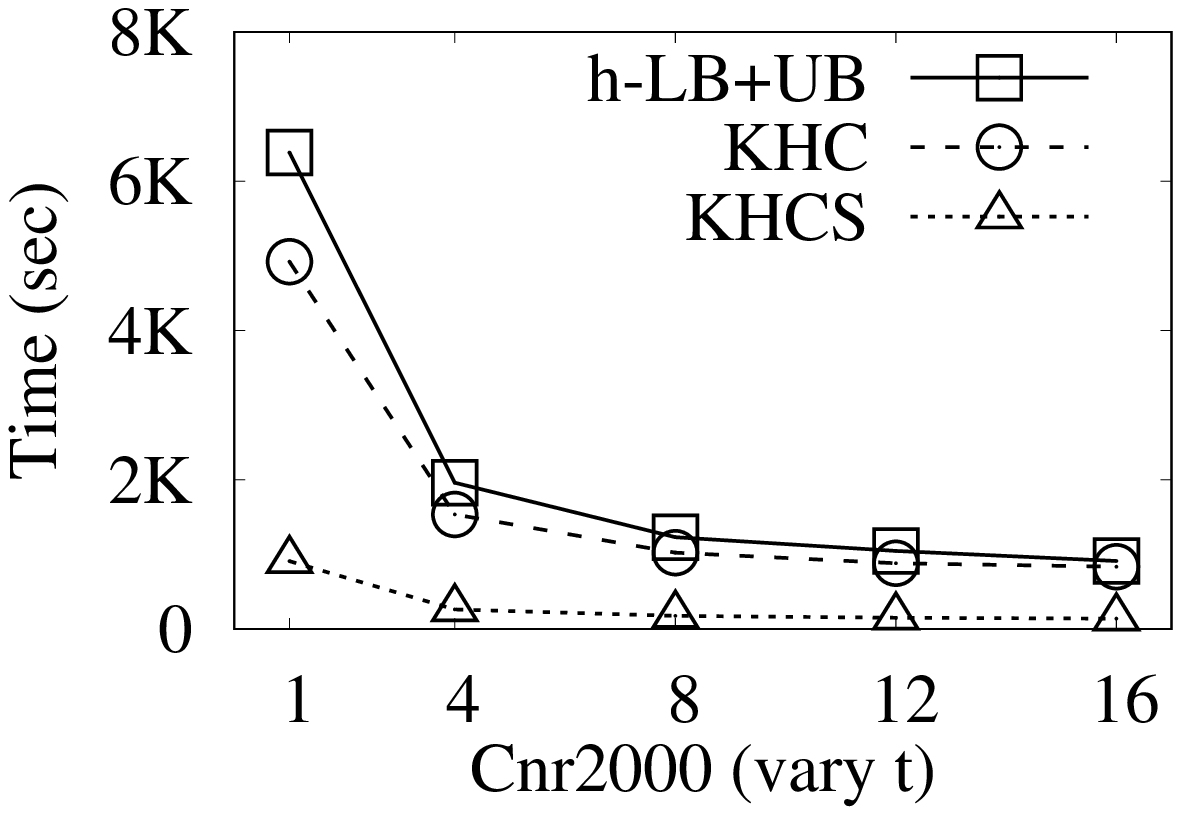}
			}\vspace*{-0.3cm}\\
			\subfigure[{\scriptsize $h=5$}]{
				\includegraphics[width=0.38\columnwidth, height=2.5cm]{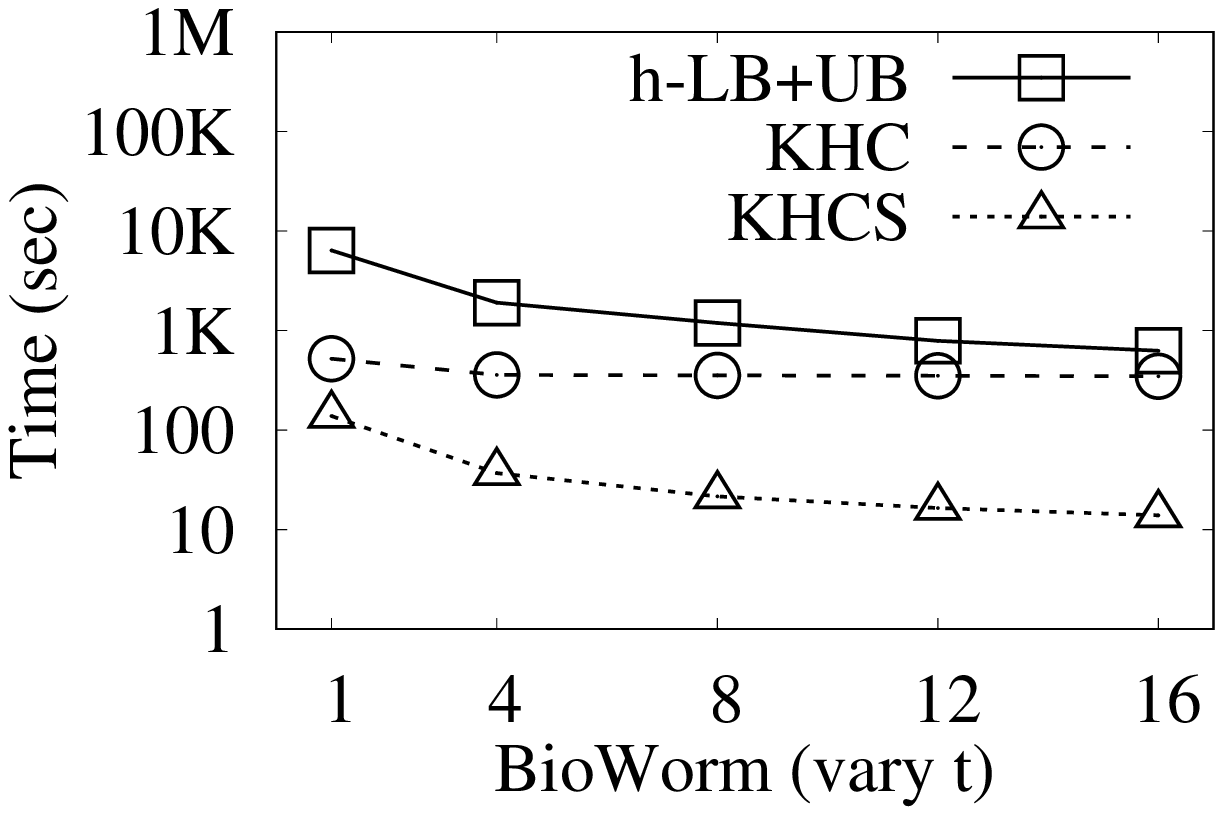}
			}
			\subfigure[{\scriptsize $h=5$}]{
				\includegraphics[width=0.38\columnwidth, height=2.5cm]{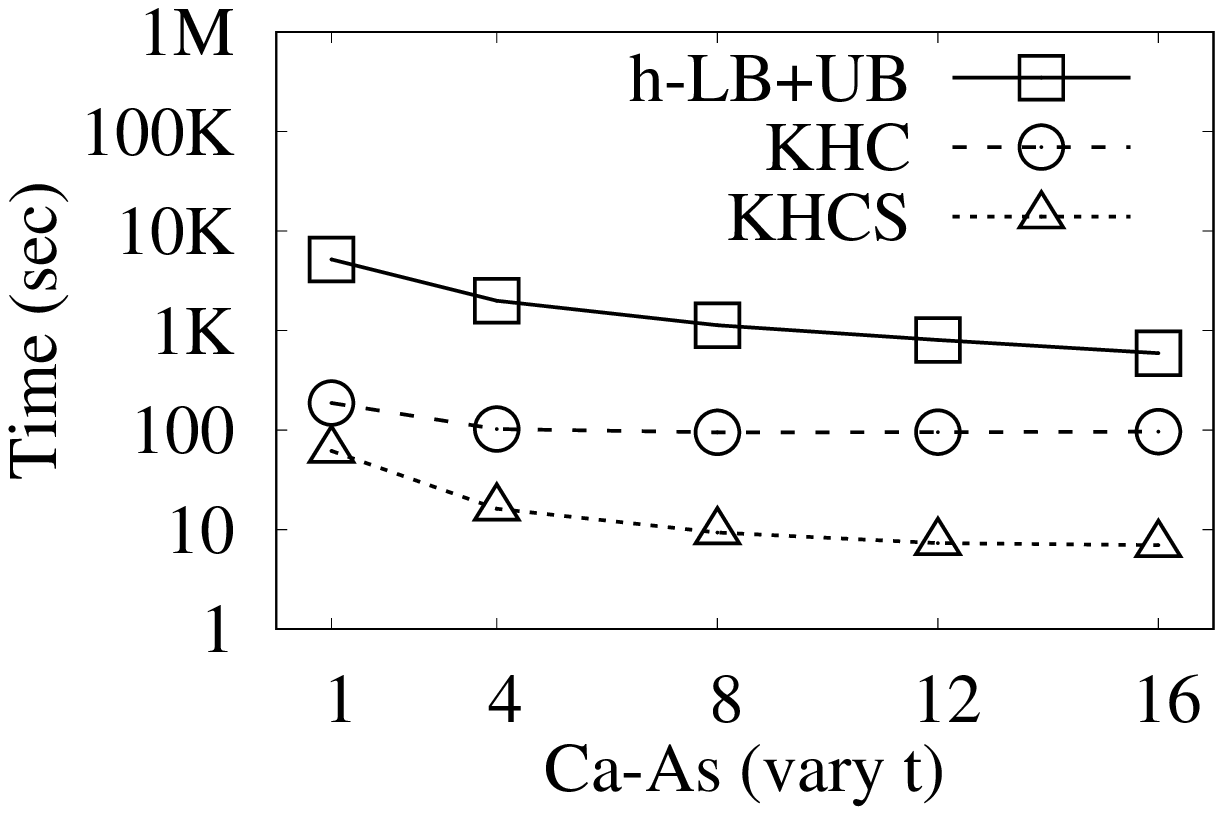}
			}\subfigure[{\scriptsize $h=5$}]{
				\includegraphics[width=0.38\columnwidth, height=2.5cm]{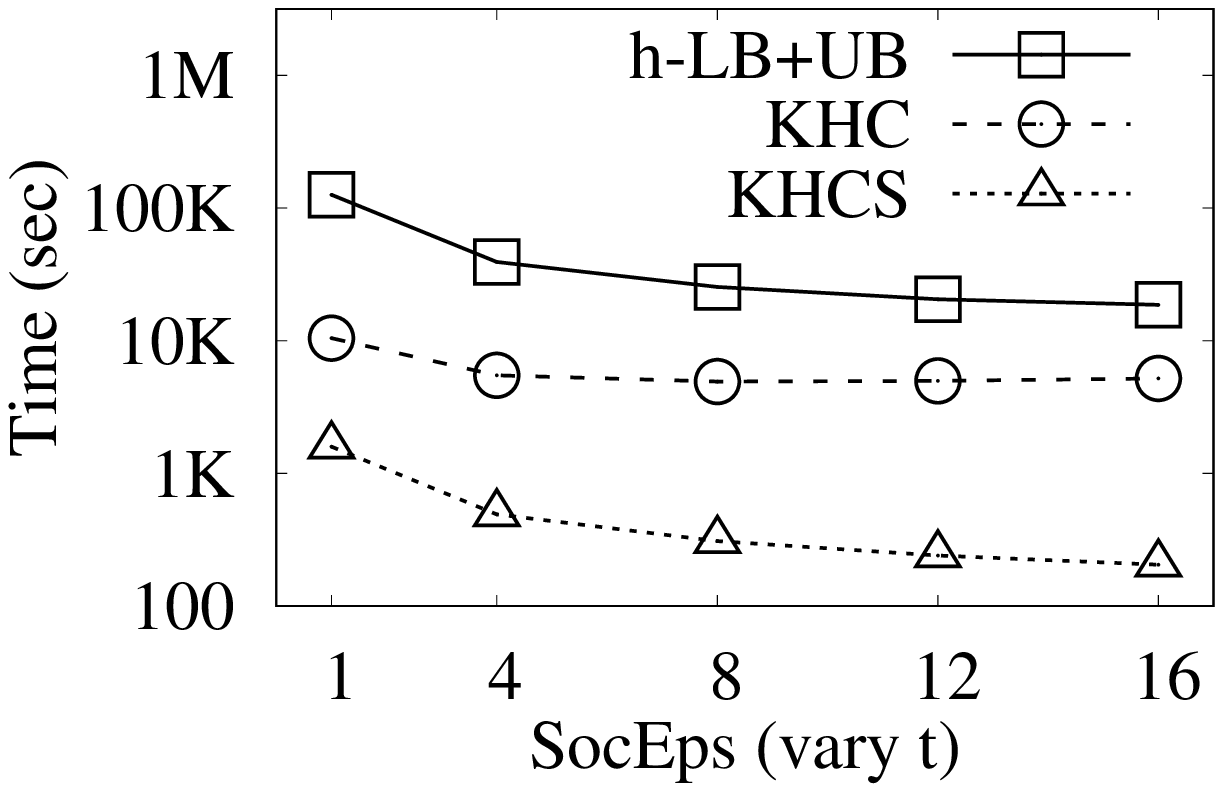}
			}
			\subfigure[{\scriptsize $h=5$}]{
				\includegraphics[width=0.38\columnwidth, height=2.5cm]{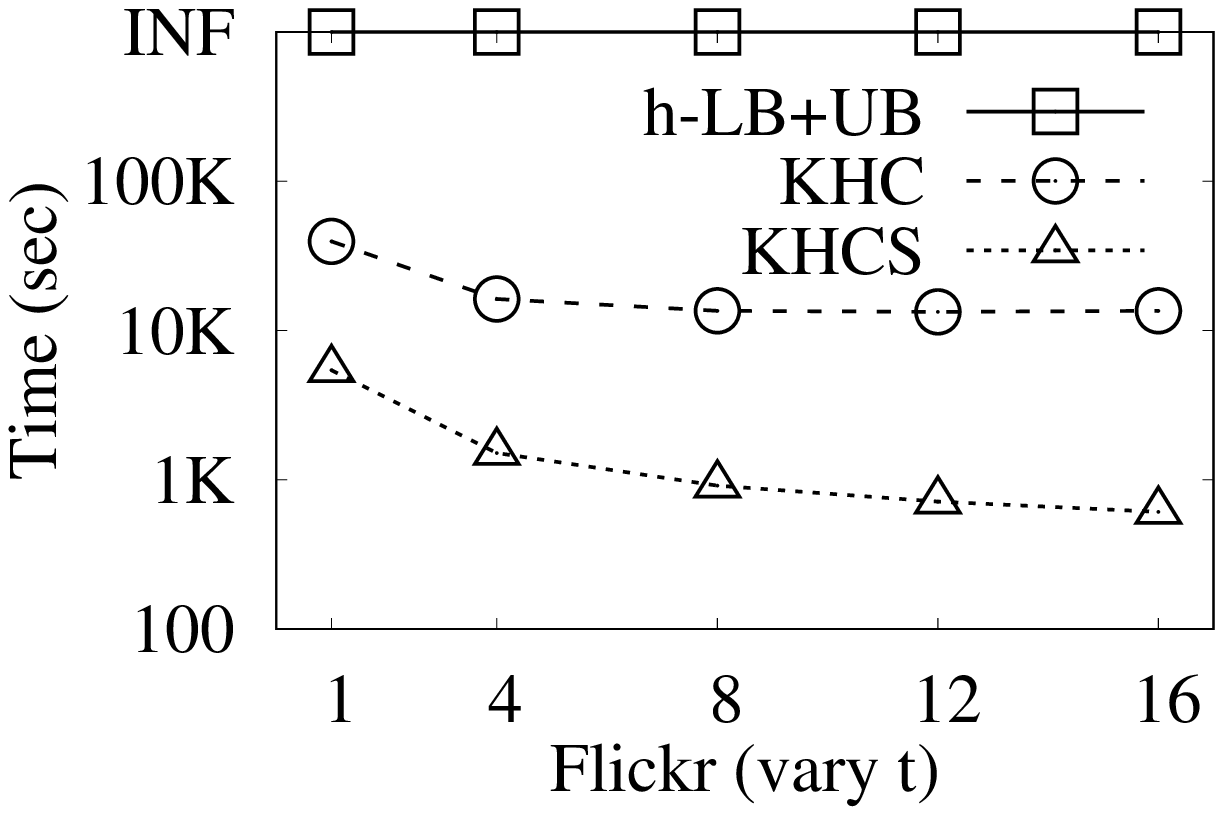}
			}
			\subfigure[{\scriptsize $h=5$}]{
				\includegraphics[width=0.38\columnwidth, height=2.5cm]{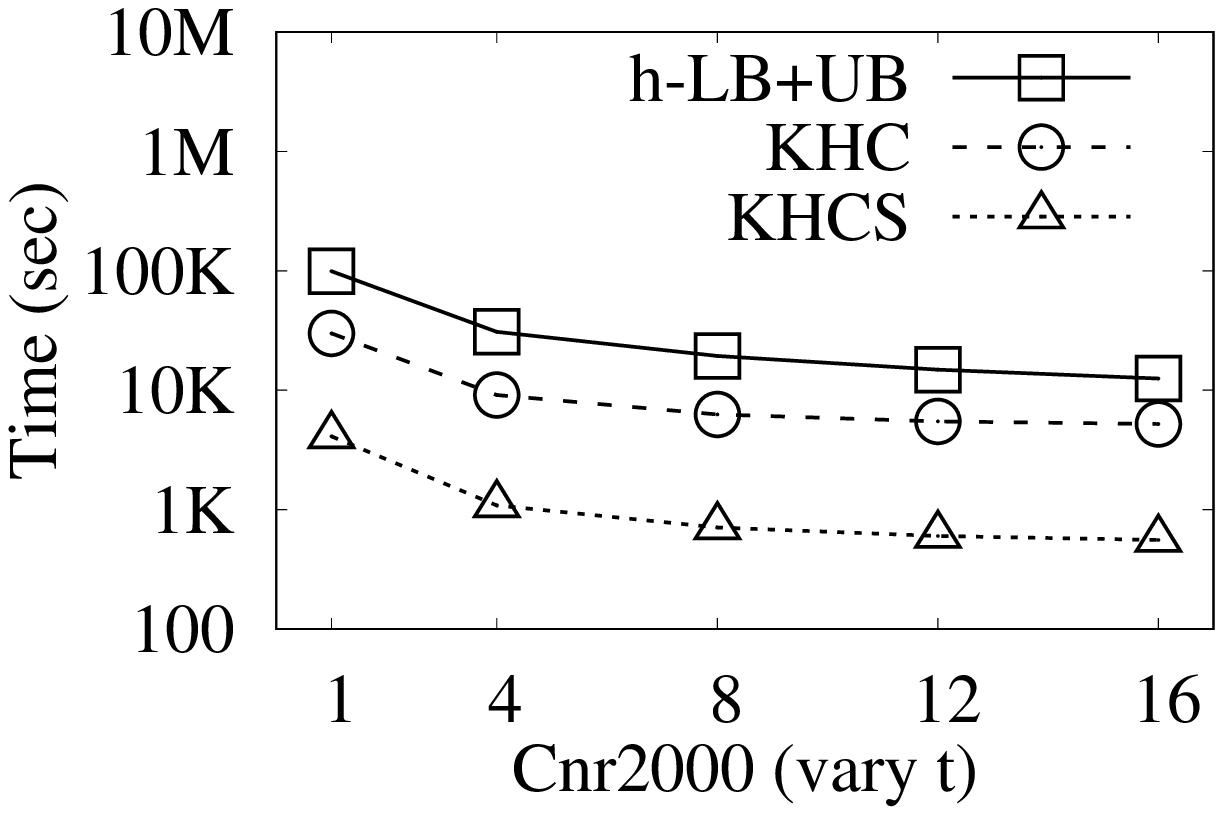}
			}
		\end{tabular}
	\end{center}
	\vspace*{-0.2cm}
	\caption{Runtime of different parallel algorithms with varying $t$ (the number of threads)}
	\vspace*{-0.3cm}
	\label{fig:exp2:multy-thread-vary-h}
\end{figure*}

\begin{figure}[t!] 
	\begin{center}
		\begin{tabular}[t]{c}\hspace*{-0.3cm}
			\subfigure[{\scriptsize $h=2$}]{
				\includegraphics[width=0.45\columnwidth, height=2.5cm]{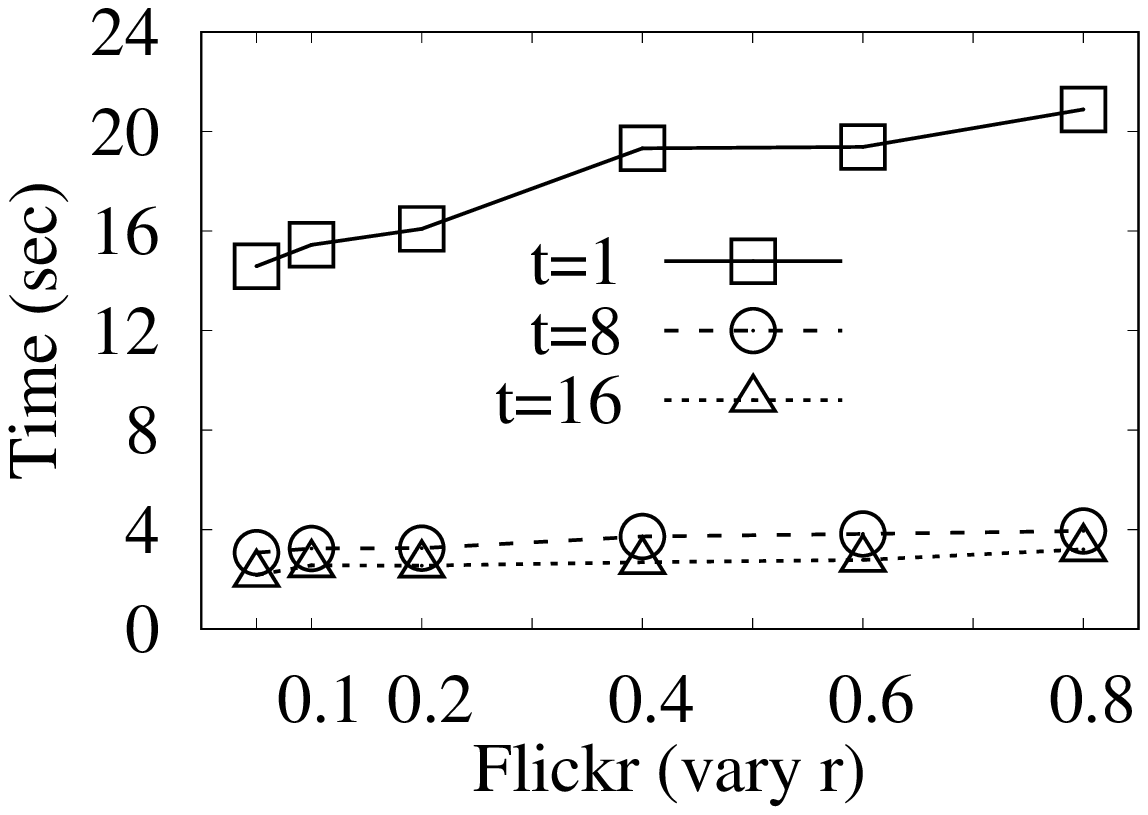}
			}\hspace*{-0.3cm}
			\subfigure[{\scriptsize $h=3$}]{
				\includegraphics[width=0.45\columnwidth, height=2.5cm]{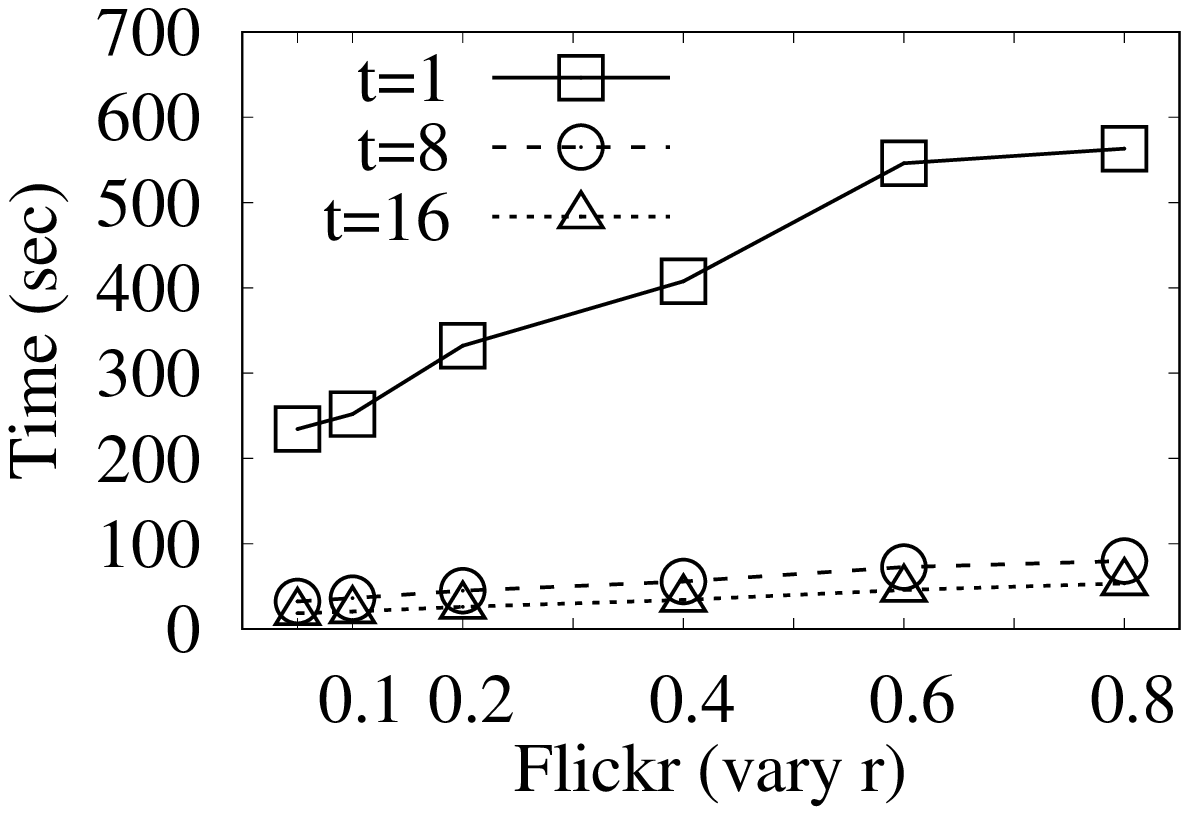}
			}
			\vspace*{-0.3cm}\\
			\subfigure[{\scriptsize $h=4$}]{
				\includegraphics[width=0.45\columnwidth, height=2.5cm]{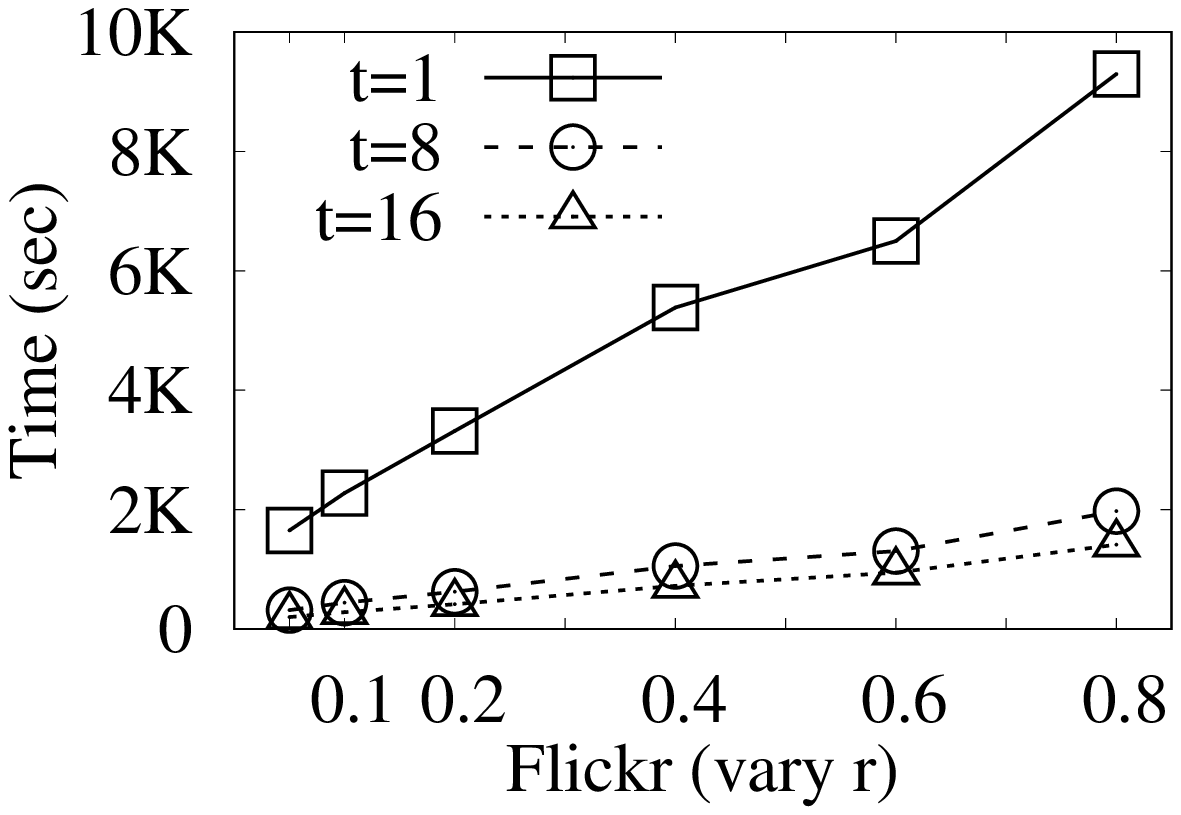}
			}	\hspace*{-0.3cm}
			\subfigure[{\scriptsize $h=5$}]{
				\includegraphics[width=0.45\columnwidth, height=2.5cm]{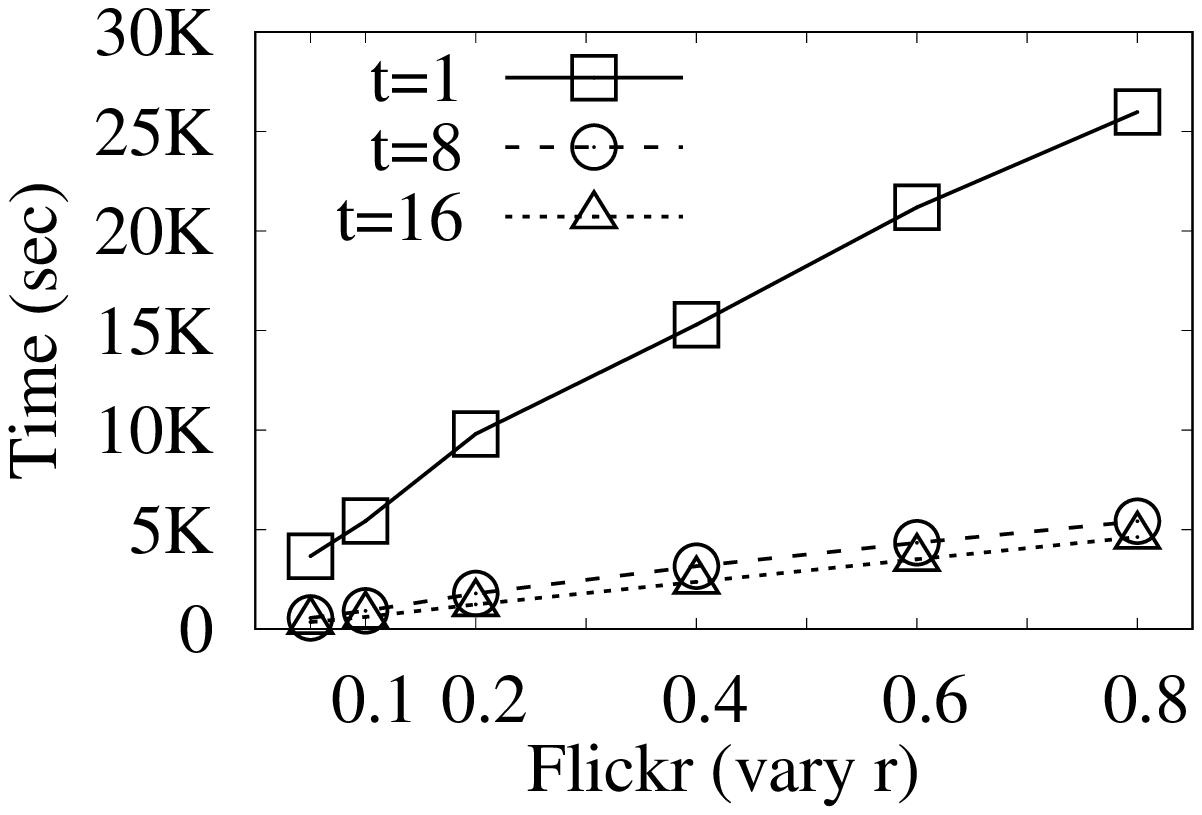}
			}
		\end{tabular}
	\end{center}
	\vspace*{-0.3cm}
	\caption{Runtime of the \khcs algorithm}
	\vspace*{-0.5cm}
	\label{fig:exp4:parallel-res-vary-r}
\end{figure}

\stitle{Exp-2: Efficiency of different parallel algorithms.} Here we evaluate the performance of the parallelized versions of \hbz, \khc, and \khcs. To this end, we vary the number of threads $t$ from 1 to 16 with different $h$ values. Fig.~\ref{fig:exp2:multy-thread-vary-h} shows the results on five datasets, and similar results can also be observed on the other datasets. As expected, the runtime of all the three algorithms decreases with increasing $t$. We also observe that if $t\ge 8$, the speedup ratios of all algorithms do not significantly increase as $t$ grows on all datasets. This is because, for all algorithms, the parallel performance mainly relies on the size of the bucket $B$ that maintains all the vertices having the minimum \hdegrees. In some iterations of each algorithm, the size of the bucket $B$ might be smaller than $t$ which limits the parallel speedup ratio of the algorithm. In addition, we also notice that the speedup ratio of \khcs is significantly higher than those of \hbz and \khc. For example, when $h=3$, the parallel \khcs algorithm with $t=16$ can achieve nearly $9\times$ speedup over the sequential \khcs algorithm on the \flic dataset (Fig.~\ref{fig:exp2:multy-thread-vary-h}(i)). However, the speedup ratios of the parallel \hbz and \khc algorithms are around 6.6 and 5.3 on \flic respectively, given $t=16$ and $h=3$. 

\stitle{Exp-3: Runtime of \khcs with varying $r$.} We evaluate the runtime of \khcs with varying $r$ (sampling rate). Fig.~\ref{fig:exp4:parallel-res-vary-r} depicts the runtime of (parallel) \khcs when $r$ varies from 0.05 to 0.8. As expected, the runtime of \khcs increases when $r$ increases, because the graph is sparser with a smaller $r$ value. In addition, we also observe that \khcs can always achieve high speedup ratios at different sampling rates. For example, when $h=3$ and $r=0.2$, \khcs takes 332 seconds to compute all $(k,h)$-cores using a single thread, while it only takes 44 seconds and 26 seconds using 8 and 16 threads, respectively. These results further confirm the high efficiency of our parallel \khcs algorithm.

\begin{figure*}[t!]
	\begin{center}
		\begin{tabular}[t]{c}\hspace*{-0.3cm}
			\subfigure[{\scriptsize $h=2$, top-$1$}]{
				\includegraphics[width=0.45\columnwidth, height=2.5cm]{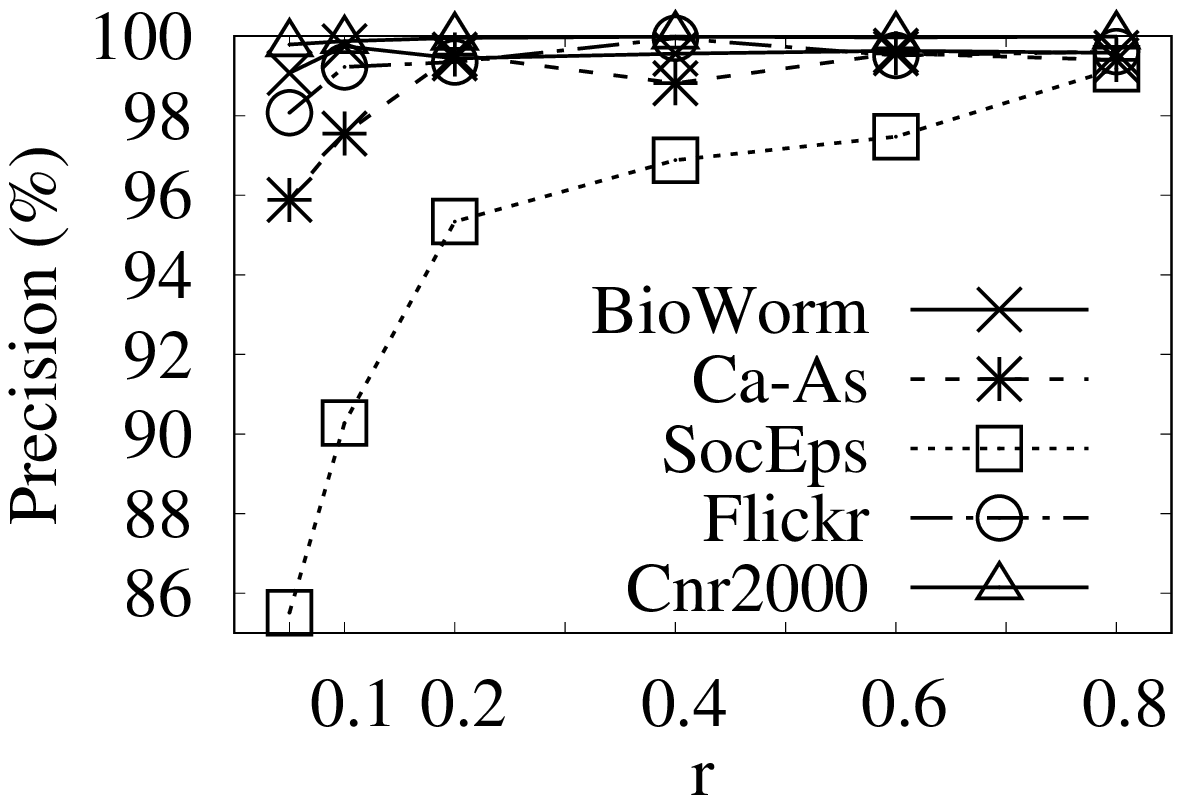}
			}\hspace*{0.3cm}
			\subfigure[{\scriptsize $h=3$, top-$1$}]{
				\includegraphics[width=0.45\columnwidth, height=2.5cm]{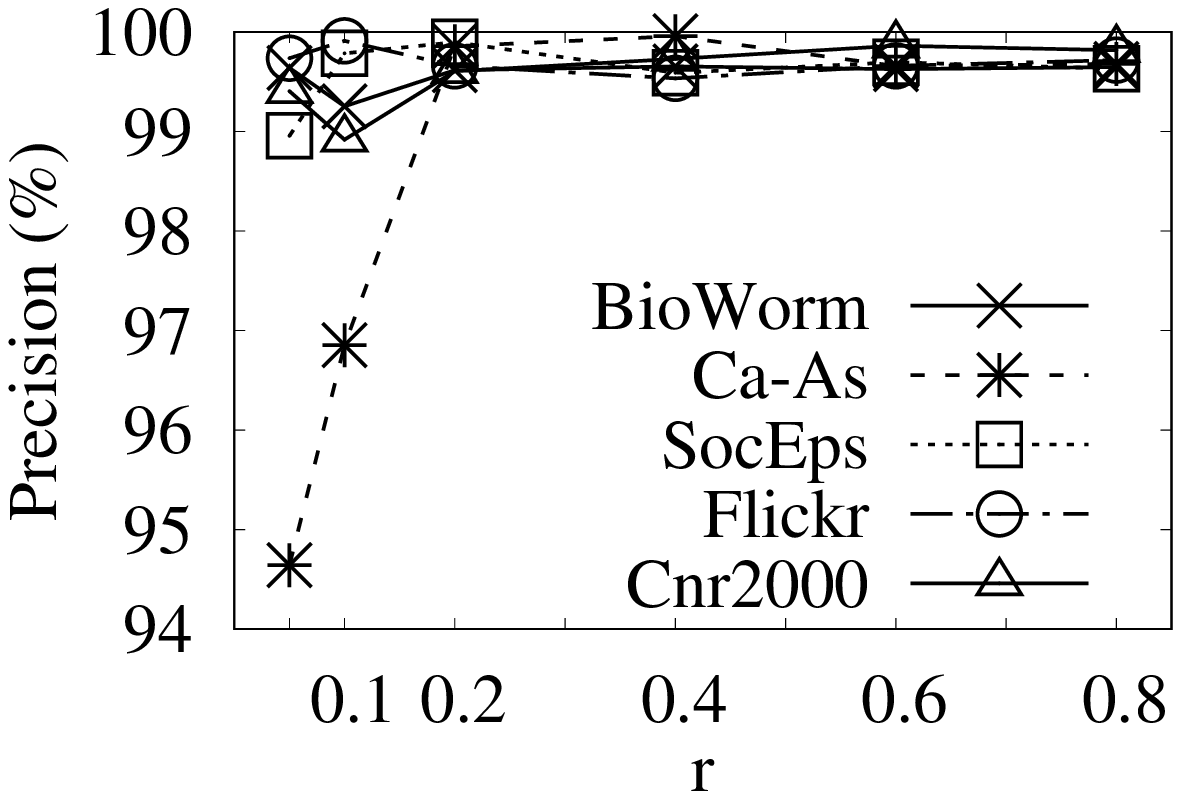}
			}\hspace*{0.3cm}
			\subfigure[{\scriptsize $h=4$, top-$1$}]{
				\includegraphics[width=0.45\columnwidth, height=2.5cm]{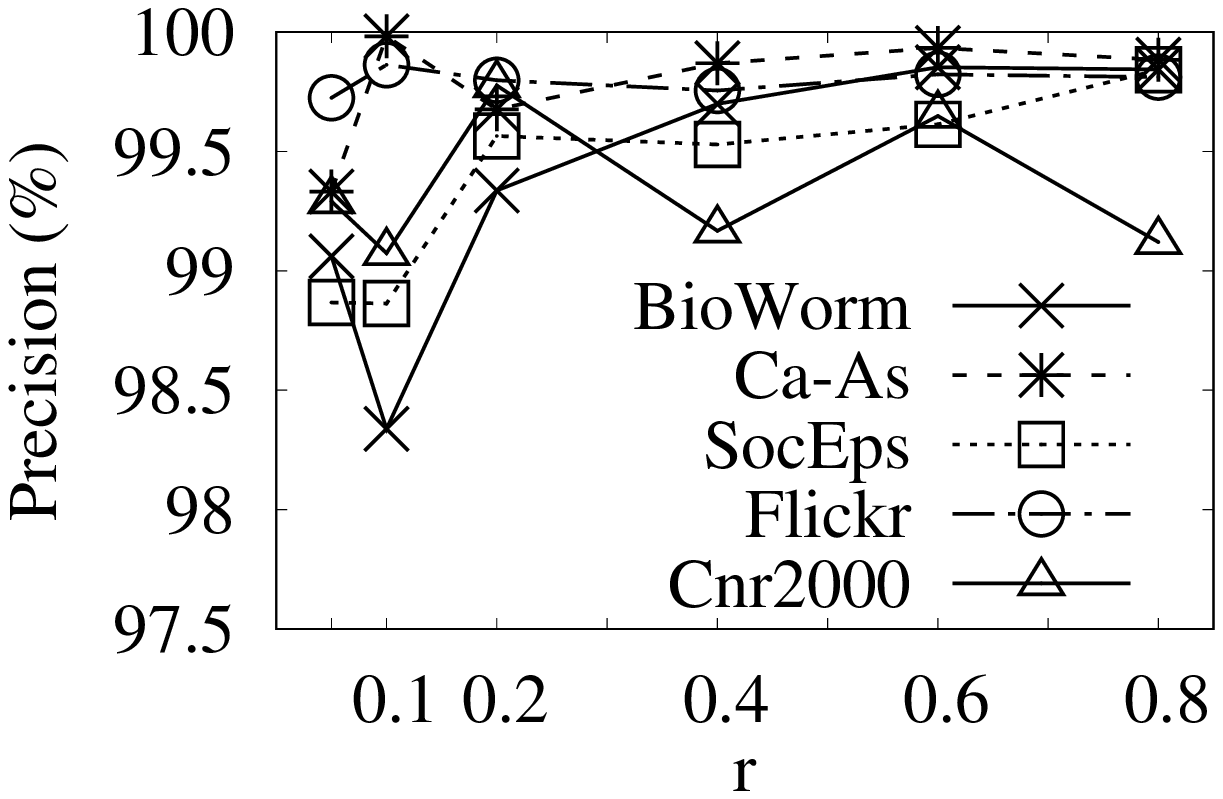}
			}\hspace*{0.3cm}
			\subfigure[{\scriptsize $h=5$, top-$1$}]{
				\includegraphics[width=0.45\columnwidth, height=2.5cm]{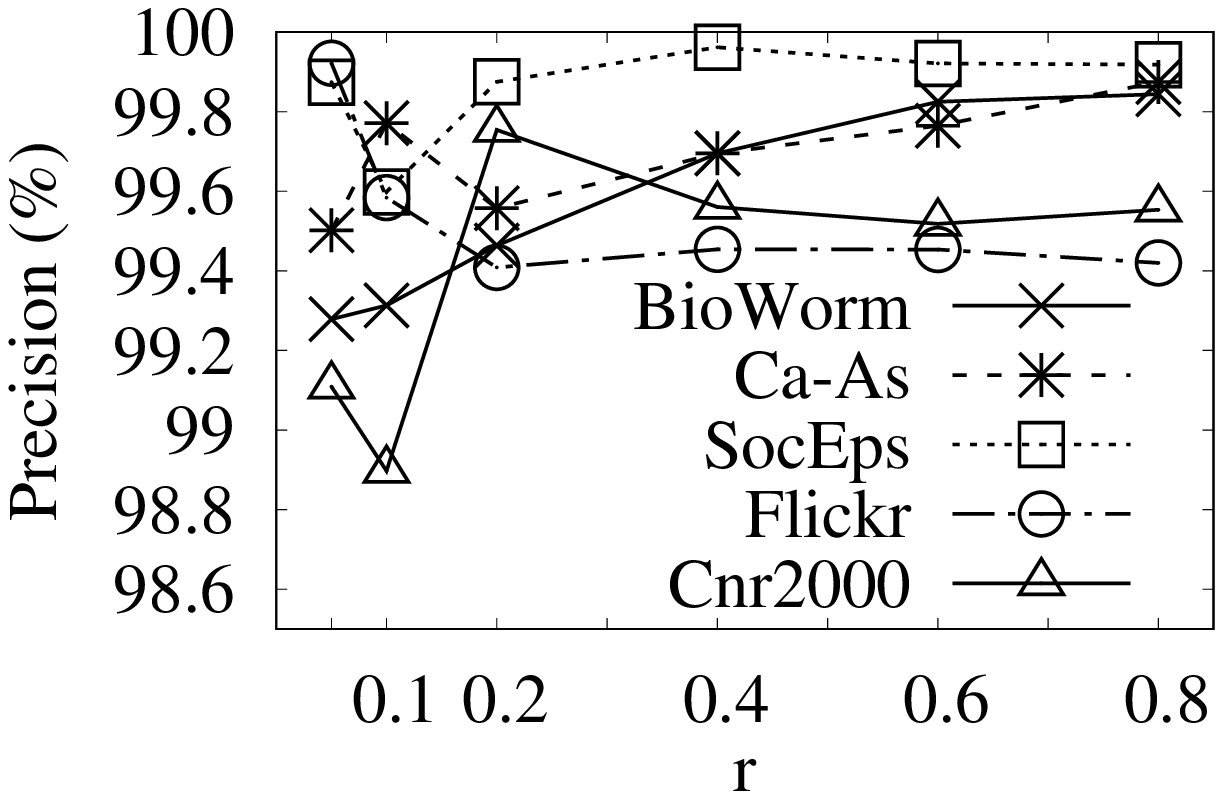}
			}
			\vspace*{-0.3cm}\\ 
			\subfigure[{\scriptsize $h=2$, top-$50$}]{
				\includegraphics[width=0.45\columnwidth, height=2.5cm]{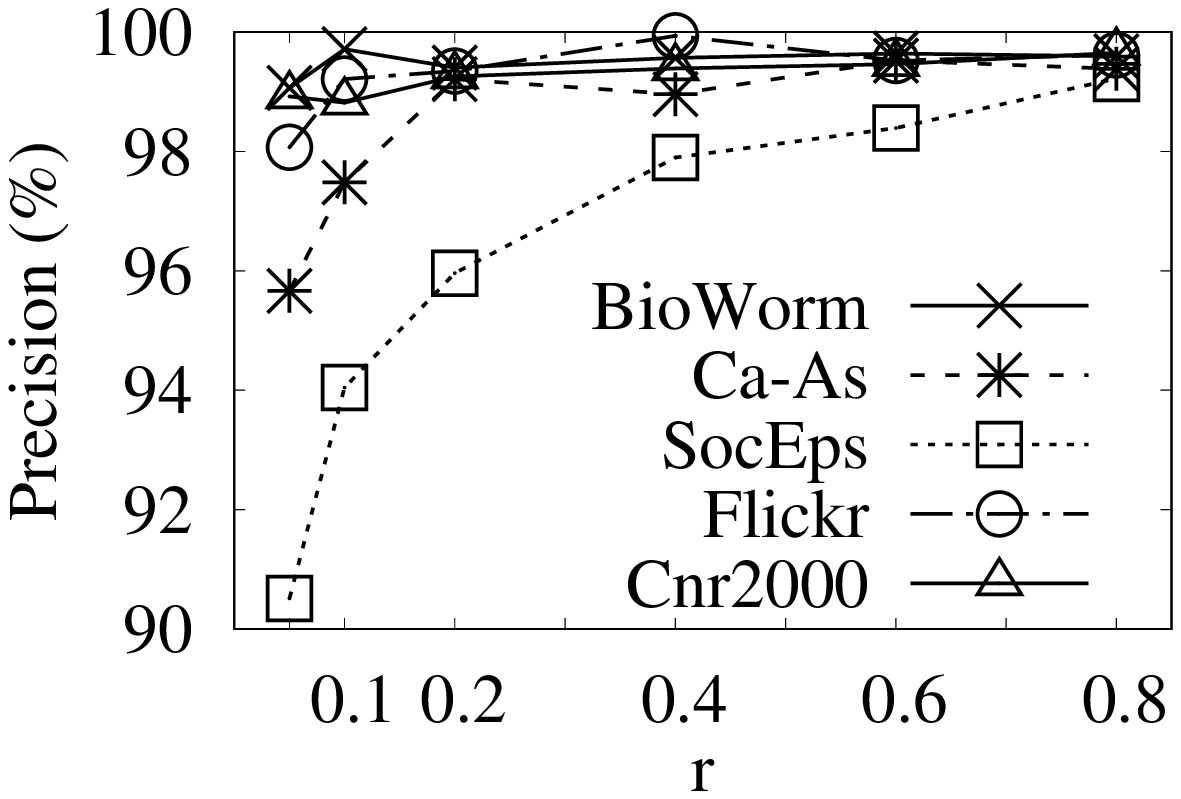}
			}\hspace*{0.3cm}
			\subfigure[{\scriptsize $h=3$, top-$50$}]{
				\includegraphics[width=0.45\columnwidth, height=2.5cm]{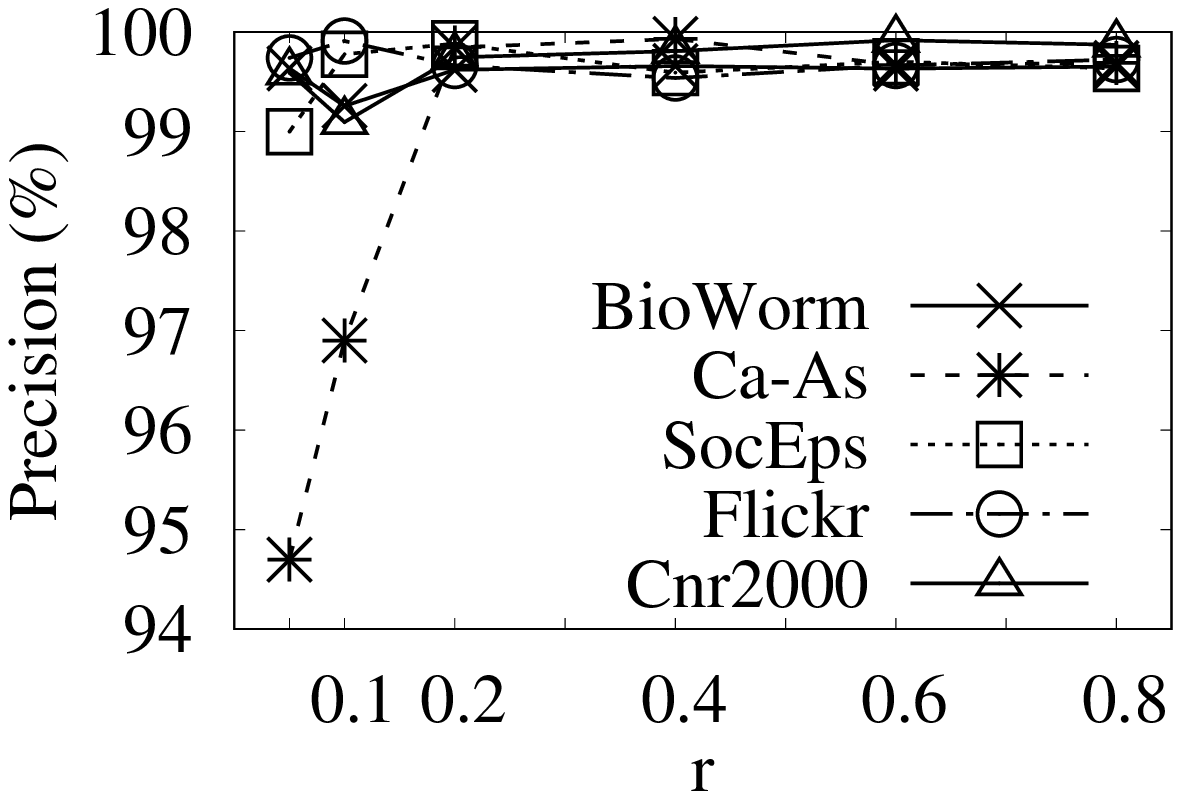}
			}\hspace*{0.3cm}
			\subfigure[{\scriptsize $h=4$, top-$50$}]{
				\includegraphics[width=0.45\columnwidth, height=2.5cm]{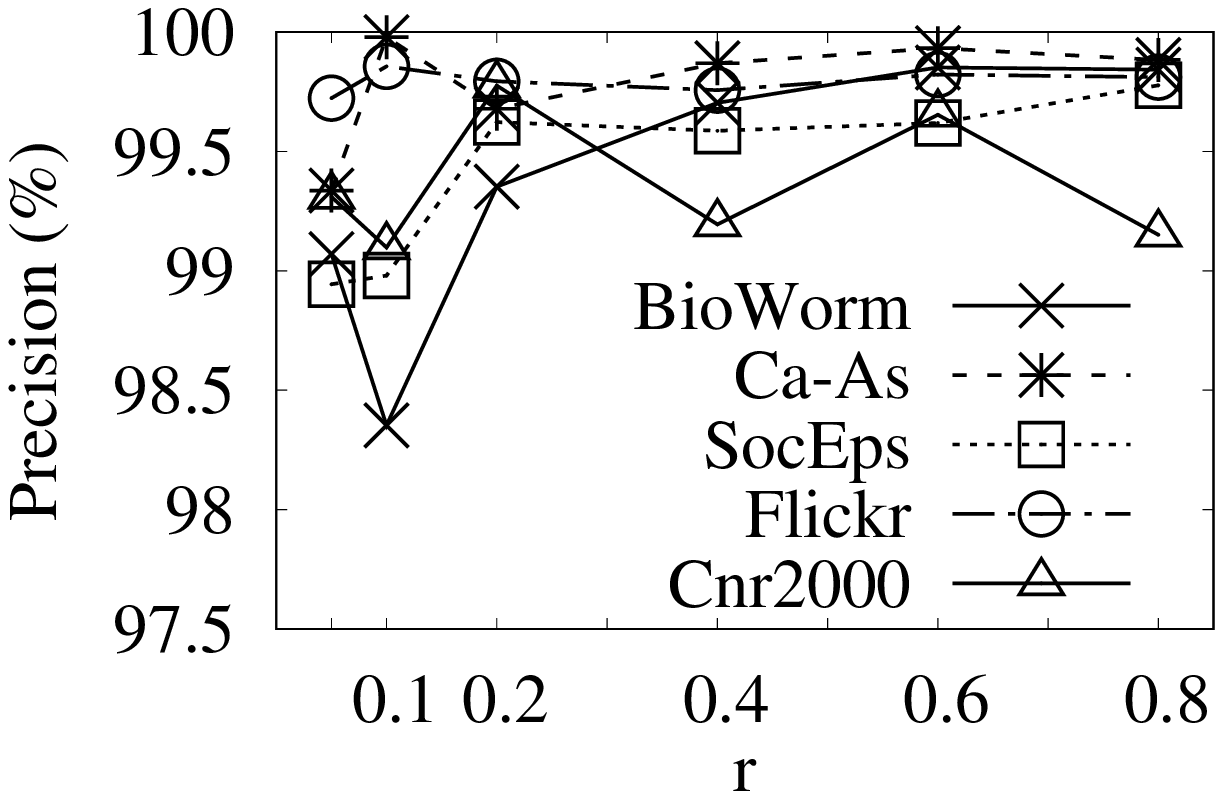}
			}\hspace*{0.3cm}
			\subfigure[{\scriptsize $h=5$, top-$50$}]{
				\includegraphics[width=0.45\columnwidth, height=2.5cm]{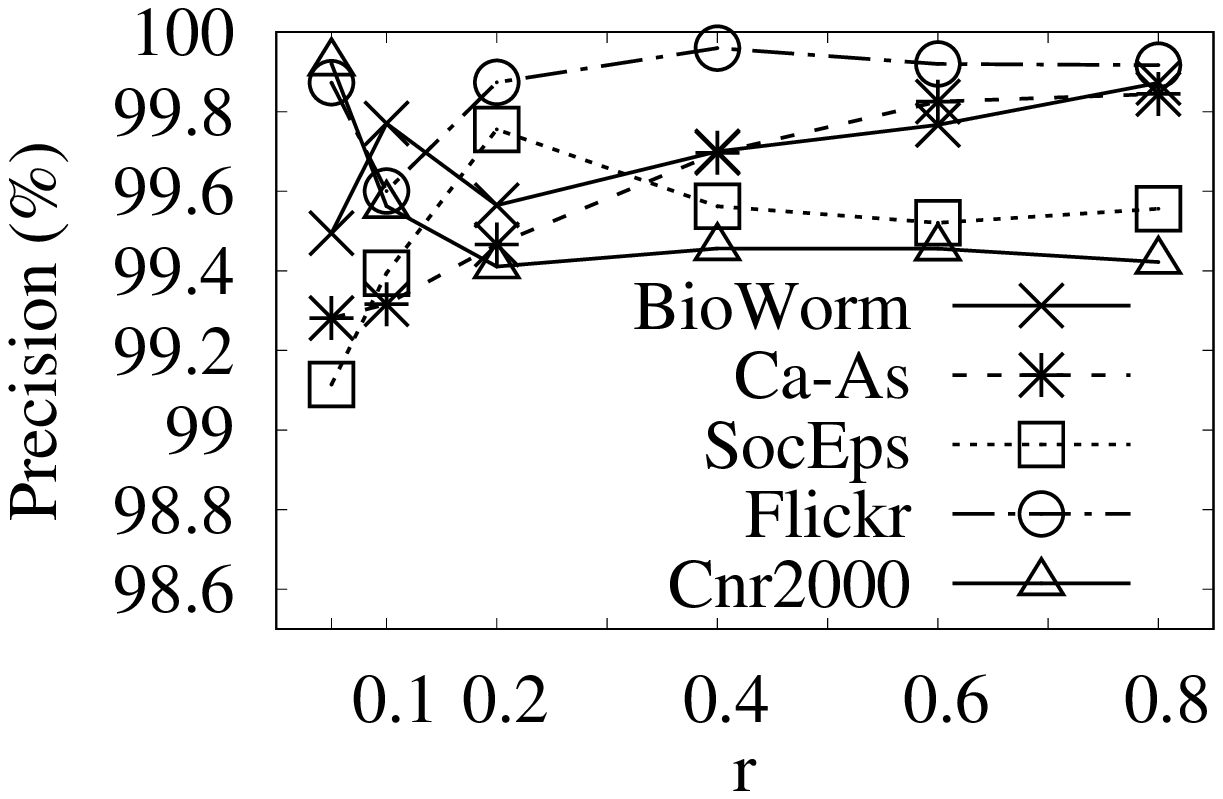}
			}
		\end{tabular}
	\end{center}
	\vspace*{-0.2cm}
	\caption{ Precisions of \khcs by only considering the top-$s$ maximal $(k,h)$-cores ($s=1$ and $s=50$)}
	\vspace*{-0.3cm}
	\label{fig:exp3:precision-vary-h-top-s}
\end{figure*}

\begin{figure}[t!]
	\begin{center}
		\begin{tabular}[t]{c}\hspace*{-0.3cm}
			\subfigure[{\scriptsize $h=2$}]{
				\includegraphics[width=0.45\columnwidth, height=2.5cm]{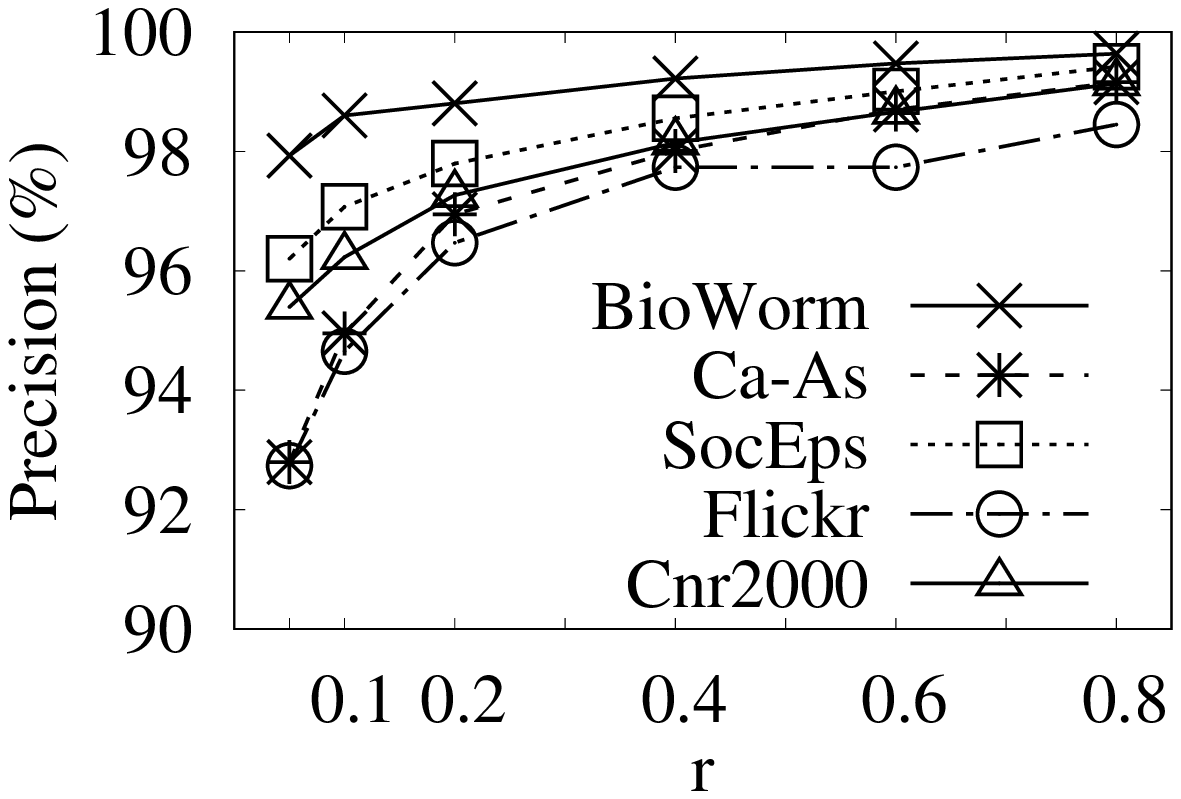}
			}\hspace*{0.3cm}
			\subfigure[{\scriptsize $h=3$}]{
				\includegraphics[width=0.45\columnwidth, height=2.5cm]{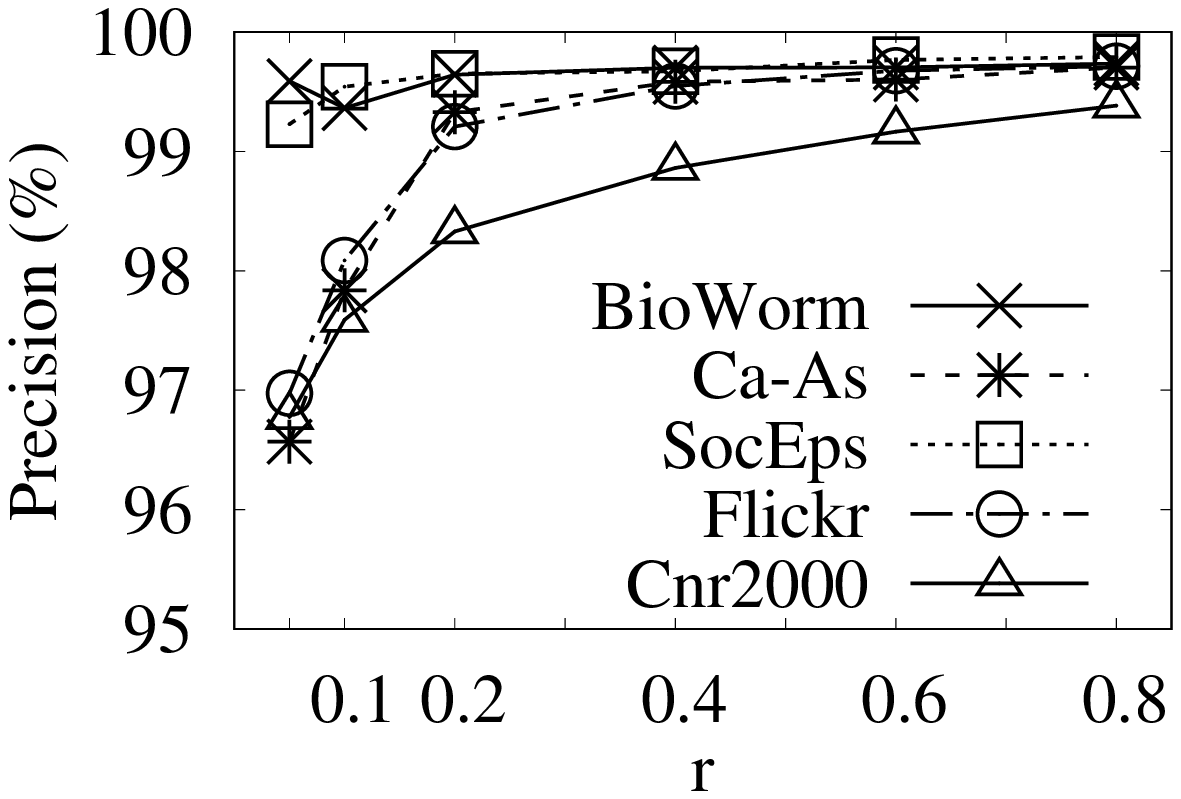}
			}\vspace*{-0.3cm}\\
			\subfigure[{\scriptsize $h=4$}]{
				\includegraphics[width=0.45\columnwidth, height=2.5cm]{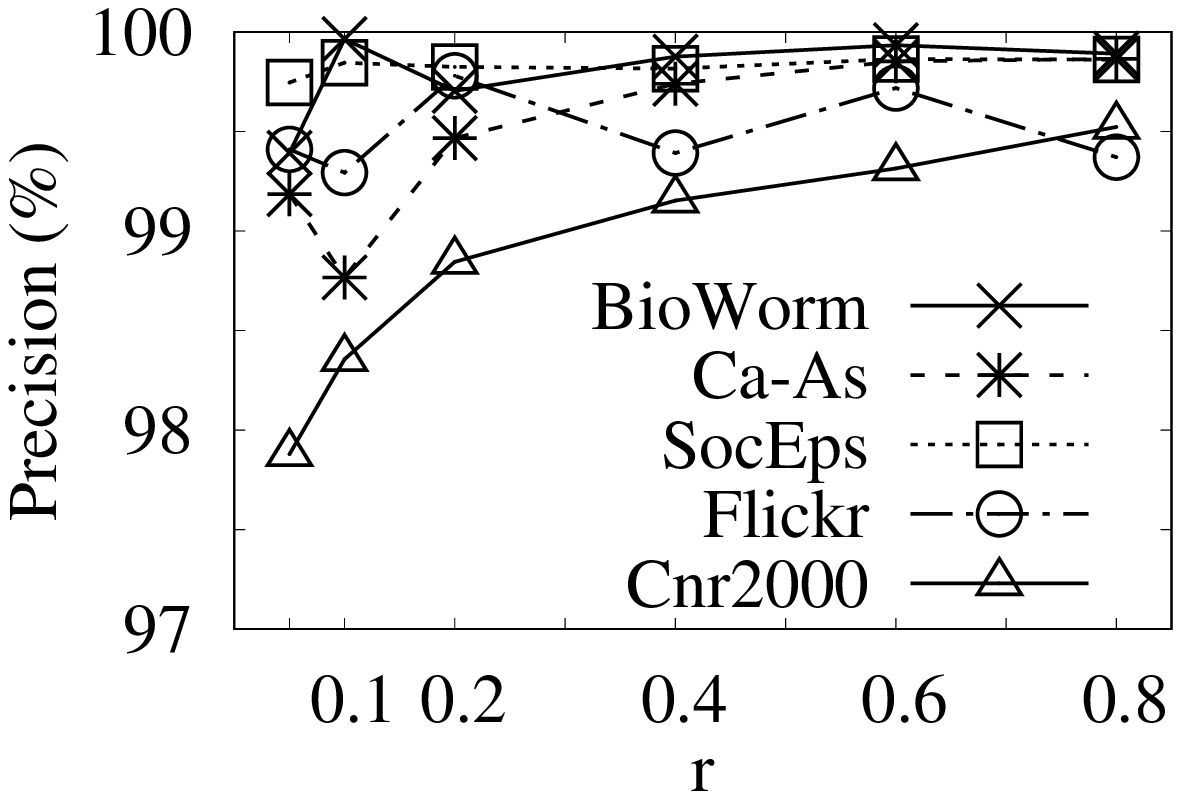}
			}\hspace*{0.3cm}
			\subfigure[{\scriptsize $h=5$}]{
				\includegraphics[width=0.45\columnwidth, height=2.5cm]{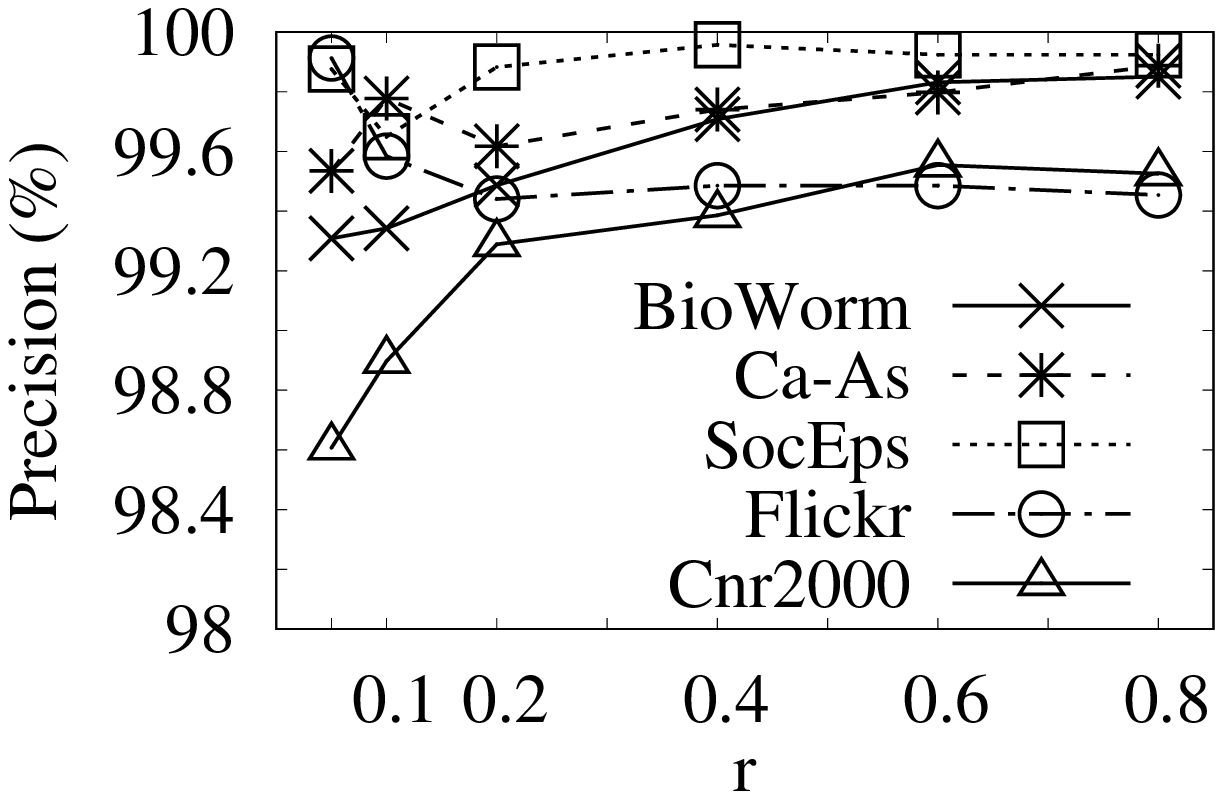}
			}
		\end{tabular}
	\end{center}
	\vspace*{-0.2cm}
	\caption{Precisions of \khcs with varying $r$}
	\vspace*{-0.3cm}
	\label{fig:exp3:precision-vary-h}
\end{figure}

\stitle{Exp-4: Precisions of \khcs with varying $r$.} In this experiment, we evaluate the precision of the \khcs algorithm with various sampling rates. Here we define the precision as follows. Let $\core_h[v]$ and $\widehat \core_h[v]$ be the exact and the estimated $(k,h)$-core number of the vertex $v$, respectively. Then, the precision of an algorithm is computed by $1-(\sum\nolimits_{v\in V}(|\core_h[v]-\widehat \core_h[v]|)/\core_h[v])/|V|$. Fig.~\ref{fig:exp3:precision-vary-h} shows the precisions of \khcs with varying $r$ on five datasets. Similar results can also be observed on the other datasets. As expected, the precisions of \khcs typically increase as $r$ increases. When $h=2$ (Fig.~\ref{fig:exp3:precision-vary-h}(a)), the precisions of \khcs are no less than 92\% on all datasets even when $r=0.05$. Moreover, with $r$ increases, the precisions can be quickly improved to 98\% on all datasets given that $h=2$. When $h \geq 3$ (Fig.~\ref{fig:exp3:precision-vary-h}(b-d)), \khcs exhibits very high precisions ($\ge 99$\%) in most cases. For example, even when $r=0.05$, the precision of \khcs is higher than 99\% with $h\ge 4$ on most datasets. These results indicate that \khcs is very accurate in practice even for a very small sampling rate (e.g., $r=0.1$).

We also evaluate the precision of the \khcs algorithm by only considering the top-$s$ maximal $(k,h)$-cores. Specifically, 
the precision of an algorithm is computed by $1-(\sum\nolimits_{v\in S}(|\core_h[v]-\widehat \core_h[v]|)/\core_h[v])/|S|$, where the $S$ is the set of vertices of the top-$s$ maximal $(k,h)$-cores. Fig.~\ref{fig:exp3:precision-vary-h-top-s} shows the precision results of \khcs on five datasets {that only considers the top-$1$ and top-$50$ maximal $(k,h)$-cores.} Similar results can also be observed on the other datasets. As expected, the precisions of {top-$s$ maximal $(k,h)$-cores also} typically increase as $r$ increases on most datasets. When $h=2$ and $s=1$ (Fig.~\ref{fig:exp3:precision-vary-h-top-s}(a)), the precisions of \khcs are no less than 95\% on all datasets except \soceps even when $r=0.05$. Moreover, with $r$ increases, the precisions can be quickly improved to 98\% on most datasets given that $h=2$. When $h \geq 3$ (Fig.~\ref{fig:exp3:precision-vary-h-top-s}(b-d)), \khcs exhibits very high precisions ($\ge 99$\%) in most cases. From Fig.~\ref{fig:exp3:precision-vary-h-top-s}(e-h), we find that the results for $s=50$ are consistent. These results further confirm that the \khcs algorithm is very accurate in practice even for a very small sampling rate (e.g., $r=0.1$).

\begin{figure}[t!] 
	\begin{center}
		\begin{tabular}[t]{c}\hspace*{-0.3cm}
			\subfigure[{\scriptsize \flic}]{
				\includegraphics[width=0.43\columnwidth, height=2.5cm]{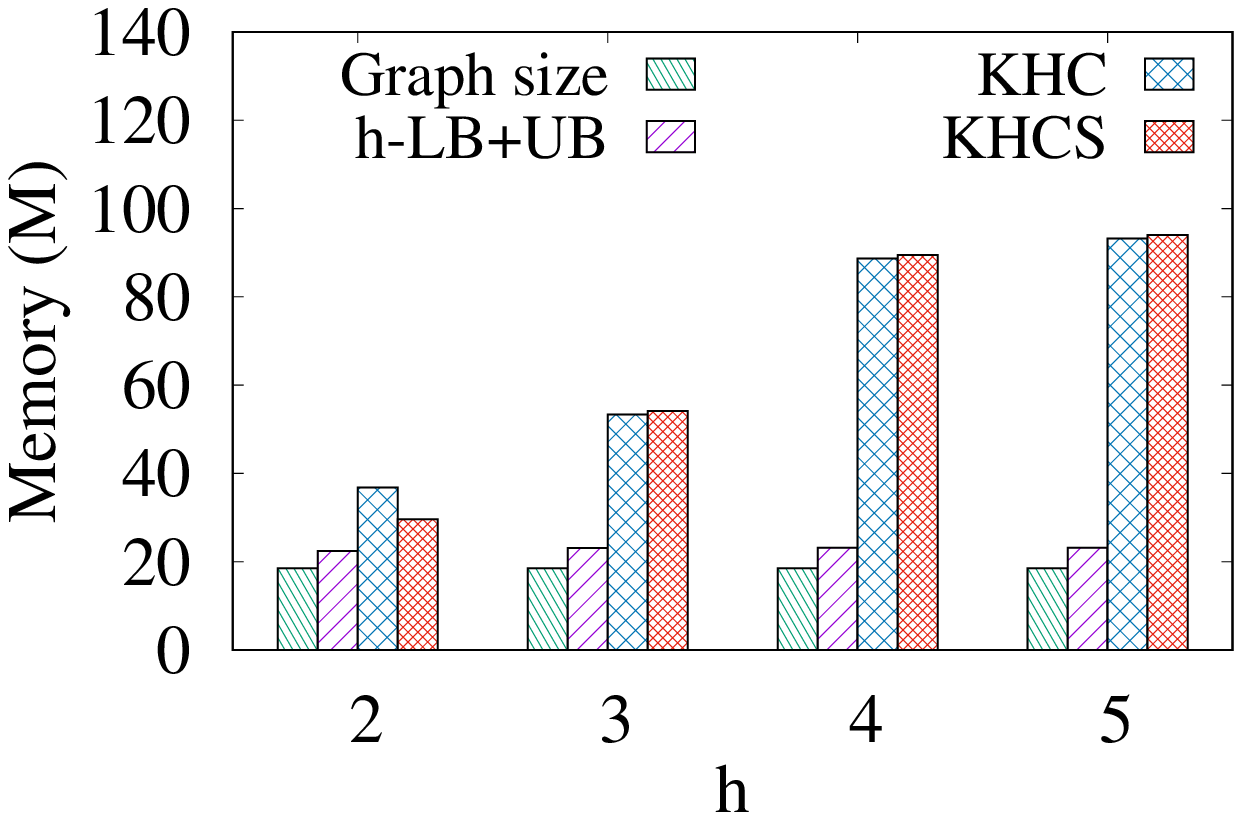}
			}		\hspace*{0.3cm}	
			\subfigure[{\scriptsize \itcnr}]{
				\includegraphics[width=0.43\columnwidth, height=2.5cm]{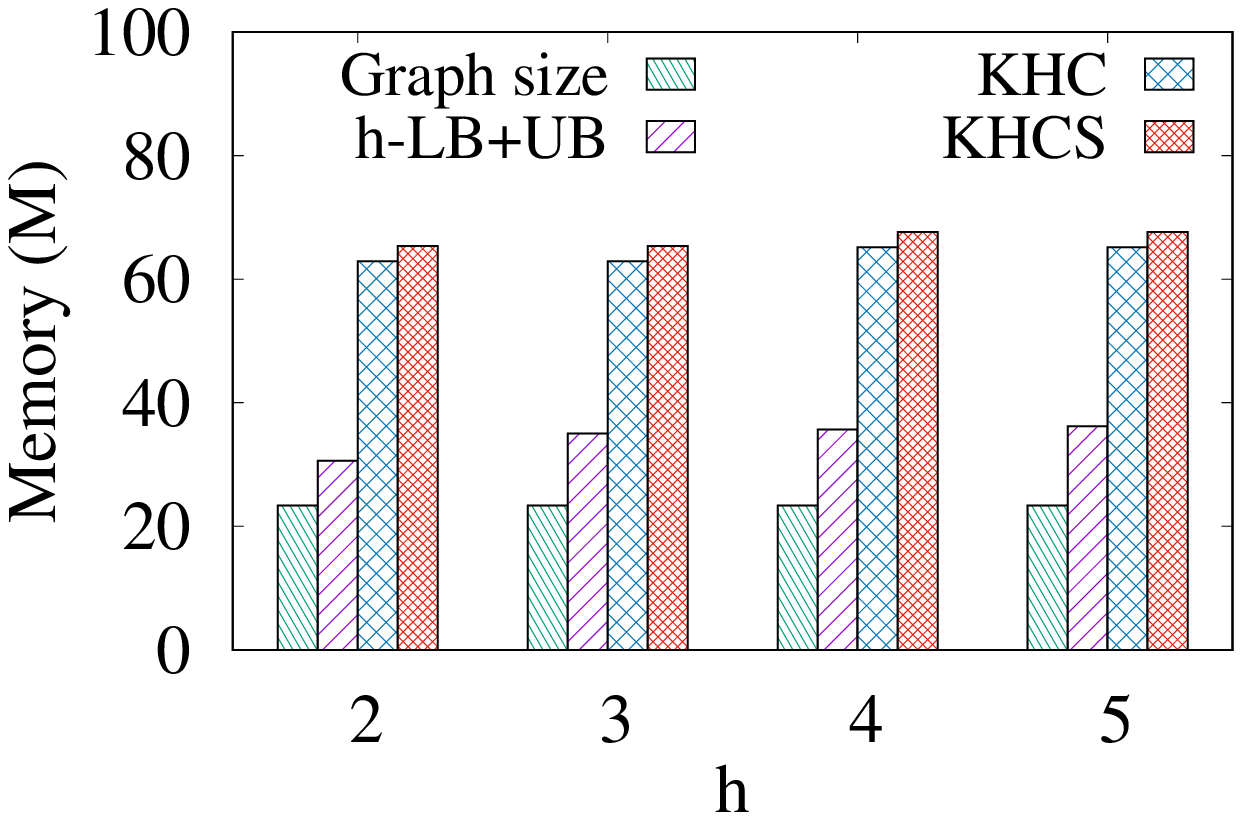}
			}
		\end{tabular}
	\end{center}
	\vspace*{-0.2cm}
	\caption{Memory overheads of various algorithms}
	\vspace*{-0.3cm}
	\label{fig:exp5:memory-overhead}
\end{figure}

\stitle{Exp-5: Memory overhead.} We compare the memory overhead of different algorithms. Fig.~\ref{fig:exp5:memory-overhead} shows the results on \flic and \itcnr, and similar results can also be obtained on the other datasets. As expected, the memory overheads of \khc and \khcs are slightly higher than that of the \hbz algorithm, because our algorithms need to maintain a \reach data structure (the \bitmaps for all vertices). Specifically, we can see that the memory usage of \hbz is less than twice of the graph size. The memory overhead of \khc and \khcs are comparable, both of which are less than 4 times of the graph size. These results indicate that our algorithms (with the \bitmap optimization technique) are space efficient for handling real-world graphs.

\begin{figure}[t!]
	\begin{center}
		\begin{tabular}[t]{c} \hspace*{-0.3cm}
			\subfigure[{\scriptsize $h=2$}]{
				\includegraphics[width=0.45\columnwidth, height=2.5cm]{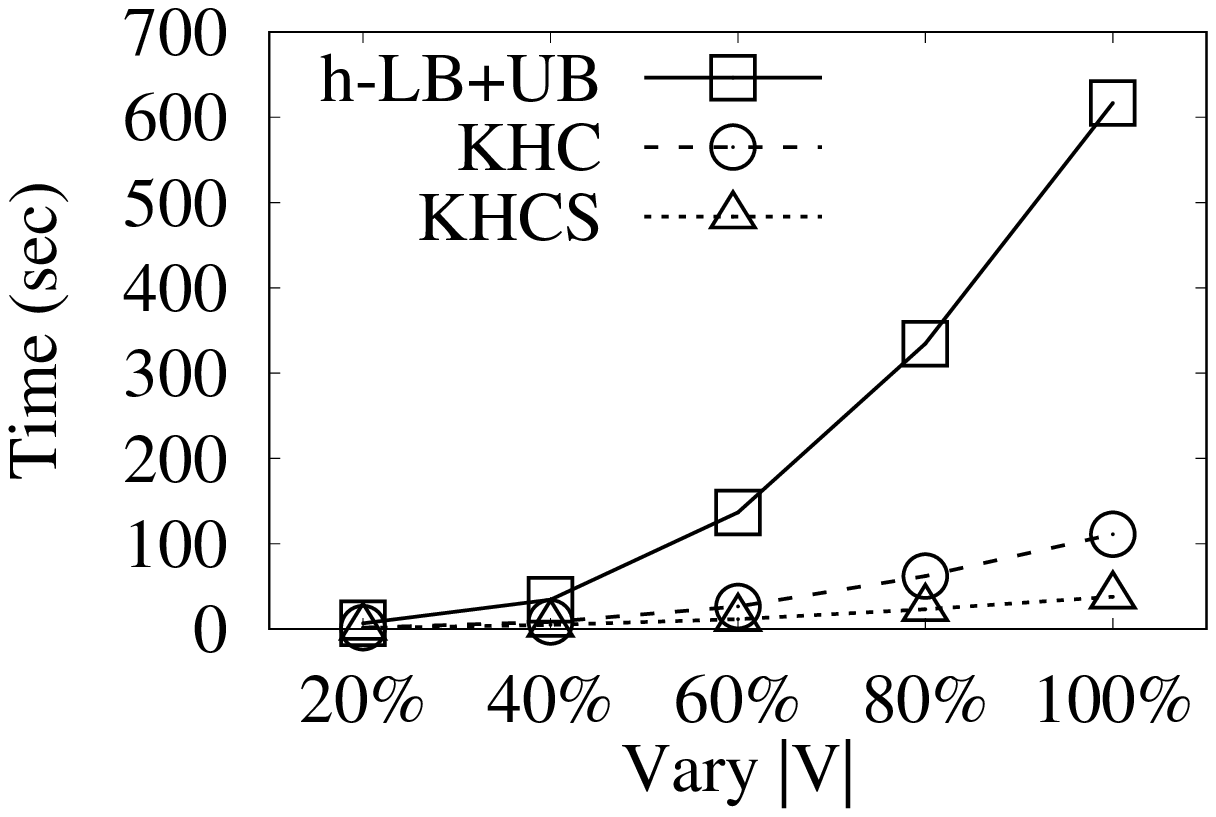}
			}\hspace*{0.3cm}
			\subfigure[{\scriptsize $h=2$}]{
				\includegraphics[width=0.45\columnwidth, height=2.5cm]{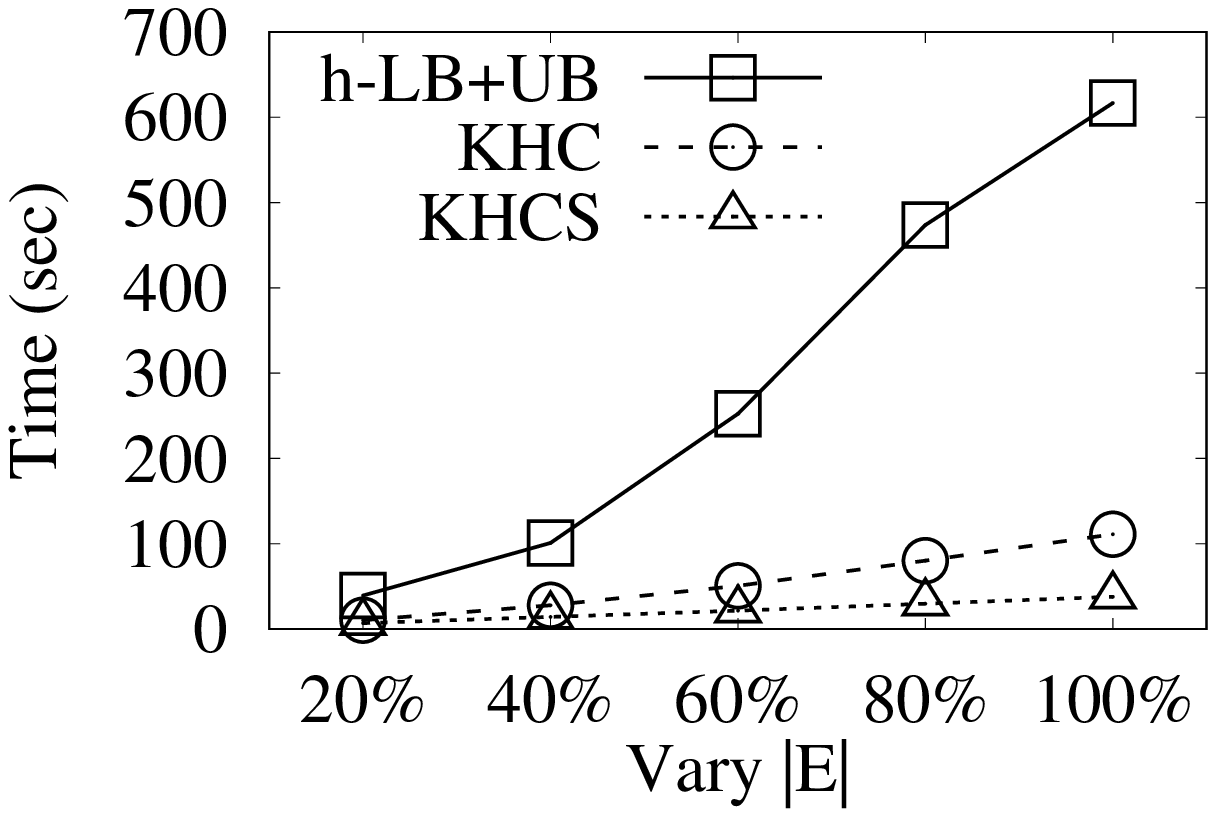}
			}\vspace*{-0.3cm}\\
			\subfigure[{\scriptsize $h=3$}]{
				\includegraphics[width=0.45\columnwidth, height=2.5cm]{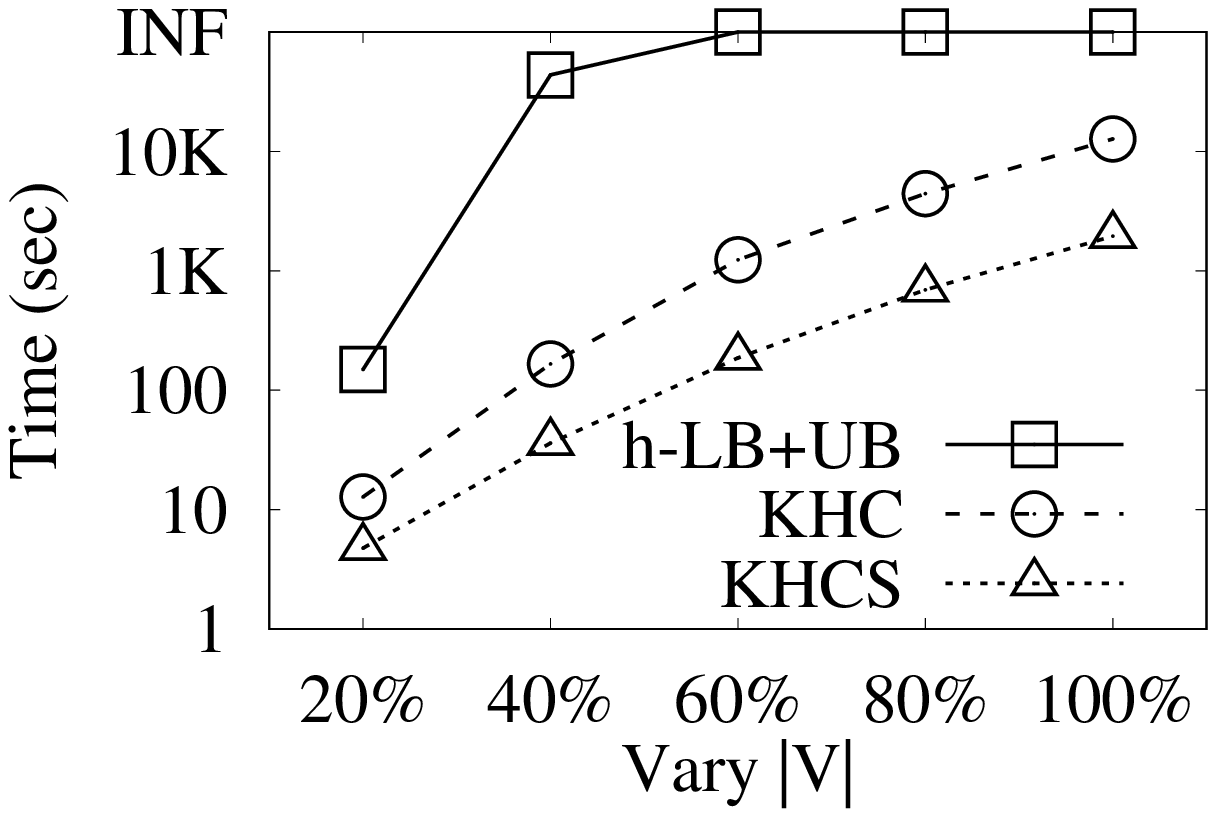}
			}\hspace*{0.3cm}
			\subfigure[{\scriptsize $h=3$}]{
				\includegraphics[width=0.45\columnwidth, height=2.5cm]{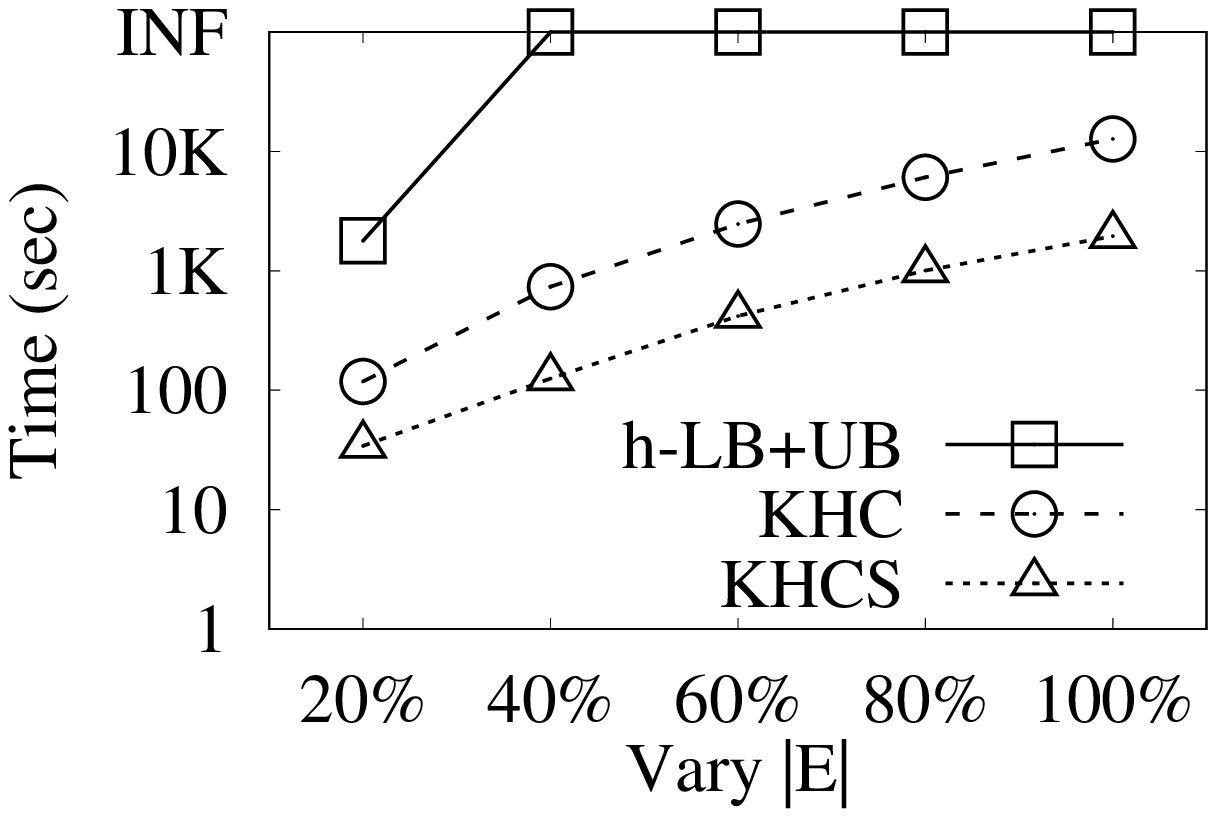}
			}
		\end{tabular}
	\end{center}
	\vspace*{-0.2cm}
	\caption{Scalability testing on the \socpokec dataset (16 threads)}
	\vspace*{-0.3cm}
	\label{fig:exp6:scalability}
\end{figure}

\stitle{Exp-6: Scalability.} Here we aim at evaluating the scalability of \hbz, \khc and \khcs, using 16 threads. To this end, we first generate eight subgraphs by randomly sampling 20-80\% of vertices and edges from the original graph respectively. Then, we evaluate the runtime of all algorithms on these subgraphs using 16 threads. The results on \socpokec with $h=2$ and $h=3$ are shown in Fig.~\ref{fig:exp6:scalability}, and the results on the other datasets and for the other $h$ values are consistent. From Fig.~\ref{fig:exp6:scalability}, we observe that the time costs of \khc and \khcs increase smoothly as $|V|$ or $|E|$ increases. The runtime of \hbz, however, increases sharply with increasing $|V|$ or $|E|$. Moreover, both \khc and \khcs significantly outperform \hbz under all parameter settings. These results suggest that both \khc and \khcs exhibit a good scalability, while \hbz shows a poor scalability when $h\ge 3$.

	\section{Related work} \label{sec:relatedwork}
	\stitle{$K$-core based models and algorithms.}  The $k$-core model was originally proposed by Seidman \cite{83kcoredef} for modeling cohesive subgraphs in an undirected network. Recently, many $k$-core based models have been proposed for modeling cohesive subgraphs on different types of networks. For example, Batagelj and Zaversnik \cite{11generalizedcoredecomposition} introduced a generalized concept of $k$-core by considering weights of the edges on weighted graphs. Bonchi et al.\ \cite{14kdduncertaincore} proposed a  $k$-core model for uncertain graphs based on a definition of \emph{reliable} degree of nodes. Li et al.\ \cite{15pvldbinfluentialcore} proposed an influential community model based on $k$-core to capture both the influence and cohesiveness of a community. Galimberti et al.\ proposed two generalized $k$-core models for multi-layer networks \cite{17cikmMultilayercore} and temporal graphs \cite{18cikmspancore}, respectively. Fang et al.\ \cite{16vldekcoreattributegraph} extended the $k$-core concept to attribute graphs. More recently, Li et al.\ \cite{18sigmodskycommtr} proposed a skyline $k$-core model for modeling communities on multi-valued networks. From the algorithmic point of view,  Batagelj and Zaversnik \cite{03omalgkcore} proposed a linear-time core decomposition algorithm. Sariy{\"u}ce et al.\ \cite{13vldbstreamcore} and  Li et al.\ \cite{14tkdecoremaintain} developed efficient algorithms for maintaining the core decomposition on dynamic graphs. Wen et al.\ \cite{16icdeioefficientcore} presented an I/O efficient core decomposition algorithm for web scale graphs. Unlike all these existing studies, we focus on developing efficient algorithms to solve the distance-generalized core decomposition problem, which was originally introduced in \cite{19sigmodBonchiKS}.
	
	\stitle{Other cohesive subgraph models.} Beyond $k$-core, there also exist many other cohesive subgraph models which have been widely used for modeling communities. Notable examples include the maximal clique model \cite{73BKalgmaximalclique,11todsmaximalclique}, the $k$-plex  model \cite{78JMSplex,15sigmodPlexes}, the  $k$-truss model \cite{05trusses,12vldbtruss,14sigmodtrusscommunity}, the nucleus model \cite{15WWWnucleus,17twebnucleus}, the locally densest subgraph (LDS) model \cite{15WWWlocaldense,15KDDlocaldense,17wwwlocaldenseconvex}, as well as the maximal $k$-edge connected subgraph ($k$-ECS) model \cite{12edbtkedgeconnected,13cikmkedgeconnected}.  Noted that the problems of enumerating all maximal cliques and all  $k$-plex subgraphs are NP-hard \cite{73BKalgmaximalclique,15sigmodPlexes}, thus they are often intractable for massive graphs. However, for the $k$-truss, the nucleus, the LDS, the $k$-ECS models, there exist polynomial-time algorithms to compute the corresponding cohesive subgraphs. Similar to these cohesive subgraph models, the $(k,h)$-core model studied in the paper can also be computed in polynomial time \cite{19sigmodBonchiKS}.
	
	\section{Conclusion} \label{sec:conclusion}
	In this paper, we propose an efficient peeling algorithm to compute the $(k,h)$-core decomposition on graphs based on a novel \hdegree updating technique. The striking feature of our algorithm is that it only needs to traverse a small induced subgraph ($G(N_v^h(G))$) to maintain the \hdegrees for all vertices after peeling a vertex $v$, instead of recomputing the \hdegrees of the vertices. We also develop an elegant \bitmap technique to efficiently implement such an \hdegree updating procedure. Additionally, we present a sampling-based algorithm and a parallelization strategy to further improve the efficiency for $(k,h)$-core decomposition. The results of extensive experiments on 12 real-world large graphs demonstrate the efficiency and scalability of the proposed algorithms. 
	
	
	\balance

\bibliographystyle{IEEEtr}
	\bibliography{khcore}
\end{document}